\documentclass[12pt,a4paper,twoside]{article}
\pdfoutput=1
\usepackage{amsmath}
\usepackage{amsfonts}
\usepackage{amssymb}
\usepackage{graphicx}
\usepackage{subfigure}
\usepackage{comment}
\usepackage{url}
\usepackage{slashed}
\usepackage{multirow}
\usepackage{threeparttable}
\usepackage{hyperref}
\usepackage[titletoc,toc,title]{appendix}
\usepackage{axodraw}
\usepackage{rotating}
\usepackage{setspace}
\usepackage[lmargin=3.5cm,rmargin=3.5cm,tmargin=3.4cm,bmargin=3.5cm]{geometry}
\usepackage{array}

\newcolumntype{L}[1]{>{\raggedright\let\newline\\\arraybackslash\hspace{0pt}}p{#1}}
\newcolumntype{C}[1]{>{\centering\let\newline\\\arraybackslash\hspace{0pt}}p{#1}}
\newcolumntype{R}[1]{>{\raggedleft\let\newline\\\arraybackslash\hspace{0pt}}p{#1}}

\hypersetup{linktocpage}

\newcommand{\MS}{M_{S}}
\newcommand{\tb}{\tan\beta}
\numberwithin{equation}{section}

\begin{document}

\begin{center}
{\Huge Signs of Susy}\\ \vspace*{10mm}
{\Large Chris Wymant}\\ \vspace*{50mm}

\vfill

{\large A thesis presented for the degree of Doctor of Philosophy,\\2013}\\ \vspace*{20mm}

Institute for Particle Physics Phenomenology\\
Department of Physics\\
University of Durham\\
UK

 \begin{figure}[!ht]
\vspace*{-15mm}
\begin{flushleft}
\includegraphics[width=0.23\linewidth]{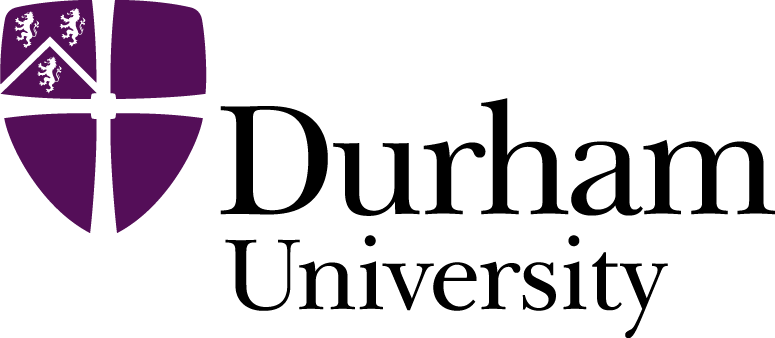}
\end{flushleft}
\vspace*{-21mm}
\begin{flushright}
\includegraphics[width=0.25\linewidth]{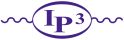}
\end{flushright}
\end{figure}
\vspace*{-5mm}
\end{center}

\newpage
\begin{center}
For my father, for giving me a map but not directions.
\end{center}

\newpage
\begin{abstract}

After a brief introduction to $21^{\rm st}$ century fundamental physics suitable for the layman with a reasonable level of mathematical competence, I introduce the concept of unnaturalness in Standard Model electroweak symmetry breaking and Supersymmetry (Susy) as a potential solution.
The {\it optimally} natural situation in Susy in light of the 2012 discovery of a Higgs boson is derived, namely that of {\it almost maximal mixing}, with the scalar top partners almost as light as can be.
The discovery is also interpreted numerically in terms of the Next-to-Minimal Supersymmetric Standard Model, with greater emphasis placed on the visibility of the Higgs boson at the observed mass, i.e. on signal strengths.
I introduce simple models of gauge-mediated Susy breaking (GMSB), and how their generalisation leads to a richer parameter space.
I then investigate the role played by the mediation scale of GMSB: this is found to be as a control of the extent to which Yukawa couplings de-tune flavour-blind relations set by gauge couplings.
Finally, issues relating to the discovery or exclusion of Susy at colliders are discussed.
Bounds are derived for the masses of new particles from Large Hadron Collider searches for excesses of jets and missing energy without leptons, and compared to constraints arising from Higgs boson searches, for models of GMSB and the Constrained Minimal Supersymmetric Standard Model.
I present a novel search strategy for new physics signatures with two neutral, stable particles, when such particles are produced by boosted decays.
(Susy examples include models with light gravitinos, pseudo-goldstinos, singlinos or new photinos.)
The method is shown to produce sharp mass peaks that enhance the visibility of the signal.

\end{abstract}

\newpage
\tableofcontents

\newpage
\section*{Declaration}
The copyright of this thesis rests with the author.
No quotation from it should be published without the author's prior written consent and information derived from it should be acknowledged.
No part of this thesis has been submitted for any other degree or qualification.
Part~\ref{CMtoQFT} and Sections~\ref{LowScaleSusyIntro}, \ref{GGMintro} and \ref{ColliderIntroduction} are introductory. 
Sections~\ref{OptimalNaturalness}, \ref{NMSSM}, \ref{RoleMessengerScale}, \ref{1/fbSearches} and \ref{MakingMostMET} are based on original research done by myself, the first of these working alone and the rest in collaboration with others, published in articles referenced prominently at the start of each Section.

\newpage
\section*{Acknowledgements}
I gratefully thank the following people.
\begin{itemize}
 \item First and foremost, my collaborators during my graduate studies: Daniel Albornoz Vasquez, Genevi\`{e}ve B\'{e}langer, C\'{e}line B\oe{}hm, Jonathan Da Silva, Christoph Englert, David Grellscheid, J\"{o}rg J\"{a}ckel, Peter Richardson, Michael Spannowsky, and especially my supervisor Valya Khoze.
 Thank you for your ideas and contributions to our shared projects, and for many enlightening discussions.
 \item Steven Abel and Ben Allanach for their detailed examination of this thesis and the resulting suggestions.
 \item Jeppe Andersen, Matt Dolan, Claude Duhr, Ilan Fridman Rojas, Hendrik Hoeth, Boaz Keren-Zur, Sabine Kraml, Daniel Ma\^{\i}tre, Alberto Mariotti, Michael Schmidt and Pietro Slavich, for helpful conversations.
 \item Howie Baer, Sven Heinemeyer, Kees Jan De Vries, Thomas Rizzo, Marc Sher, David Shih, Oscar St\aa{}l, Carlos Wagner, and particularly Olivier Mattelaer, Tim Stefaniak and Karina Williams, for useful correspondence.
 \item Adrian Signer for teaching me Susy with a healthy scepticism from the start.
 \item Frank Krauss for a casual comment taken seriously: theorists wanting to be listened to by experimentalists should suggest new signals.
 \item Mike Hobson, Julia Riley and David Tong for inspiring undergraduate teaching, and Lukas Witkowski for a shared enthusiasm.
 \item Other participants and particularly the organisers of both the ``Implications of a 125 GeV Higgs boson'' workshop at LPSC Grenoble January 2012, and the Carg\`{e}se International School 2012, for stimulating environments that gave birth to ideas in this thesis.
 \item Trudy, Linda and Mike for help with logistics and computing.
 \item Jiri, for keeping me sane while I was running Durham men's volleyball, and Tom for keeping me sane while I wasn't.
 \item All of my family, particularly my sister, mother and father for much support over many years (and my Nana for a postgraduate course in cooking!).
 \item Surtout Catherine.
 \end{itemize}
This work was supported by the STFC.
 
\section*{An Apology}
Throughout this work I will generally refer to the massive scalar boson associated with electroweak symmetry breaking as {\it the Higgs} rather than {\it the Higgs boson}, simply because the latter sounds awkward to my ears in passages when references to this particle come thick and fast.
My apologies to Peter Richardson and any others who take exception to this.

\newpage
\section*{Lists Of Figures And Tables}
Figures which benefit from being printing in colour are found on pages
\pageref{fig:ConstHiggsMasses}
\pageref{fig:HiggsMassesNLO}
\pageref{fig:NMSSMHiggsmass}
\pageref{fig:gggg_nmssm}
\pageref{fig:mh12}
\pageref{fig:gggg_all}
\pageref{unifiedLambdas}
\pageref{fig:VaryingThirdGenMmess}
\pageref{fig:theor}
\pageref{fig:bands}
\pageref{1fbMainResults}
\pageref{fig:validate}
\pageref{fig:CMSSM-GGM-comb}
\pageref{fig:GGM-LS-LG-7}
\pageref{fig:1TeV}
\pageref{fig:BoostedOrNot} and
\pageref{fig:gluino}.
Other pages in this thesis can happily be printed in black and white with no loss of information.

\begin{table}[!h]
\centering
\begin{tabular}[t]{C{1.2cm} C{0cm} C{2.5cm} }
Figure & & Page Number \\
\hline
\ref{fig:RatioNLO} &  & \pageref{fig:RatioNLO} \\
\ref{fig:ConstHiggsMasses} &  & \pageref{fig:ConstHiggsMasses} \\
\ref{fig:HiggsMassesNLO} &  & \pageref{fig:HiggsMassesNLO} \\
\ref{fig:NMSSMHiggsmass} &  & \pageref{fig:NMSSMHiggsmass} \\
\ref{fig:fatjets} &  & \pageref{fig:fatjets} \\
\ref{fig:gggg_nmssm} &  & \pageref{fig:gggg_nmssm}  \\
\ref{fig:mh12} &  & \pageref{fig:mh12}  \\
\ref{fig:SignalStrength_V_BSMdecays} &  & \pageref{fig:SignalStrength_V_BSMdecays} \\
\ref{fig:distrib} &  & \pageref{fig:distrib} \\
\ref{fig:gggg_all} &  & \pageref{fig:gggg_all} \\
\ref{fig:heavyneut} &  & \pageref{fig:heavyneut}  \\
\ref{unifiedLambdas} &  & \pageref{unifiedLambdas}  \\
\ref{MatchingLambdasCartoon} &  & \pageref{MatchingLambdasCartoon}  \\
\ref{fig:VaryingThirdGenMmess} &  & \pageref{fig:VaryingThirdGenMmess}  \\
\ref{fig:theor} &  & \pageref{fig:theor}  \\
\ref{fig:bands} &  & \pageref{fig:bands} \\
\ref{1fbMainResults} &  & \pageref{fig:GGM-sq-gl-plane} \\
\ref{fig:validate} &  & \pageref{fig:validate}  \\
\ref{fig:CMSSM-GGM-comb} &  & \pageref{fig:CMSSM-GGM-comb}  \\
\ref{fig:GGM-LS-LG-7} &  & \pageref{fig:GGM-LS-LG-7}  \\
\ref{InvisDecays} &  & \pageref{InvisDecays}  \\
\ref{MT2eg} &  & \pageref{MT2eg} \\
\ref{fig:1TeV} &  & \pageref{fig:1TeV}  \\
\ref{fig:Horns} &  & \pageref{fig:Horns} \\
\ref{fig:Diagram} &  & \pageref{fig:Diagram} \\
\ref{fig:BoostedOrNot} &  & \pageref{fig:BoostedOrNot} \\
\ref{fig:gluino} &  & \pageref{fig:gluino} \\
\ref{fig} &  & \pageref{fig}
\end{tabular}
\hspace*{2cm}
\begin{tabular}[t]{C{1.2cm} C{0cm} C{2.5cm} }
Table & & Page Number \\
\hline
\ref{SMcontent} &  & \pageref{SMcontent} \\
\ref{MSSMcontent} &  & \pageref{MSSMcontent} \\
\ref{SU5whole} &  & \pageref{SU5whole} \\
\ref{2/3split} &  & \pageref{2/3split} \\
\ref{RGIsTable} &  & \pageref{RGIsTable} \\
\ref{tab:signal_regions} &  & \pageref{tab:signal_regions}
\end{tabular}
\end{table}

\newpage
\subsection*{Abbreviations Used In The Main Text}
BSM -- beyond the Standard Model\\
c.c. -- complex conjugate\\
CMSSM -- Constrained Minimal Supersymmetric Standard Model\\
Eq. -- Equation\\
EWSB -- electroweak symmetry breaking\\
Fig. -- Figure\\
GGM -- general gauge mediation\\
GMSB -- gauge mediated supersymmetry breaking\\
GUT -- grand unified theory\\
irrep -- irreducible representation (of a group)\\
LEP -- Large Electron-Positron (Collider)\\
LHC -- Large Hadron Collider\\
LHS -- left-hand side (of an equation)\\
LO -- leading order (usually in the sense of a perturbative Feynman-diagrammatic calculation)\\
LSP -- lightest supersymmetric particle\\
MET -- missing transverse energy, \mbox{$\slashed{E}_T = |\slashed{\mathbf{p}}_T|$}\\
MSSM -- Minimal Supersymmetric Standard Model\\
NGB -- Nambu-Goldstone boson\\
NLO -- next-to-leading order (usually in the sense of a perturbative Feynman-diagrammatic calculation)\\
NLSP -- next-to-lightest supersymmetric particle\\
NMSSM -- Next-to-Minimal Supersymmetric Standard Model\\
pNGB -- pseudo-Nambu-Goldstone boson\\
QCD -- quantum chromodynamics\\
QFT -- quantum field theory\\
RG -- renormalisation group\\
RGI -- renormalisation group invariant\\
RHS -- right-hand side (of an equation)\\
SM -- Standard Model\\
Susy -- Supersymmetry\\
UV -- ultraviolet (in the sense of high energy scales)\\
VEV -- vacuum expectation value\\
WMAP -- Wilkinson Microwave Anisotropy Probe\\

\newpage
\subsection*{A Non-Selective Glossary Of Important/Confusing Terms For BSM/Collider Phenomenology}
Terms explained elsewhere in the glossary are italicised.\\ \\
{\it Acceptance} -- the fraction of {\it events} which pass the {\it cuts} in a given experimental analysis.\\ \vspace*{-3mm}\\
{\it Background} -- anything which is not the physics one is trying to see (usually new physics), but unfortunately looks like it.\\ \vspace*{-3mm}\\
{\it Background, reducible} -- a {\it background} consisting of misidentified or mismeasured particles, which could therefore be eliminated (in principle) by perfect measurement.\\ \vspace*{-3mm}\\
{\it Background, irreducible} -- a {\it background} with truly the same {\it final state}, though produced via a different intermediate particle/particles.\\ \vspace*{-3mm}\\
{\it Cuts} -- the set of requirements we choose to impose on the {\it final state} and {\it phase space} in order to significantly reduce the number of uninteresting {\it events} without decreasing the number of interesting {\it events} too much.\\ \vspace*{-3mm}\\
{\it A Decade (of RG-running)} -- a factor of ten in between two energy scales.
For example after two decades of running starting at an energy scale $Q$ one would be at a scale $10^2 Q$ (or $10^{-2} Q$).\\ \vspace*{-3mm}\\
{\it Decay cascade} -- the set of one or more decay steps from the initially produced particles to the {\it final state}.\\ \vspace*{-3mm}\\
{\it Event} -- a single collision of particles (specifically protons at the LHC) together with what happens as a result of the collision.
For simulated collisions, what happens at an intermediate stage can be forced; for example one may simulate $x$ events where particles collide and produce a Higgs boson which decays arbitrarily.
Compare with experiment where requirements can only be imposed on detected particles, not intermediates, by definition.
For example in $x$~fb of data (which corresponds to a certain number of collisions $y$, i.e. $y$ total events), after focussing only on events where the detected particles meet certain interesting criteria one is left with $z$ events ($z<y$).
In this case unlike the simulated case one cannot say with certainty what happened in any given event, and only statistical statements can be made.
Observed events which are more likely to have proceeded via intermediate new physics rather than established physics (i.e. the {\it background}) are described as candidate events.\\ \vspace*{-3mm}\\
{\it Final state} -- the set of outgoing particles (i.e. those that don't decay further before being detected) resulting from a collision.
Can be understood either in a precise sense more appropriate the calculation of amplitudes, such as a final state of one electron, one electron neutrino and a $u\bar{u}$ quark pair; or in a broad sense more appropriate at the detector level, such as one charged lepton, missing energy and (a perhaps unspecified number of) jets.
In either of these senses the term final state usually refers just to the identity and {\it multiplicity} of the particles involved.
However sometimes the term is taken to include some information about the {\it phase space} as well, e.g. `a final state of {\it hard} jets'.\\ \vspace*{-3mm}\\
{\it Hard} -- energetic.
A hard scattering is one which involves a large exchange of energy, for example the production of heavy particles in a collision; a hard object is an object with large three-momentum (or simply large transverse momentum at hadron colliders where $z$ momentum is less relevant).
In both cases the antonym is {\it soft.}\\ \vspace*{-3mm}\\
{\it High-level objects} -- isolated leptons, isolated photons, jets (as opposed to the individual hadrons inside), missing momentum, all defined within the spatial coverage of the detector.\\ \vspace*{-3mm}\\
{\it K-factor} -- the ratio of a cross-section (or possibly some other calculable observable) calculated at a given order in perturbation theory to the same quantity calculated at the order below.
Usually the ratio of NLO to LO is implied.\\ \vspace*{-3mm}\\
{\it Matrix element} -- an element of the $S$ matrix, i.e. the amplitude for a given initial state to interact and produce some specific particles.\\ \vspace*{-3mm}\\
{\it Multiplicity} -- number of. For example, jet multiplicity = number of jets.\\ \vspace*{-3mm}\\
{\it Phase space} -- the space of all possible three-momenta for the (on-shell) {\it final state} particles.\\ \vspace*{-3mm}\\
{\it Signal strength (in a given final state)} -- production cross-section multiplied by branching ratio (into the given {\it final state}), possibly also multiplied by the relevant {\it acceptance}, and possibly normalised to some expectation.\\ \vspace*{-3mm}\\
{\it Soft} -- see {\it hard}.


\newpage
\part{Prelude: From Classical Mechanics To Quantum Field Theory} \label{CMtoQFT}
\vspace*{-2mm}
{\Large (A Sketch For The Lay-Reader Familiar With Calculus And Vectors)}\\

\begin{center}
\begin{tabular}{ c c c }
\multicolumn{3}{ c }{\hspace*{15mm}\Large{$\xrightarrow{v\; \slashed{\ll} \;c}$}} \\
\multirow{2}{*}{$S \: \slashed{\gg} \:\hbar$ \rotatebox[origin=c]{270}{$\longrightarrow$}} & Classical Mechanics & Special Relativity \\
& Quantum Mechanics & Quantum Field Theory 
\end{tabular}
\end{center}

\subsection{Classical Mechanics} \label{ClassicalMech}

A reformulation of Newtonian mechanics, where an object of mass $m$ experiencing a force $F$ undergoes an acceleration $a=F/m$, is as follows.
The {\it Lagrangian} $L$ is defined to be the difference between the kinetic energy $T$ and the potential energy $V$, and is a function of the coordinate\footnote{
We consider systems with more than one coordinate, e.g. multiple particles or one particle able to move through more than one dimension, in the same way as presented here but replacing $q\rightarrow \mathbf{q} = (q_1,q_2,\ldots)$.
I will focus on a single $q$ for clarity when introducing functionals.
} $q$ of the system (e.g. the position of a particle) and how rapidly that coordinate is changing with time $\dot{q}$ (the dot denoting one time derivative):
\begin{equation}
L(q,\dot{q})=T-V
\end{equation}
The {\it Action} $S$ is the integral of the Lagrangian over the time interval we are interested in -- from $t_1$ to $t_2$.
It is therefore a {\it functional} of $q(t)$ -- an infinite-dimensional function in a sense, in that it depends on the value of the function $q(t)$ at each of the (infinitely many) instants between $t_1$ and $t_2$.
\begin{equation} \label{action}
S[q(t)]=\int_{t=t_1}^{t_2} L(q,\dot{q})\,dt
\end{equation}
Of all possible time-evolutions of the system $q(t)$, the one chosen by our universe is the one which minimises the action: the functional derivative\footnote{
To understand what a functional derivative is, a vector example helps.
Take a single number which is a function of a vector $\mathbf{x}$, e.g. the dot product of the vector with some other vector: $\sum_i x_i b_i$.
We can differentiate this single number with respect to the vector $x_i$ and the result is a vector: $b_i$.
Writing it this way, with the index $i$ unspecified, allows us to express simultaneously the different results that arise when we choose to differentiate with respect to different components of the vector $\mathbf{x}$.
A function $q(t)$ is an infinite-dimensional vector of sorts -- there is a different value associated with each `index' $t$, with $t$ continuous.
By analogy with the vector example, when we differentiate a functional with respect to a function, the result -- a functional derivative -- is a function.
}
of $S$ with respect to $q(t)$ vanishes:
\begin{equation} \label{ActionExtremised}
\frac{\delta S}{\delta q}= 0
\end{equation}
The solution of~\eqref{ActionExtremised} -- the {\it equation(s) of motion} -- can be obtained by solving the {\it Euler-Lagrange equation}:
\begin{equation} \label{EulerLagrange}
\frac{\partial L}{\partial q} - \frac{d}{dt}\frac{\partial L}{\partial \dot{q}} = 0,
\end{equation}
where $q$ and $\dot{q}$ are considered independent for the partial differentiation.
As an example, the familiar dynamics of a single particle of mass $m$ in a potential $V$ in one dimension $x$ are obtained from the following Lagrangian:
\begin{gather}
L = \tfrac{1}{2} m \dot{x}^2 - V(x) \\
\xrightarrow{\eqref{EulerLagrange}} \quad m\ddot{x} = -\frac{dV}{dx}
\end{gather}
This shows how spatial variation of the potential forces the particle to accelerate towards regions of lower energy.

\subsection{Special Relativity}
Special relativity tells us that when velocities become {\it relativistic}, i.e. non-negligible fractions of the speed of light $c$, we leave the regime of Newtonian mechanics.
A further startling prediction relevant even at non-relativistic velocities is that mass is merely another form of energy -- the now familiar relation
\begin{equation}
 E = mc^2
\end{equation}
We can study the relativistic dynamics of individual objects, approximated as particles: we find that time in a reference frame in motion relative to our observations passes more slowly, and that space contracts along the direction of motion.
By analogy with the mixing in the directions called `left', `right', `in front' and `behind' as one spins -- these quantities are not absolute but depend on the direction one is facing -- time and space are not absolute, indeed they are not even separate: they are mixed together by motion.
We can thus think of them as separate parts of the same thing: {\it spacetime}, $x^{\mu=0,1,2,3} = (t,\mathbf{x}) = (t,x,y,z)$.

The context of special relativity is a good one to introduce the concept of a {\it field}.
Essentially, a field is just something which is a function of position in space: in our three dimensions of space, it's a function of $x$, $y$ and $z$, and it can also change with time.
A {\it scalar field} is a field characterised by a single number as a function of space and time.
For example, one could describe the spatial and temporal variation of temperature with a scalar field -- one number at each point, giving the local temperature at that moment.
A {\it vector field} is a field characterised by a vector -- that is to say a single number with an associated direction, or equivalently a set of numbers -- as a function of space and time.
An example is the gravitational field: everywhere in the space around a planet, or a star (everywhere at all in fact), there is a number characterising how strong gravity is there, and a direction that the force acts in.
A localised disturbance in a field may propagate elsewhere as a wave, or it may simply slowly recover to how it was before.
Compare prodding the surface of water to prodding the surface of treacle.

The behaviour of fields can be captured quantitatively using a Lagrangian description as in Section~\ref{ClassicalMech}.
Where previously the Lagrangian was a simple function of the system's coordinate $q$ and its time derivative $\dot{q}$, now it is an integral over the spatial region of interest $V$ of a quantity defined at each point in space: the {\it Lagrangian density} $\mathcal{L}$, with density meaning per unit volume.
$\mathcal{L}$ is a function of the field and its spacetime derivatives; the latter are obtained with the operator $\partial_{\mu}=\frac{\partial}{\partial x^{\mu}}$.
For example, for a {\it scalar} (just one number) field $\phi$ in three spatial dimensions $\mathbf{x}$:
\begin{equation}
\begin{split}
\phi = &\; \phi(x^{\mu}) \\
\mathcal{L} = &\; \mathcal{L}(\phi, \partial_{\mu} \phi) \\
 L= &\;\int_{\mathbf{x}\in V} \mathcal{L}\,d^3\mathbf{x}
\end{split}
\end{equation}
Otherwise the equations of motion are obtained almost exactly as in the previous section:
\begin{gather}
S[\phi]= \int_{t = t_1}^{t_2} L\,dt=\int_{t=t_1}^{t_2} \int_{\mathbf{x}\in V}\mathcal{L}\:d^4x, \\
\frac{\delta S}{\delta \phi} = 0 \\
\implies \frac{\partial\mathcal{L}}{\partial \phi} - \sum_\mu \partial_{\mu} \frac{\partial\mathcal{L}}{\partial (\partial_{\mu} \phi)} = 0 \label{EulerLagrangeFields}
\end{gather}
As we take $t_1\rightarrow -\infty, t_2 \rightarrow +\infty$ and $V$ to be all of three-dimensional space $V\rightarrow\mathbb{R}^3$, in other words if we are interested in all spacetime, the action is a {\it Lorentz invariant} -- the same number for all observers regardless of their relative motion.

An Example: a {\it free} (non-interacting), massive scalar field $\phi$:
\begin{gather}
 \mathcal{L} = \sum_{\mu,\nu} \tfrac{1}{2} g^{\mu\nu} \partial_{\mu} \phi \:\partial_{\nu} \phi - \tfrac{1}{2} m^2 \phi^2 \\
\xrightarrow{\eqref{EulerLagrangeFields}} \sum_{\mu,\nu} g^{\mu\nu} \partial_{\mu} \partial_{\nu} \phi + m^2 \phi = 0 \label{EulerLagrangeForScalar},
\end{gather}
where $g^{\mu\nu}$ is the {\it metric} of spacetime, which defines how to contract two spacetime vectors together to obtain a Lorentz invariant\footnote{
A more familiar example of this kind of object is the metric of ordinary three-dimensional space, the three-by-three unit matrix $g_{ij} = \text{diag}\{+1,+1,+1\}$.
When we contract two vectors together with this metric, $\mathbf{a}.\mathbf{b} = \sum_{i,j} g_{ij}a^i b^j = \sum_i a^i b^i$, the result is invariant under rotations of the space, i.e. $SO(3)$ transformations.
}: it is the diagonal $4\times4$ matrix $g^{\mu\nu} = g_{\mu\nu} = \text{diag}\{+1,-1,-1,-1 \}$.
We can solve~\eqref{EulerLagrangeForScalar} by first taking the {\it Fourier transform} of $\phi(x^{\mu})$: expressing it as an arbitrary linear superposition (actually integrating rather than summing) of $\exp(i\sum_{\mu,\nu}g_{\mu\nu}p^{\mu}x^{\nu})$ terms with $i$ is the imaginary number $\surd\! -\!1$.
Plugging this into~\eqref{EulerLagrangeForScalar} we find the constraint that $p^{\mu} = (\surd(m^2 + \mathbf{p}^2), \mathbf{p})$, with $\mathbf{p} = (p_x,p_y,p_z)$ arbitrary and the linear superposition of terms still arbitrary.
Each term in this superposition taken singly represents a wave with wave-vector/momentum $\mathbf{p}$ and an associated energy $E=\surd(m^2 + \mathbf{p}^2)$.

Note that the energy does not vanish as the momentum vanishes: \mbox{$E\xrightarrow{\mathbf{p}\rightarrow0}m$}.
This energy gap between the vacuum (zero energy) and the lowest energy mode (the limit $\mathbf{p}\rightarrow0$) is the mass of the field.
The situation is the same as for particles: there is a minimum energy $E=mc^2$ needed to create a stationary particle, and further energy becomes its kinetic energy.

\subsection{Quantum Mechanics}
If all speeds that we encounter in our problem are well below $c$, we do not need to worry about extending classical mechanics with special relativistic effects.
However if we consider actions (recall Eq.~\eqref{action}) so small that they're comparable to the fundamental constant of nature $\hbar$ -- the {\it reduced Planck constant} -- we enter the realm of quantum mechanics.
Here we discover that there is an inherent uncertainty at the microscopic level; for example a particle cannot possess a well-defined position and momentum simultaneously.
The state of a system is described by a {\it wave function}, often denoted $|\psi\rangle$.
For a single particle this could be a probability distribution for its position and momentum.
We associate with each physical observable an {\it operator}\footnote{
A function is something which takes a number and returns another number; e.g. ``$2x$'' takes any number and returns that number doubled.
An operator, simply understood, takes a function and returns another function; e.g. ``$d/dx$'' takes any function of $x$ and returns the first derivative.
This definition of an operator makes sense in the quantum mechanical context when we consider wave functions $|\psi\rangle$ that are simple functions, such as position/momentum probability distribution functions: the operator acting on $|\psi\rangle$ returns some other function.
We may consider wavefunctions that are less readily understood as simple functions, however, such as those describing a particle's intrinsic spin.
In this context the more general definition of an operator -- a mapping from one vector space to another -- is appropriate: a quantum mechanical operator is a mapping from the space of all possible wavefunctions onto that same space (but not necessarily the same point in that space!).
The {\it eigenvectors} of an operator are points in the relevant space which are mapped back onto themselves, multiplied by a constant called the {\it eigenvalue}.
For example all points on the $z$ axis are eigenvectors of the operator in three-dimensional space ``rotate around the $z$ axis'' with eigenvalue $1$ (note that points away from the $z$ axis are {\it not} eigenvectors).
Another example is the operator ``$d/dx$'', whose eigenvectors are $e^{kx}$ (with $k$ any constant) with eigenvalue $k$.
}.
It is a law of quantum mechanics that any time we measure a property of a system, we can only observe states of the system that are eigenvectors of the corresponding operator, and the result of our measurement is the associated eigenvalue.
As an example, the intrinsic spin of a fermion such as an electron may be aligned with whatever direction we decide to call the positive $z$ axis -- the state $|\uparrow\rangle$ -- or in the opposite direction -- $|\downarrow\rangle$ -- or it may be a linear combination of these two possibilities.
$|\uparrow\rangle$ and $|\downarrow\rangle$ are eigenstates of the $z$-axis spin operator $\hat{s}_z$:
\begin{itemize}
\item $\hat{s}_z|\uparrow\rangle = +\tfrac{1}{2}|\uparrow\rangle$,
\item $\hat{s}_z|\downarrow\rangle = -\tfrac{1}{2}|\downarrow\rangle$,
\item but $\hat{s}_z(|\uparrow\rangle+|\downarrow\rangle) = +\tfrac{1}{2}|\uparrow\rangle-\tfrac{1}{2}|\downarrow\rangle \neq \text{constant}\times(|\uparrow\rangle+|\downarrow\rangle)$, so this state is not an eigenstate of $\hat{s}_z$ and in measuring the $z$-axis spin we will never observe this state.
If the system really is in this state, then when we measure the $z$-axis spin we force it to change state either into $|\uparrow\rangle$ or into $|\downarrow\rangle$: this forced change is referred to as the {\it collapse of the wavefunction}.
How (or indeed if) this really happens has been the subject of much debate, known as the {\it measurement problem}.
\end{itemize}

In general when the system is in a state $|\psi\rangle$ and we want to know the probability of finding it in the particular state $|\phi\rangle$ (a probability which is not necessarily $0$ by definition, since $|\psi\rangle$ may be a superposition of states, one of which is $|\phi\rangle$), this probability is $|\langle\phi|\psi\rangle|^2$.
The quantity $\langle\phi|\psi\rangle$ is highly analogous to taking the dot product of two vectors to quantify how much they overlap; indeed $|\psi\rangle$ and $|\phi\rangle$ are vectors in the (possibly infinite-dimensional) space of all possible states the system can be in.

Time evolution of a state $|\psi\rangle$ occurs according to the {\it Schr\"{o}dinger equation}:
\begin{equation} \label{Schrod}
 \frac{\partial}{\partial t} |\psi\rangle = -\frac{i}{\hbar} \hat{H} |\psi\rangle,
\end{equation}
where $\hat{H}$ is the {\it Hamiltonian} -- the operator whose eigenvalues are the possible energies the system (which are sometimes discrete, e.g. the energy levels of electrons in atoms).
From Eq.~\eqref{Schrod} we see that if the system is in a state of definite energy -- it is an eigenstate of $\hat{H}$ with eigenvalue $E$ -- its dependence on time is simply given by a factor $\exp(-iEt)$.
If $|\psi\rangle$ is not a state of definite energy, we can still write down a solution to Eq.~\eqref{Schrod} with the correct time dependence by inspection:
\begin{equation}
 |\psi(t=T)\rangle = \exp\left(-\frac{i}{\hbar} \hat{H} T\right) |\psi(t=0)\rangle
\end{equation}
However understanding what this solution really means may be highly non-trivial because in general $\hat{H}$ is the sum of non-commuting\footnote{
If operators $\hat{A}$ and $\hat{B}$ do not commute, this means $\hat{A}\hat{B} \neq \hat{B}\hat{A}$.
This is a strange concept at first because we're mostly used to normal numbers, and $2\times3 = 3\times2$, but $\hat{A}$ could be the operation `rotate an object $90^{\circ}$ left' and $\hat{B}$ the operation `rotate an object $90^{\circ}$ forwards'.
You can see for yourself with whatever object you have to hand that the order in which these operations are performed matters.
In the context of quantum mechanics, a bit of simple maths shows that two operators representing two physical properties that cannot both take definite values simultaneously, such as position and momentum, must not commute; and indeed the position and momentum operators generally feature in the Hamiltonian.
} operators.

If at a time $t=0$ a particle has definite position $q_1$, a state we call $|q_1\rangle$, then at a later time $T$ the state of the system is
\begin{equation}
 |\psi(t=T)\rangle = \exp\left(-\frac{i}{\hbar} \hat{H} T\right) |q_1\rangle
\end{equation}
The probability to observe the particle at a definite, different position $q_2$ at this later time is
\begin{equation}
 \left|\langle q_2 | \psi(t=T)\rangle \right|^2 =  \left|\langle q_2 |\exp\left(-\frac{i}{\hbar} \hat{H} T\right) |q_1\rangle\right|^2
\end{equation}
A gorgeous piece of mathematics
shows that contents of the $|\ldots|^2$ can be evaluated as a {\it path integral}:
\begin{equation} \label{PathIntegral}
\langle q_2 |\exp\left(-\frac{i}{\hbar} \hat{H} T\right) |q_1\rangle = \int_{q(t=0)=q_1}^{q(t=T)=q_2} e^{iS[q]/\hbar}\:\mathcal{D}q,
\end{equation}
where we integrate over the infinite number of possible functions of time $q(t)$ in the interval $t\in[0,T]$ subject to the boundary conditions $q(t=0)=q_1$ and $q(t=T)=q_2$.
In a sense this is an infinite-dimensional integral, since the space of possible functions is infinite dimensional.

\subsection{Quantum Field Theory}
Non-relativistic quantum mechanics describes the time evolution of quantum systems -- particles moving, changing their intrinsic spin etc. -- without allowing for the creation or destruction of particles.
Constant particle number is hard-coded in the theory.
However special relativity tells us that mass is merely another form of energy, and therefore particles can be created and destroyed.
Quantum field theory (QFT) extends quantum mechanics to include this new feature, and makes the physical laws Lorentz invariant.
Central to the idea of QFT is that there exists a field, permeating the whole of space, for each kind of fundamental particle; a single particle is simply a localised excitation of its field, able to propagate through space.

The surface of a body of water can be described by a field (height as a function of position) which can have two opposite kinds of excitation -- peaks and troughs -- both with positive energy but which are able to cancel each other out.
The same is true of the fundamental particle fields -- they may support both particle and antiparticle excitations, and pairs of these may annihilate into pure energy or be produced out of pure energy.
Note that this picture of antiparticles is fundamentally different from (though often gives similar answers to) the older description in terms of a {\it Dirac sea}.
The latter consists of a vacuum which is an infinite number of negative energy particle states all filled, and no positive states filled; antiparticles are then interpreted as available negative energy states, which can annihilate particles in an intuitive way.

In QFT the kinds of objects we wish to compute resemble those of normal quantum mechanics -- probability amplitudes for a given initial state to later be observed as a given {\it final state} (see the glossary).
A result familiar from the description of mundane every-day waves is that energy is inversely related to distance -- higher energy waves have smaller wavelengths -- and therefore to understand nature at ever smaller length scales, the relevant processes for our calculation of probability amplitudes are those in which large amounts of energy are exchanged.
Of course we must actually carry out these processes as well, so that our calculations, and the theories that define them, can be tested.
In practice this is done by colliding particles together as hard as we can, then recording the relative number of occurrences of the different final states that result (i.e. establishing probability distributions for the final states).
Hence the initial states for our calculations generally consist of two particles travelling towards each other at a high relative velocity; the final states considered will be everything that those two particles can produce after interacting with each other in an arbitrary way\footnote{
Note that not {\it everything} is possible -- conservation rules such as the conservation of energy, momentum and electric charge imply that the final state must have the same values for these quantities as the initial state.
}.
As with the quantum mechanical path integral, the probability amplitude for a transition between given states can be calculated as the integral of $\exp(iS/\hbar)$ over the space of all possible things that could have happened during the transition.
However previously this object was simpler: the integral was over all possible functions of time the coordinate could take, $q(t)$.
Now our transition is from two localised excitations of fields, to some other number of localised excitations, and the associated fields could do anything in between times: for each of these fields (and indeed any other field that interacts with them), we must integrate over all possible behaviours as a function of time {\it for each point in space}!
In other words we have an infinite-dimensional integral of the form Eq.~\eqref{PathIntegral} associated with each of the infinitely many points in continuous space.

These tremendously daunting mathematical objects have been computed exactly in certain theories simpler than those of direct relevance to our universe and our current particle colliders.
In the latter cases, we typically have to resort to an {\it expansion} of the probability amplitude (that is, a systematic grouping of the infinite number of contributing terms into a series consisting first of the largest then the second largest etc. {\it ad infinitum}) followed by a truncated calculation of as much of the series as we are able to do in a given number of man-hours.
Each term in the expansion can be represented by a {\it Feynman diagram}, with the translation between diagram and term established by a set of universal {\it Feynman rules}.
Crudely speaking, the simpler the Feynman diagram -- the smaller the number of intermediate, localised excitations of the fields connecting the initial configuration of fields (two incoming particles) to the final one -- the larger the term it represents.
This is because connecting the initial and final states with an increasing number of interactions between fields usually means, via the Feynman rules, that the term has an increasingly small pre-factor.
Sometimes this is not the case, when the coupling between fields is `strong' (meaning strong enough to compensate for the small pre-factor); then the terms at each step of the expansion are as big as those preceding them, and without the ability to sum the entire infinite series we lose calculability.
In the author's opinion, this is the greatest unsolved problem in mathematical physics.

An important difference between QFT and classical field theory is that the parameters appearing the in the Lagrangian density $\mathcal{L}$ no longer relate directly to physical observables in the same way.
For example, the dimensionless number multiplying a term containing two or more different fields (thus allowing the respective particles to interact), known as a {\it coupling} or {\it coupling constant}, does not quantify the interaction strength, or at least not in a simple way.
The relations between the parameters of $\mathcal{L}$ and their corresponding observables now contain infinities, which have to be subtracted by hand in such a way as to leave a set of parameters that agree with the experimentally established values by construction.
Note that choosing the parameters of a model of nature in order to reproduce what is known to be true is often regarded as a bad thing, as science should make predictions that allow for falsification.
However predictivity is only lost when there is enough flexibility in the model parameters to accommodate {\it any} experimental result; in the case of QFT there are considerably more observables than those corresponding directly to the parameters of $\mathcal{L}$, and these are genuine predictions, which match known measurements to a simply stunning level of precision.

The removal of infinities to set the parameters of $\mathcal{L}$, known as {\it renormalisation}, must be done at a particular energy scale.
An extremely deep feature of QFT is that, once this has been done, a different set of parameters is appropriate for describing particle interactions at a different energy scale; in other words, the renormalised (infinity-subtracted) parameters of $\mathcal{L}$ {\it run} with energy scale.
The coupled differential equations controlling this running -- the renormalisation group (RG) equations -- are an example of something predicted by a given QFT, once each of the parameters of $\mathcal{L}$ has been defined at one scale.
The physical picture often used to summarise the positive running of the electromagnetism coupling is the following.
In probing a charged point-like particle with a low-energy photon, one is really probing only the rough area of space the charge is sitting in, due to the photon's long wavelength.
This space also contains virtual charged particle-antiparticle pairs by virtue of quantum uncertainty, with the opposite charges tending to lie closer to the real physical particle (in the usual manner of polarisable media), {\it shielding} the charge that is effectively seen.
This effect decreases the smaller the wavelength of the photon, hence the increasing strength of the electromagnetic coupling with energy.


\newpage
\part{Weak-Scale Susy's Raison d'Etre: The Higgs}

\section{Introduction} \label{LowScaleSusyIntro}
\subsection{Motivation} \label{SusyMotivation}

The Standard Model is the mathematical description of our most fundamental understanding of particle physics.
It is a quantum field theory in $3+1$ dimensions of spacetime with Poincar\'{e} invariance, the gauge group $SU(3)_c\times SU(2)_L\times U(1)_Y$, and the field content shown in Table~\ref{SMcontent}.

\begin{table}[!ht]
\begin{center}
\begin{tabular}{c c c | c}
scalars & fermions & vector bosons & gauge group\\
& & & representation \\ \hline
$\vphantom{ \dfrac{w}{w} } H$ & & & $(\mathbf{1},\mathbf{2},\tfrac{1}{2})$\\
&$\vphantom{ \dfrac{w}{w} } Q_i$ & & $(\mathbf{3},\mathbf{2},\tfrac{1}{6})$ \\
&$\vphantom{ \dfrac{w}{w} } \overline{u}_i$ & & $(\mathbf{\overline{3}},\mathbf{1},-\tfrac{2}{3})$ \\
&$\vphantom{ \dfrac{w}{w} } \overline{d}_i$ & & $(\mathbf{\overline{3}},\mathbf{1},\tfrac{1}{3})$ \\
&$\vphantom{ \dfrac{w}{w} } L_i $ & & $(\mathbf{1},\mathbf{2},-\tfrac{1}{2})$ \\
&$\vphantom{ \dfrac{w}{w} } \overline{e}_i$ & & $(\mathbf{1},\mathbf{1},1)$ \\
& &$\vphantom{ \dfrac{w}{w} } g$ & $(\mathbf{8},\mathbf{1},0)$\\
& &$\vphantom{ \dfrac{w}{w} } W$ & $(\mathbf{0},\mathbf{3},0)$ \\
& &$\vphantom{ \dfrac{w}{w} } B$ & $(\mathbf{1},\mathbf{1},0)$
\end{tabular}
\end{center}
\caption{The Standard Model field content, organised by spin and representation under the $SU(3)_c\times SU(2)_L\times U(1)_Y$ gauge group.
$i=1,2,3$ denotes the generation/family/flavour of the fermion.}
\label{SMcontent}
\end{table}

The Standard Model Lagrangian $\mathcal{L}_{SM}$ specifies the physical behaviour of the particle excitations of these fields, by quantifying their masses, mixings and interaction strengths.
A point of great interest concerning $\mathcal{L}_{SM}$ is that all terms except one are {\it marginal} or {\it exactly renormalisable}, being the product of fields with total mass dimension $4$ and a dimensionless coefficient.
The single {\it relevant} or {\it super-renormalisable} term, having dimensionful coefficient, is the mass-squared for the Higgs field: $\mathcal{L}_{SM} \supset -m_H^2 H^{\dag}H$.
(Masses for all the other particles arise from dimensionless couplings to the Higgs field, which acquires a non-zero vacuum expectation value (VEV) $v$.)
This term is at the heart of a problem the Standard Model is widely believed to suffer from: unnaturalness.

't Hooft argued that a small parameter of the Lagrangian is {\it naturally} small if the Lagrangian has an enhanced symmetry when the parameter vanishes~\cite{'tHooft:1979bh}.
In that limit, the symmetry forces the radiative corrections to vanish, and so with the parameter non-zero the corrections must be proportional to the parameter itself.
This is sometimes referred to as {\it technical naturalness}.
For example, small electron masses in quantum electrodynamics are technically natural because chiral symmetry in the massless limit enforces
\begin{equation} \label{FermionNoQuadDiv}
\delta m \propto m
\end{equation}
However if we have a fundamental scalar $\phi$ then, outside of conformally invariant theories, allowing its mass-squared to vanish does not enhance the symmetry; indeed explicitly calculating the radiative corrections one finds terms like
\begin{equation}
 \delta m_\phi^2 \supset \frac{g^2}{16\pi^2}\Lambda^2,
\end{equation}
where $g$ is the coupling of $\phi$ to a particle that can run in a loop of the two-point correlator $\langle \phi(-p)\phi(p)\rangle$, and $\Lambda$ is an ultraviolet (UV) cutoff for the divergent integral.
The {\it hierarchy problem} can be loosely phrased in the following way: if $\Lambda^2$ is parametrically larger than the renormalised mass-squared $m_{\phi}^2 = m_{\phi,\rm{bare}}^2 + \delta m_\phi^2$, the cancellation on the right-hand side (RHS) of this equation between the bare value and radiative corrections requires parametrically large fine-tuning; otherwise we would expect $m_{\phi}^2 = \mathcal{O}(\,(g^2/16\pi^2)\:\Lambda^2)$.
However, if $\Lambda$ is to be understood as nothing more than a non-physical regulator to be taken to infinity, our question is not well posed.
This leads us to the more precise {\it technical hierarchy problem}, which we have when $\Lambda$ is interpreted as a physical cutoff, below which our theory is an {\it effective} theory.
When we enter the regime of the latter by integrating out heavier degrees of freedom at the scale $\Lambda$, there will generically be corrections of this size to any scalar masses, suppressed by however many loop-factors are necessary to couple the scalar to these heavy particles\footnote{
One might consider new physics at a scale $\Lambda$ which does not couple to Higgs at any order in perturbation theory; it would therefore not couple to any part of the Standard Model -- a fairly uninteresting scenario.
Gravity at least must couple to the Higgs, as the latter has mass and energy.
}.
Unless there is no new physics coupling to the Higgs at any higher energy scale, we then return to the conclusion of the simple hierarchy problem that the Higgs mass should be $\mathcal{O}(\,(g^2/16\pi^2)\:\Lambda^2)$.
More concretely, in the Standard Model we have a mass-squared $\mathcal{O}(100~\text{GeV})^2$ and a correction $(y_t^2/8\pi^2)\,\Lambda^2$ from the large coupling $y_t$ of the Higgs to the top-quark.
We would therefore require a cutoff $\Lambda \sim \surd(8\pi^2/y_t^2)\:100~\text{GeV} =\mathcal{O}(1~\text{TeV})$.
Taking the cutoff instead to be the Planck mass $M_P\sim10^{18}$~GeV -- the highest scale at which the Standard Model without quantum gravity could be valid -- requires fine-tuning to one part in $\sim\!\!10^{30}$.

The requirement for new physics {\it fully} explaining naturalness can be pushed from $1$~TeV to higher energies if the Higgs is a pseudo-Nambu-Goldstone boson (pNGB); I briefly review the idea of these {\it little Higgs} models in Appendix~\ref{littleHiggs}.
(Note however that this still requires some new physics at the TeV scale -- specifically top partners -- to cancel the top quark correction to the Higgs mass.)
Full explanations of naturalness, traditionally considered at the TeV scale (rather than at the higher scale permitted by little Higgs) generally fall into one of the three following camps.
For the first two I will merely state the idea without explaining details; a nice introduction to all three can be found at~\cite{naturalEWCBnotes}, and there are several different sets of TASI lecture notes available on these topics.

Firstly, quantum-gravitational effects effects may become relevant, requiring extra dimensions.
The Higgs could be the fifth component of a five-dimensional gauge field; masslessness in the limit of the fifth dimension becoming large and flat means the Higgs is naturally light.
Alternatively it could be a fundamental scalar confined to a physical four-dimensional subspace -- a {\it brane} -- with an effective momentum cutoff resulting from the extra dimensions being warped~\cite{Randall:1999ee}.

Secondly, the Higgs may be composite, i.e. (extensions of) the idea of {\it Technicolour}.
It is a bound state of a fermion and an anti-fermion ({\it techniquarks}), which are confined by a new gauge group.
Techniquark condensation (i.e $\langle \bar{q}_{TC} q_{TC} \rangle$ becoming non-zero) spontaneously breaks the global $SU(2)_L\times SU(2)_R$ symmetry of the techniquarks to the diagonal $SU(2)_V$ giving three would-be Nambu-Goldstone bosons (NGBs) -- technipions, by close analogy with regular pions -- which the hungry massless $W$ and $Z$ bosons eat instead of the three would-be NGBs in the $H$ doublet in the Standard Model.
Technicolour in particular is more attractive with a little Higgs setup than without~\cite{Contino:2003ve,Agashe:2004rs}.

Thirdly, and finally moving to the topic of this thesis, a light fundamental scalar may be protected by Supersymmetry (see~\cite{Martin:1997ns} for an extensive introduction, review and many references; also~\cite{Drees:1996ca,Signer:2009dx} for preliminaries), henceforth {\it Susy}.
Susy generators convert bosons into fermions and vice-versa; a supersymmetric theory must therefore be constructed from objects called superfields, in which half the degrees of freedom are bosonic and half are fermionic; each half is said to be the {\it superpartner} of the other, with exactly the same quantum numbers excepting spin.
In such theories, radiative corrections to the mass-squareds of scalars vanish at every order in perturbation theory: bosonic and fermionic loops always come in pairs, with the same magnitude but opposite sign.
With Susy broken, but only {\it softly} -- that is, where superpartners have different dimensionful `couplings' (including mass) but identical dimensionless couplings -- quadratic divergences still vanish, since the associated coupling constant must be dimensionless for the diagram to have dimension two.
There are still logarithmically divergent corrections associated with the effective theory where one particle has been integrated out but its superpartner has not; such corrections are proportional to the scale of soft Susy breaking.
In this manner we may have a light Higgs naturally.

Motivations for considering our universe to be supersymmetric are as follows.
\begin{enumerate}
  \item \label{Hierarchy} The (technical) hierarchy problem as discussed.
  \item \label{ColemanMandula} The symmetry group for our physical laws increased in size until we arrived at the Poincar\'{e} group.
Coleman and Mandula's no-go theorem~\cite{PhysRev.159.1251} `proved' (for a while) that there are no non-trivial extensions of this group, i.e. that any extension would be the direct product of the Poincar\'{e} group and {\it internal} (non-spacetime) symmetries.
Haag, \L{}opusza\'{n}ski and Sohnius~\cite{Haag1975257} found the loop-hole: non-trivial extensions are allowed but only with fermionic symmetry generators, i.e. Susy.
Susy is thus unique, in a fairly profound way, among possible extensions of the Standard Model.
Intimately tied up with this feature is the widely believed uniqueness of Susy (or more precisely its generalisation from a global to a local symmetry -- {\it Supergravity}) as a setting in which a spin-$\frac{3}{2}$ particle may exist (see e.g.~\cite{Boulware1979141,Hack:2011yv}).
\item \label{StringTheory} Susy is widely believed to be a requirement of consistent string theory, and the latter is widely believed to be the correct framework for a description of gravity at the quantum level.
The necessary existence of the latter is thus a motivation for Susy.
\item The three independent gauge couplings of the Standard Model, when evolved using the renormalisation group (RG) to an energy scale \linebreak\mbox{$\mathcal{O}(10^{16}~\text{GeV})$}, {\it almost} meet at a point.
This suggests the tempting possibility that with extra charged matter added at lower scales, they really do meet (which tends to happen in Susy) and there is only one fundamental gauge group: {\it Grand Unification}.
There is strong historical motivation for us to explain more phenomena with fewer principles.
Proposing a much enlarged field content, as we'll see shortly is necessary, to solve the one-parameter problem of making three lines meet at a point sounds like overkill.
However in Grand Unified Theories (GUTs) there is an extremely large scale with new physics that couples directly to the Higgs, thus removing any doubt about the applicability of the technical hierarchy problem.
Grand Unification thus motivates Susy for two reasons:
\begin{enumerate}
 \item \label{GUT1} the superpartners modify the RG running of the gauge couplings in a way that generally improves the extent to which they meet at a single scale; and
 \item \label{GUT2} we fall squarely into the scope of the technical hierarchy problem, and with Susy the unavoidable, physical, $\mathcal{O}(10^{16}~\text{GeV})^2$ contributions to the Higgs mass-squared sum to zero.
\end{enumerate}
\item \label{DarkMatter} There is overwhelming evidence from astrophysical observations for the existence of a new neutral particle which is stable on cosmological time-scales -- {\it dark matter}, see for example~\cite{Bertone:2004pz} for a review.
Often we impose a discrete $Z_2$ symmetry called $R$~parity on Susy theories in order to suppress terms which would strongly violate experimental bounds on proton stability.
With $R$~parity, the lightest new particle with opposite charge to the Standard Model particles -- the Lightest Supersymmetric Particle (LSP) -- is stable; if it is also neutral it is a dark matter candidate.
However introducing Susy and the plethora of necessary particles solely for one of them to be dark matter is overkill -- this problem may be addressed more minimally than the (technical) hierarchy problem, one example being with the Peccei-Quinn axion which also solves the strong CP problem~\cite{Peccei:1977hh,Peccei:1977ur}.
\item \label{HiggsDiscovery} On July 4$^{\rm th}$ 2012 the ATLAS and CMS Collaborations announced their independent discoveries at roughly $5\sigma$ confidence level of a new bosonic resonance, with properties as measured so far coinciding closely with the Higgs of the Standard Model, and mass $\sim\!\!126~\text{GeV}$~\cite{:2012gk,:2012gu}.
Now, Susy is a broad collection of different theories with different properties; however what's common to almost all of them is that the Higgs mass $m_h$ is connected to the electroweak gauge boson masses $\sim\!\!M_Z$, and enjoys a special protection from radiative corrections.
The authors of~\cite{Giudice:2011cg} studied $m_h$ in both {\it split Susy}~\cite{Giudice:1998xp,Wells:2003tf,ArkaniHamed:2004fb,Giudice:2004tc} and {\it high-scale}/{\it supersplit Susy}~\cite{Fox:2005yp}, both of which are versions of the Minimal Supersymmetric Standard Model (MSSM, see Section~\ref{MSSM}).
Split Susy decouples all scalar superpartners of the Standard Model fermions while leaving all fermionic superpartners light (with the possible exception of those of Higgs, according to taste); high-scale Susy decouples all non-Standard Model particles.
The finding of~\cite{Giudice:2011cg} was that even sending superpartner masses to the Planck scale, $m_h$ remains below $157$ ($143$)$~\text{GeV}$ in split (high-scale) Susy.
The importance of this result, beyond showing which models have $m_h \approx 126~\text{GeV}$ and which do not, is as an illustration of the robustness of the MSSM prediction for $m_h = \mathcal{O}(M_Z)$: low-scale Susy gives both electroweak naturalness and $m_h = \mathcal{O}(M_Z)$; in decoupling Susy we lose the former and thus part of the motivation, but we are still left with the latter.
This is to be contrasted with the Standard Model, for which the prediction was $m_h\lesssim 700~\text{GeV}$ for perturbative unitarity of $WW$ scattering at the TeV scale.
From this point of view, Susy (or more precisely the MSSM and its not-too-distant cousins) gave a slightly better prediction for $m_h$ than the Standard Model.
\end{enumerate}

Note that points~\ref{Hierarchy} and~\ref{GUT2} motivate weak-scale Susy; points~\ref{GUT1} and~\ref{DarkMatter} have a slight preference for at least some Susy particles at the weak scale; points~\ref{ColemanMandula},~\ref{StringTheory} and~\ref{HiggsDiscovery} motivate Susy but without a preference for its breaking scale, which could be anywhere up to $M_P$.

Having motivated Susy, I now give a brief explanation of what it is.

\subsection{The Superpotential} \label{SuperPotSusyBreaking}
(See~\cite{Signer:2009dx,Martin:1997ns} for much more thorough explanations than the sketch given here.)
As I have mentioned, a supersymmetric theory should be built from superfields.
Superfields can be considered simply as book-keeping devices, as they pair up bosonic and fermionic fields with otherwise identical quantum numbers.
More formally they are functions of {\it superspace}, $X \equiv (x^\mu,\theta^\alpha,\bar{\theta}^{\dot{\alpha}})$, where $\theta^{\alpha=1,2},\bar{\theta}^{\dot{\alpha}=1,2}$ are two-component spinors of anti-commuting Grassman variables.
The most general function of $X$, expanded in powers of $\theta$ and $\bar{\theta}$ with $x^\mu$-dependent coefficients, is of finite length: Grassman variables are nilpotent -- they vanish when squared.
This polynomial is a reducible representation of the group of Susy transformations, and irreducible representations (irreps) can be made by taking subsets of the finite number of terms.
Two such irreps are left-handed and right-handed chiral superfields, each of which contains a single complex scalar, a fermion of the denoted chirality, and an unphysical/auxiliary scalar $F$.
Expanding a left-handed superfield $\Phi$ in powers of $\theta$ and $\bar{\theta}$, the coefficient of the $\theta^\alpha\theta_\alpha$ term is $F$; then following the rules of Grassman integration,
\begin{equation}
 \int d^2 \theta\: \Phi = F
\end{equation}
Under a Susy transformation $F$ can be seen to change by a total derivative, and so an action defined from a Lagrangian consisting of $F$-terms will be invariant under Susy transformations.
The product of left- (right-) handed superfields transforms itself like a left- (right-) handed superfield under the Susy group, and so the following {\it Wess-Zumino} Lagrangian is supersymmetric:
\begin{gather} \label{superpotential}
 \mathcal{L} = \int d^2\theta\: W(\Phi) \:+\: {\rm c.c.},\\
 {\rm where} \quad W(\Phi) \equiv \chi_i \Phi_i  + M_{ij}\Phi_i \Phi_j + y_{ijk}\Phi_i \Phi_j\Phi_k
\end{gather}
The quantity $W(\Phi)$ is the {\it superpotential} -- a holomorphic dimension-three function of all the left-handed chiral superfields $\Phi_i$ in the model.
Writing the scalar in $\Phi_i$ as $\phi_i$, and with $W(\phi)$ the same function of these scalar fields as $W(\Phi)$ is of the corresponding superfields, the contribution of the superpotential to the regular potential $V(\phi)$ is
\begin{equation} \label{Fterms}
 V(\phi) \supset \sum_i \left|\frac{\partial W(\phi_i)}{\partial \phi_i}\right|^2
\end{equation}
where this has come from Eq.~\eqref{superpotential} followed by a replacement of each non-physical $F$ field by the solution of its (algebraic) equation of motion; the terms on the RHS are thus referred to as $F$-terms.

The other contributions to the potential come from so-called $D$-terms, which arise from supersymmetric gauge interactions.
Spin-one vector bosons associated with gauge symmetries live inside a different representation of the Susy group called a vector superfield, with a fermionic partner called a gaugino and an auxiliary scalar $D$ field.
The coefficients in front these different component fields in the Lagrangian are fixed by Susy; however in the specific case of a vector superfield for an {\it abelian} gauge symmetry, we can add to Lagrangian an arbitrary extra amount of the associated $D$ field: $\mathcal{L} \supset kD$.
This is the Fayet-Iliopoulos term~\cite{Fayet:1974jb}.
Arranging all of the $\phi_i$ in the theory into a vector transforming in a (possibly reducible) representation of the full gauge group with generators $T^a$, the $D$-terms are
\begin{equation}\label{Dterms}
 V(\phi) \supset \tfrac{1}{2} \sum_a (g_a\, \phi_i^\dagger T^a_{ij} \phi_j + k^a)^2,
\end{equation}
where the gauge coupling $g_a$ is of course common to different generators of the same gauge symmetry, but if the full gauge group is a product of different groups there is one coupling for each of these groups.

\subsection{The MSSM} \label{MSSM}

To supersymmetrise the Standard Model we must place each of its fields (shown in Table~\ref{SMcontent}) into a superfield, and include one more superfield~\cite{Dimopoulos1981150,SakaiMSSM}, expanding the set of degrees of freedom to those shown in Table~\ref{MSSMcontent}.
Each row (i.e. each superfield) in Table~\ref{MSSMcontent} except the second one contains a field as in the Standard Model, with a new bosonic or fermionic partner.
The second row, $H_d$ and $\tilde{H}_d$, is the new superfield.
This is required, forcing us into a two-Higgs-doublet model, for two reasons.
Firstly the Standard Model mechanism of generating masses for the charged leptons and down-type quarks through a coupling with complex conjugate of the Higgs doublet is not possible due to the requirement that the superpotential be holomorphic.
Secondly, our introduction of the $\tilde{H}_u$ fermion violates the $SU(2)_L\times U(1)_Y$ gauge anomaly cancellation conditions; they are restored with an extra fermion of opposite hypercharge $Y$, such as $\tilde{H}_d$.

\begin{table}[!ht]
\begin{center}
\begin{tabular}{c c c | c}
scalars & fermions & vector bosons & gauge group\\
& & & representation \\ \hline
$\vphantom{ \dfrac{w}{w} } H_u$ & $\tilde{H}_u$ & & $(\mathbf{1},\mathbf{2},\tfrac{1}{2})$\\
$\vphantom{ \dfrac{w}{w} } H_d$ & $\tilde{H}_d$ & & $(\mathbf{1},\mathbf{2},-\tfrac{1}{2})$\\
$\vphantom{ \dfrac{w}{w} } \tilde{Q}_i$ &$Q_i$ &  &$(\mathbf{3},\mathbf{2},\tfrac{1}{6})$ \\
$\vphantom{ \dfrac{w}{w} } \tilde{\overline{u}}_i$ & $\overline{u}_i$ &  &$(\mathbf{\overline{3}},\mathbf{1},-\tfrac{2}{3})$ \\
$\vphantom{ \dfrac{w}{w} } \tilde{\overline{d}}_i$ & $\overline{d}_i$ &  &$(\mathbf{\overline{3}},\mathbf{1},\tfrac{1}{3})$ \\
$\vphantom{ \dfrac{w}{w} } \tilde{L}_i $ & $L_i $ &  &$(\mathbf{1},\mathbf{2},-\tfrac{1}{2})$ \\
$\vphantom{ \dfrac{w}{w} } \tilde{\overline{e}}_i$ & $\overline{e}_i$ &  &$(\mathbf{1},\mathbf{1},1)$ \\
& $\vphantom{ \dfrac{w}{w} } \tilde{g}$ &$g$ & $(\mathbf{8},\mathbf{1},0)$\\
& $\vphantom{ \dfrac{w}{w} } \tilde{W}$ &$W$ & $(\mathbf{1},\mathbf{3},0)$ \\
& $\vphantom{ \dfrac{w}{w} } \tilde{B}$ &$B$ & $(\mathbf{1},\mathbf{1},0)$
\end{tabular}
\end{center}
\caption{The Minimal Supersymmetric Standard Model field content, organised by spin and representation under the $SU(3)_c\times SU(2)_L\times U(1)_Y$ gauge group.
$i=1,2,3$ denotes the generation/family/flavour of the (s)fermion.}
\label{MSSMcontent}
\end{table}

The superpotential of the $R$~parity conserving MSSM is
\begin{equation} \label{MSSMsuperpot}
 W_{\rm MSSM} = y_{u,ij} \bar{u}_i Q_j H_u - y_{d,ij} \bar{d}_i Q_j H_d - y_{e,ij} \bar{e}_i L_j H_d + \mu H_u H_d,
\end{equation}
where, with the standard abuse of notation, the same symbol is used for a superfield and its scalar component (for the Higgs fields) or fermionic component (for the Standard Model fermions).
In Eq.~\eqref{MSSMsuperpot} $SU(2)_L$ and $SU(3)_c$ gauge indices within each term are implicitly contracted to form a singlet.

The fermionic partners of the Higgs scalars and vector bosons are called Higgsinos and gauginos respectively, with the latter consisting of a bino, winos and a gluino.
The scalar partners of Standard Model fermions take the name of the corresponding fermion with the letter `s' prepended: for example the left- and right-handed {\it stops} $\tilde{t}_{L,R}$ are the scalar partners of the left- and right-handed chiralities of the top quark, and we define the {\it sfermions} and {\it sleptons} etc. similarly.
Non-Standard Model particles are collectively called {\it sparticles}.

In the two, complex, scalar Higgs doublets there are eight degrees of freedom.
Three of these are the would-be NGBs that become the longitudinal components of the W and Z bosons; the other five are the physical particles $h$ and $H$ ($CP$-even and electrically neutral, $h$ defined to be the lighter), $A$ ($CP$-odd and neutral) and $H^{\pm}$ ($CP$-even and charged).
The couplings of $h$ to Standard Model particles become Standard Model-like in the {\it decoupling limit}, where $H,A,H^{\pm}$ are all several times heavier than the $Z$ boson; my earlier point in the final bullet point of Section~\ref{SusyMotivation} about `the' Higgs of the MSSM being robustly of mass $\mathcal{O}(M_Z)$ referred specifically to $h$.
In this work, motivated by the Standard Model-like couplings seen so far and non-observation of $A$ or $H^{\pm}$, in the context of the MSSM I will always consider the $\sim\!\!126~\text{GeV}$ resonance to be $h$.
At the time of writing, the possibility that this resonance is $H$ is still being debated -- see~\cite{Bechtle:2012jw,Arbey:2012bp,Bechtle:2013gu}.

With unbroken Susy, the boson and fermion in the same superfield have the same mass.
Non-observation of any superpartner degenerate with its Standard Model partner means we have to add in masses for all the superpartners by hand.
At first glance this seems a lot to swallow, rather than simply conceding that the model is false.
However the subset of all particles in the MSSM that we have discovered so far is exactly the subset of particles whose masses only arise from electroweak symmetry breaking (EWSB), and the undiscovered subset is exactly the group of particles whose masses are not tied to EWSB.
This fact, which we have not put in by hand, is highly encouraging.

The general tree-level vanishing~\cite{PhysRevD.20.403} of the {\it supertrace}
\begin{equation} \label{eq:supertrace}
 {\rm STr}M^2 \equiv \sum_{\text{all particles}} (-1)^s (2s+1) M^2 = 0
\end{equation}
where $M$ is a particle's mass and $s$ its spin, means that spontaneous breaking of Susy from within the MSSM will be phenomenologically unacceptable -- some superpartners will be lighter than their Standard Model partners, whereas we need them all to be heavier.
Susy breaking must therefore be a higher-order/quantum effect, mediated to the MSSM from elsewhere -- the {\it hidden} sector.
The three main approaches to achieving the mediation are through gravitational interactions, extra dimensions, or gauge interactions (see Part~\ref{GGM}).

Susy breaking results in mass-squareds $m_{\phi}^2$ and Majorana masses $M_{\tilde{\lambda}_i}$ for all of the scalars $\phi=H_u.H_d,\tilde{Q}_i,\tilde{\overline{u}}_i,\tilde{\overline{d}}_i,\tilde{L}_i,\tilde{\overline{e}}_i$ and gauginos $\tilde{\lambda}_i = \tilde{B}, \tilde{W}, \tilde{g}$ respectively in Table~\ref{MSSMcontent}.
It also gives the trilinear mixing terms
$\mathcal{L} \supset -\tilde{\overline{u}}_i a_{u,ij} \tilde{Q}_j H_u + \tilde{\overline{d}}_i a_{d,ij}\tilde{Q}_j H_d +\tilde{\overline{e}}_i a_{e,ij}\tilde{L}_j H_d + {\rm c.c.} $; each $a_{ij}$ matrix is usually assumed to be proportional to the corresponding Yukawa matrix.
In the basis where these are diagonal, we have for example $a_{33} = y_t A_t$; the stop trilinear mixing parameter $A_t$ gives potentially large corrections to the physical Higgs mass as we will see in the following section.
The final term arising from Susy breaking is the dimension-two mixing term for the Higgs scalars $\mathcal{L} \supset -b H_u H_d + {\rm c.c.}$.

In addition to the Susy breaking masses there is a single supersymmetric mass parameter in the MSSM, for the two Higgs superfields: the $\mu$ term.
This completes the list of dimensionful terms in the MSSM.
Those relevant for the two $CP$-even electrically neutral scalars $H_u^0$ and $H_d^0$ are:
\begin{equation} \label{HiggsMassMatrix}
V\quad\supset \quad(H_u^0 \quad(H_d^0)^*)
\begin{pmatrix}
  |\mu|^2 + m_{H_u}^2 & -b\\
  -b & |\mu|^2 + m_{H_d}^2
 \end{pmatrix}
\begin{pmatrix}
(H_u^0)^* \\
H_d^0  
 \end{pmatrix}
\end{equation}
For EWSB one linear combination of $H_u^0$ and $H_d^0$ must have a negative mass-squared to destabilise the origin.
The resulting VEV then lies partly along the $H_u^0$ direction and partly along the $H_d^0$ direction, and is characterised by $\tan\beta = v_u / v_d$.
It is instructive (and motivated -- see Section~\ref{OptimalNaturalness}) to consider large $\tan\beta \gg 1$; in this limit $H_u^0$ alone corresponds to the single $CP$-even and neutral degree of freedom in the Standard Model $H$ doublet.
Its (negative) mass-squared -- the upper-left entry of the matrix in Eq.~\eqref{HiggsMassMatrix} -- then leads to all Standard Model masses.
For example we have 
\begin{equation} \label{REWSB1}
-\tfrac{1}{2} M_Z^2 = m^2_{H_u} + |\mu|^2 + \mathcal{O}((\tan\beta)^{-2}).
\end{equation}
An appealing feature of Susy, once it is broken, is the possibility of {\it radiative} EWSB.
Even if $m^2_{H_u}$ is positive at the high-scale $\Lambda$ where mediation of Susy breaking to the visible sector takes place, $H_u$ couples to the top sector via the large top Yukawa coupling resulting in a strong tendency for $m^2_{H_u}$ to be pushed negative by RG evolution from $\Lambda$ to the electroweak scale.
This strong radiative correction of $m_{H_u}^2$ motivates us to write this explicitly it as\begin{equation} \label{REWSB2}
-\tfrac{1}{2} M_Z^2 = m^2_{H_u}(\Lambda) + \delta m^2_{H_u} + |\mu|^2 + \mathcal{O}((\tb)^{-2}).
\end{equation}
With this equation we are ready to discuss naturalness.

\subsection{Naturalness Under Pressure} \label{NaturalnessUnderPressure}
Probably the most common measure of naturalness or fine-tuning is the Barbieri-Giudice measure~\cite{Barbieri198863}, which can be calculated for UV-complete models where a set of fundamental parameters $p_i$ set all of the masses at the scale $\Lambda$.
For a given point in $p_i$ space that results in the observed value of $M_Z$ (through Eq.~\eqref{REWSB2}), one calculates derivatives of $\log M_Z$ with respect to $\log p_i$; $M_Z$ is taken to be natural if all such derivatives are $\lesssim \mathcal{O}(1)$ -- doubling a fundamental parameter at most doubles the resulting $M_Z$.
If one of the derivatives is considerably larger than this, the associated $p_i$ needs to be finely tuned to produce the observed $M_Z$.
However this measure does not penalise a situation that we should still regard as unnatural.

If a single parameter $p$ sets both $m_{H_u}^2(\Lambda)$ and all of the (most important) masses involved in the radiative correction $\delta m_{H_u}^2$, it could set these in such a pattern that $m^2_{H_u}(\Lambda)$ and $\delta m^2_{H_u}$ happen to cancel each other out even if each term separately is very large, and we have insensitivity of $M_Z$ to the parameter $p$.
This is known as {\it focus-point Susy}.
However the cancellation depends sensitively not only on this mass-setting pattern but also on the value of the top Yukawa\footnote{
The squared top Yukawa is a prefactor to the $\delta m^2_{H_u}$ -- see Section~\ref{One-loop} -- so cancellation between $m^2_{H_u}(\Lambda)$ and  $\delta m^2_{H_u}$ to $1$ part in $N$ happens only for an {\it ad hoc} tuning of the top quark mass to $1$ part in $2N$.
}, and weakly on the scale $\Lambda$; unless these three are linked by some symmetry the cancellation is accidental.
A natural theory, by contrast, does not have large cancellations except those enforced by symmetries.
(High sensitivity to the imprecisely known top mass also makes it uncertain whether such models have EWSB at all~\cite{Allanach:2012qd}.)

A stricter criterion for naturalness is simply to ask that none of the terms contributing to the right-hand side of Eq.~\eqref{REWSB2} are dramatically larger than $\tfrac{1}{2} M_Z^2$, in the manner of Kitano and Nomura~\cite{Kitano:2005wc}.
This has the further advantage of allowing bottom-up deductions to be made, i.e. without knowing the underlying high-scale theory, for example as was done in~\cite{Papucci:2011wy} to lay out requirements on a natural spectrum.

The stop is the chief contributor to $\delta m^2_{H_u}$, and thus we require light stops for naturalness.
We can also see this intuitively -- the Standard Model Higgs couples most strongly to the top, so to protect it from strong radiative corrections we want Susy broken as weakly as possible in the top sector.
Searches at the Large Hadron Collider (LHC), however, exclude squarks up to masses exceeding $1$~TeV in the strongest cases~\cite{ATLAS-CONF-2012-109}; these are when all three generations of squarks are degenerate and decay to {\it hard} (see the glossary) jets and a light LSP carrying away large missing transverse energy $\slashed{E}_T$.

One way to ease the tension between such bounds on squarks and the desire for light stops is to change the way the squarks decay, for example to an LSP which is only a little lighter rather than a lot lighter.
In this case the jet which is also produced in the decay is forced to be very {\it soft} (see the glossary), and may fail to be detected or simply be less visible over the large {\it backgrounds} (see the glossary) for soft jets.
The {\it final state} (see the glossary) then has no large visible or invisible transverse energy at leading order, but may still be constrained due to the possibility of recoil against hard initial state radiation.
See~\cite{Dreiner:2012gx} (\cite{Belanger:2012mk}) where hard emission of a jet (photon) by the initial state in such circumstances is studied.

A second method of weakening these bounds is, rather than removing the visible energy, to remove the $\slashed{E}_T$ by having the LSP decay through an $R$ parity violating coupling: see for example~\cite{Allanach:2012vj}.
Hadronic $R$ parity violating decays are a prime example of a signal with rich jet substructure -- giving {\it fat jets} with large masses and containing many {\it subjets}~\cite{Hedri:2013pvl}.
In~\cite{Curtin:2012rm} such decays of boosted hadronising gluinos are found to show soft-radiation patterns as expected from colour singlets, with generalisations of the $N$-{\it subjettiness}~\cite{Thaler:2010tr} and {\it pull}~\cite{Gallicchio:2010sw} variables able to exploit this.

Thirdly, rather than hide the decay of the squarks one can suppress their production cross-section while still keeping them light, by having the gluino be a Dirac fermion rather than Majorana (requiring an extension of the MSSM field content).
Squark pair-production by t-channel gluino exchange is then suppressed -- see~\cite{Kribs:2012gx}.

Finally, what is undoubtedly the most effective way to keep stops light is to drop the assumption of approximate mass degeneracy between the three generations of squarks, having the first two generations heavy and the third generation light.
Constraints on the latter alone are considerably weakened due to direct production cross-sections suppressed by parton distribution functions, and the less distinctive final state signals that may result, typically being too similar to the large Standard Model top backgrounds.
Indeed stops decaying to tops and stable neutralinos can still be as light or lighter than the top quark~\cite{Chatrchyan:2013lya,CMS-PAS-SUS-12-023,ATLAS-CONF-2013-037,ATLAS-CONF-2012-166}, and a large number of alternative decays are possible, particularly if one extends the MSSM.

The more minimal MSSM in which we keep light only the particles important for naturalness (including the stops) and decouple the rest (including the first two generations of squarks) is referred to as {\it Effective} or {\it Natural} Susy, introduced in~\cite{Dimopoulos:1995mi,Cohen:1996vb} and revisited more recently in~\cite{Papucci:2011wy,Brust:2011tb}.
The authors of~\cite{Papucci:2011wy} argued that the inter-generational squark mass splitting should be a feature of the mediation of Susy breaking rather than an RG-running effect, since same coupling that drives the latter effect also gives strong running $m^2_{H_u}$ -- precisely what we are trying to avoid\footnote{
An exception would be a heavy right-handed sbottom at large $\tan\beta$, which would cause the left-handed stop to run lighter than the left-handed sup and scharm without driving the running $m^2_{H_u}$.
However it is the {\it integral} of the running stop masses that gives $\delta m^2_{H_u}$, so their starting heavy but running light only half solves the problem; furthermore we need both stops to be light, not just one.
I therefore do not consider this possibility to contradict the argument of~\cite{Papucci:2011wy}.
}.
Models which achieve such mediation include~\cite{Craig:2012di,Craig:2012hc} where Susy breaking occurs via gauge mediation but with some non-trivial interaction with flavour.
In~\cite{Craig:2012di}, the Standard Model gauge group is supplemented by a gauged flavour symmetry broken progressively from $SU(3)$ to $SU(2)$ to nothing; when this happens above the Susy-breaking mediation scale $\Lambda$ and both gauge groups are involved in the mediation, a light third generation results.
In~\cite{Craig:2012hc} the Standard Model gauge group splits, at high scales, into one $SU(5)$ group which mediates Susy breaking and one which does not (with a bifundamental link field obtaining a VEV to break to the diagonal group at low scales).
The first two generations are charged only under the mediating group; the third generation and Higgs fields are charged only under the non-mediating group, and thus couple less strongly to the source of Susy breaking.

The conclusion is that light-stop scenarios are desirable for naturalness, can be realised in concrete models, and are not excluded by direct searches.
They are, however, in some tension with another experimental result -- the putative Higgs signal of mass $\sim\!\!126~\text{GeV}$.
I discuss this in the following section.

\newpage
\section{Optimal Naturalness} \label{OptimalNaturalness}
{\it This section is based on my single-authored work~\cite{Wymant:2012zp}; the text here follows it closely.}\\

In the MSSM the tree-level $m_{h}$ is bounded from above by $M_Z\cos 2\beta$, and saturates this bound in the aforementioned decoupling limit.
Moderate to large $\tan\beta$ (say $5$ or greater) helpfully raises $\cos 2\beta$ (to $0.92$ or greater).
Even then, as has long been known, some substantial combination of stop mass- and stop mixing-induced corrections to $m_{h}$ is needed to lift it above the lower bound of $114.4~\text{GeV}$ set by the Large Electron-Positron Collider (LEP).
A $\sim\!\!126~\text{GeV}$ Higgs requires these corrections to be even more substantial, with correspondingly worse implications for naturalness.
Investigations of supersymmetric Higgs bosons in light of the $\sim\!\!126~\text{GeV}$ discovery has become a field in its own right; at the time of writing~\cite{Wymant:2012zp}, the interplay of parameters for such a Higgs mass when looking agnostically at the MSSM had been studied in \cite{Hall:2011aa,Baer:2011ab,Heinemeyer:2011aa,Arbey:2011ab,Draper:2011aa,Carena:2011aa,Cao:2011sn,Kang:2012sy,Desai:2012qy,Cao:2012fz,Lee:2012sy,Christensen:2012ei,Brummer:2012ns,Badziak:2012rf,CahillRowley:2012rv,Arbey:2012dq,Baer:2012up,Antusch:2012gv}.
Other works at that time had shown the implications of such a Higgs mass in particular models of Susy breaking, in extensions of the MSSM, or else had focused predominantly on issues relating to the decays of such a Susy Higgs into different final states; the literature concerning all of these topics has continued to grow since.

\subsection{Leading-Order Analysis} \label{One-loop}

The dominant radiative correction to the physics Higgs mass is
\begin{equation} \label{HiggsMass}
 \delta m_{h}^2 \approx \frac{3}{4\pi^2} \frac{m_t^4}{v^2} \left[ \log\left(\frac{\MS^2}{M_t^2}\right) + \frac{X_t^2}{\MS^2}\left(1 - \frac{X_t^2}{12\MS^2} \right) \right],
\end{equation}
where $v=174~\text{GeV}$, $X_t=A_t - \mu \cot\beta$ and $\MS$ is an average of the two stop masses.
The second term in square brackets is the threshold correction to the Higgs self coupling from integrating out both stops, and the first is the Standard Model Higgs self coupling beta function integrated (at leading log order) from that threshold down to the top mass scale, where the running Higgs mass coincides closely with the pole Higgs mass.
From the first term we see how large stop masses (which we don't want for naturalness) help to boost the Higgs mass (which we do want for $m_{h}\sim126~\text{GeV}$).
The second term -- the mixing term -- comes to our aid: it too can be used to raise $m_{h}$.
{\it Maximal mixing} refers to this term being maximal; it therefore allows minimal stop masses for a given $m_{h}$ and thus is naively the most natural arrangement, motivating much attention in the aforementioned literature and elsewhere.
However the mixing term, like the stop masses, also contributes to unnaturalness as we will see and so {\it a priori} it is not clear that maximising it is the best thing to do.

Unnaturalness arises from excessive running of $m^2_{H_u}$.
At one loop, the latter is~\cite{Martin:1993zk}:
\begin{equation} \label{mHuBetaFunction}
\begin{split}
16 \pi^2 \frac{d}{dt}m_{H_u}^2 = & \quad 6 y_t^2 (m_{\tilde{Q}_3}^2 + m_{\tilde{u}_3}^2 +A_t^2) \\
& + 6 y_t^2 m_{H_u}^2 -6g_2^2 M_2^2 -\frac{6}{5}g_1^2 M_1^2 + \frac{3}{5}g_1^2 \text{Tr}[Y_{\tilde{f}} m_{\tilde{f}}^2],
\end{split}
\end{equation}
where $t=\log(Q/\Lambda)$, with $\Lambda$ the high/mediation scale at which the soft Susy-breaking mass terms are generated.
One can roughly neglect the terms of the second line\footnote{
The effect of $m_{H_u}^2$ on its own running is small if the leading log approximation is valid (i.e. $(\mbox{one-loop factor}) \times \log(\Lambda/\MS) < 1)$. Then, since the overall radiative correction must be substantial enough to turn $m_{H_u}^2$ negative, the $m_{H_u}^2$ term in the beta function must be appreciably smaller than the other terms.
The electroweak couplings are somewhat smaller than $y_t^2$.
While the trace term is a sum over all scalars, it couples only through $g_1$ and is `relatively small in most known realistic models'~\cite{Martin:1997ns}.
For example it vanishes at the high scale in all models of General Gauge Mediation~\cite{Meade:2008wd}, and all models with universal scalar masses (such as minimal supergravity) since $\text{Tr}[Y] = 0$.
Furthermore the running of the trace is proportional to the trace itself.
The wino term on the other hand may be appreciable~\cite{CahillRowley:2012rv}, but here I will be differentiating with respect to stop-sector terms, so this effect drops out.
}; keeping only the large stop-sector terms, taking these to be constant and integrating gives the leading log expression
\begin{equation} \label{deltamHuLL}
 \delta m_{H_u}^2 \approx -\frac{3}{8\pi^2}\; y_t^2 \: (m_{\tilde{Q}_3}^2 + m_{\tilde{u}_3}^2 +A_t^2)\log\left(\frac{\Lambda}{\MS}\right),
\end{equation}
at a scale $\MS$ -- the scale at which Eq.~\eqref{REWSB1} holds most accurately~\cite{Gamberini1990331,PhysRevD.46.3981,deCarlos:1993yy}.

Before connecting Eq.~\eqref{deltamHuLL} to the physical Higgs mass $m_h$, I note that it tells us something about stop naturalness on its own.
It can be re-written in terms of the stop mass eigenvalues: taking the tree-level stop mass matrix without the subdominant electroweak $D$-term contributions, we have
\begin{equation} \label{deltamHuLL2}
\delta m_{H_u}^2 \approx -\frac{3}{8\pi^2}\; y_t^2 \: \left[m_{\tilde{t}_1}^2 + m_{\tilde{t}_2}^2 - 2m_t^2 + \frac{(m_{\tilde{t}_1}^2 - m_{\tilde{t}_2}^2)^2}{m_t^2} \cos^2\theta_{\tilde{t}}\,\sin^2\theta_{\tilde{t}} \right] \log\left(\frac{\Lambda}{\MS}\right)
\end{equation}
where $\theta_{\tilde{t}}$ is the stop mass mixing angle.
In~\cite{Lee:2012sy} it was argued that the final term in square brackets motivates $m_{\tilde{t}_1} \sim m_{\tilde{t}_2}$ for naturalness; then since the left-handed stop shares a mass with the left-handed sbottom ($m_{\tilde{Q}_3}$), non-observation of sbottoms translates into constraints on both stops.
However in Eq.~\eqref{deltamHuLL} we can define the average stop mass by $2\MS^2=m_{\tilde{Q}_3}^2 + m_{\tilde{u}_3}^2$, and there is explicit insensitivity to $m_{\tilde{Q}_3}^2 - m_{\tilde{u}_3}^2$ which will split the mass eigenvalues.
The discrepancy arises from the neglected $\cos^2\theta_{\tilde{t}}\sin^2\theta_{\tilde{t}}$ factor in Eq.~\eqref{deltamHuLL2}, which goes to zero as we pull apart $m_{\tilde{Q}_3}^2$ and $m_{\tilde{u}_3}^2$.
We see that in fact the two mass eigenvalues can be arbitrarily split without naturalness penalty.

I now want to find what Eq.~\eqref{deltamHuLL} tells us in conjunction with the physical Higgs mass-squared -- Eq.~\eqref{HiggsMass} plus the tree level value $\sim\!\!(M_Z\cos 2\beta)^2$.
Firstly, note from Eq.~\eqref{REWSB2} that the value of $|\mu|$ required for the correct $M_Z$ depends on the unknown high-scale value of $m_{H_u}^2$, and $|\mu|$ enters the physical Higgs mass expression through $X_t=A_t - \mu \cot\beta$.
However, (a) the aim for natural Susy is $|\mu| / (100~\text{GeV}) \lesssim \mbox{a few}$, (b) a large Higgs mass $\sim\!\!126~\text{GeV}$ needs\footnote{
Unless one enters the realm of split or high-scale Susy $\MS \gtrsim \mathcal{O}(10^{4,5}\,\mbox{GeV})$,~\cite{Giudice:2011cg}.
} $\tan\beta \gtrsim \mathcal{O}(5)$, and (c) later we will arrive at $A_t\gtrsim \mathcal{O}(1~\text{TeV})$.
Thus we expect $X_t$ to be very close to $A_t$ without knowing the precise value of $\mu$.

Secondly, we see that while the physical Higgs mass depends only on the average stop mass $\MS$, $\delta m_{H_u}^2$ depends on both $\MS$ and the precise linear combination $m_{\tilde{Q}_3}^2 + m_{\tilde{u}_3}^2$.
We then must choose a definition of $\MS$.
Often this is taken to be a geometric mean; the minimum $(m_{\tilde{Q}_3}^2 + m_{\tilde{u}_3}^2)$ for constant $(m_{\tilde{Q}_3}^2 \times m_{\tilde{u}_3}^2)^{1/2}$ then provides weak motivation for $m_{\tilde{Q}_3}^2 = m_{\tilde{u}_3}^2 $.
If instead the linear average $\MS^2 {\equiv} \tfrac{1}{2}(m_{\tilde{Q}_3}^2 + m_{\tilde{u}_3}^2)$ is chosen, the orthogonal linear combination is entirely free as previously mentioned.
A further alternative would be to take an average of the mass eigenvalues $m_{\tilde{t}_{1,2}}$: the dependence of $\delta m_{H_u}^2$ on the underlying parameters $m_{\tilde{Q}_3}^2,m_{\tilde{u}_3}^2,A_t$ then shifts very slightly but becomes much less transparent, as we have already seen.
We can appeal to the limit\footnote{
$\Lambda/\MS$ is very large in all but the most extreme cases; $m_{\tilde{t}_{2}}/m_{\tilde{t}_{1}}$ cannot be large if we integrate out both stops together to calculate the Higgs mass.
} $\log(\Lambda/\MS) \gg \log(m_{\tilde{t}_{2}}/m_{\tilde{t}_{1}})$, in which the former log and thus $\delta m_{H_u}^2$ has no sensitivity to how $\MS$ is defined.
We can thus take the aforementioned linear average, so that the functions $\delta m_h^2$ and $\delta m_{H_u}^2$ depend on the stop sector simply through $\MS$ and $A_t$.
Note that though other particles besides the stop make smaller contributions to both the physical Higgs mass and unnaturalness, below I will differentiate with respect to stop-sector parameters and so this effect drops out.

Having now made $\delta m_{h}^2$ and $\delta m_{H_u}^2$ functions of the stop sector through the parameters $\MS$ and $A_t$ only, we can find {\it optimal naturalness} -- maximal $\delta m_{h}^2$ for minimal $\delta m_{H_u}^2$ -- with Lagrange constrained optimisation.
The solution of
\begin{equation} \label{LagrangeMin}
 \frac{\partial}{\partial (\MS^2)} \left(\delta m_{h}^2 - \lambda \, \delta m_{H_u}^2 \right) = \frac{\partial}{\partial (A_t^2)} \left(\delta m_{h}^2 - \lambda \, \delta m_{H_u}^2 \right) = 0,
\end{equation}
where $\lambda$ is the unspecified Lagrange multiplier, gives the most natural ratio $x\equiv A_t^2/\MS^2$, with the scale of one of these two dimensionful parameters freely chosen thereafter.
Using $\delta m_h$ at one-loop~\eqref{HiggsMass} and $\delta m_{H_u}^2$ at leading log~\eqref{deltamHuLL}, I find
\begin{equation} \label{MostNaturalRatio1loop}
x_{\text{natural}} \equiv \left(\frac{A_t^2}{\MS^2}\right)_{\text{natural}} = 2 + \sqrt{4+ \frac{6(L-2)}{L-1}} \quad\sim\: 5,
\end{equation}
with $L=\log(\Lambda^2 / \MS^2)$.
The solution is real for $L>\tfrac{8}{5}$, asymptotes to $2+\surd 10 \approx 5.16$ as $L\rightarrow \infty$, and is already $5$ for $L=7$ (i.e. $\Lambda / \MS = 33$) -- thus it is essentially constant over phenomenologically interesting mediation scales and stop masses.
That the optimal $x$ should be {\it close} to six is not surprising: using the logarithmic stop mass term to boost the Higgs mass requires exponentially heavy stops and thus exponentially bad fine-tuning; whereas the stop mixing term contribution to $m_h$ can be large even for small $A_t^2$ and $\MS^2$, provided their ratio is favourable.
However the optimal $x$ must in fact be {\it less} than the maximal mixing value $x=6$: decreasing it from $6$ to $6-\delta$ reduces the physical Higgs mass by $\mathcal{O}(\delta^2)$ but increases naturalness by $\mathcal{O}(\delta)$.
We see that {\it almost maximal mixing} is optimal.

\subsection{Higher-Order Effects} \label{Two-loop}

\begin{figure}[!ht]
\centering
\includegraphics[width=0.55\linewidth]{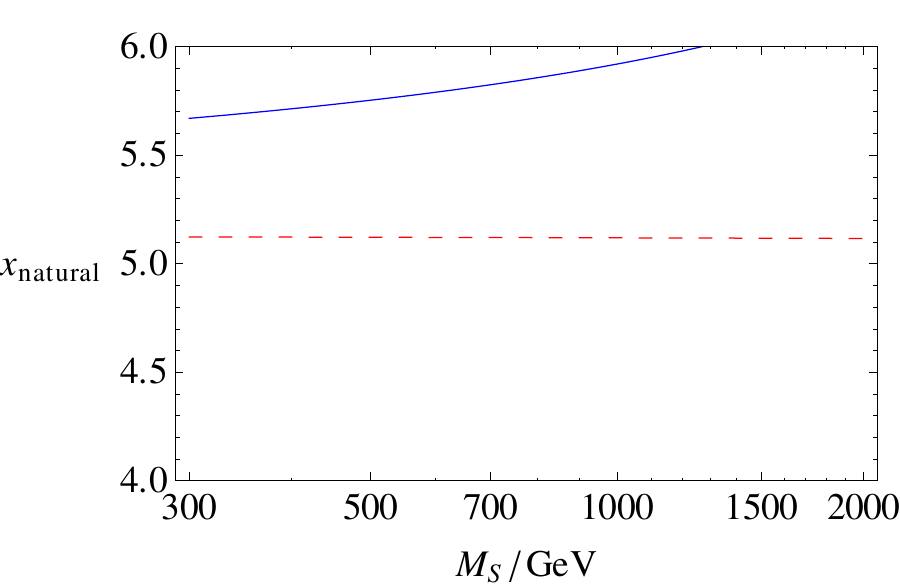}
\caption{The most natural ratio $x\equiv A_t^2/\MS^2$ obtained from maximising the Higgs mass at one loop (red, dashed) and two-loop (blue, solid) for constant electroweak symmetry breaking term $\delta m_{H_u}^2$, as a function of the average stop mass $\MS$.}
\label{fig:RatioNLO}
\end{figure}

Higher order effects of the stop on the physical Higgs mass can be taken into account with the two-loop expression of~\cite{Carena:1995bx}:
\begin{gather} 
 \delta m_{h}^2 = \frac{3}{4\pi^2}\frac{m_t^4}{v^2}\left[ \frac{1}{2}\tilde{X}_t + \left( 1+D \right) T +\epsilon\left(\tilde{X}_t T+T^2\right) \right]\,, \label{CarenaWagner} \\
\mbox{with} \quad m_t = \frac{M_t}{1+\frac{4}{3\pi}\alpha_3(M_t)}, \notag \\
\alpha_3(M_t) = \frac{\alpha_3(M_Z)}{1+\frac{23}{12\pi}\alpha_3(M_Z)}, \notag \\
T=\log\frac{ \MS^2}{M_t^2}, \notag \\
D = -\frac{M_Z^2}{2m_t^2}\cos^2 2\beta, \notag \\
\tilde{X}_{t} = \frac{2A_t^2}{\MS^2} \left(1 - \frac{A_t^2}{12 \MS^2} \right), \notag \\
\mbox{and} \quad \epsilon = \frac{1}{16\pi^2}\left(\frac{3}{2}\frac{m_t^2}{v^2}-32\pi\alpha_3(M_t) \right) \notag
\end{gather}
(which also includes the smaller, soft-mass independent, one-loop $D$-term $\mathcal{O}(M_Z^2 m_t^2)$ of~\cite{Brignole:1992uf}).
The optimisation, Eq.~\eqref{LagrangeMin}, goes through exactly as before.
The solution is the positive root of the following equation (which recovers Eq.~\eqref{MostNaturalRatio1loop} as $D,\epsilon \rightarrow 0$)
\begin{multline} \label{MostNaturalRatio2loop}
   \left[  1 + 2\epsilon T + L (-1 + \epsilon -2\epsilon T)\right] x_{\text{natural}}^2 \\
+ 4\left[ -1 - 2\epsilon T + L (1 -3\epsilon +2\epsilon T) \right] x_{\text{natural}} \\
- 6\left[  2 + 4\epsilon T + L (-1 + D -2\epsilon T) \right]= 0
\end{multline}
I show the variation of this solution with $\MS$ in Fig.~\ref{fig:RatioNLO}; dependence on \linebreak\mbox{$\tan\beta \in [5,45]$} and the top quark mass uncertainty is negligible.

Two other approaches are trivially equivalent to using Eq.~\eqref{LagrangeMin} to find $x_{\text{natural}}$.
Firstly, one could invert the $\delta m_{H_u}^2$ expression to find the function $\MS(x)|_{\delta m_{H_u}^2}$ for how the stop mass must vary as a function of $x$ in order to keep $\delta m_{H_u}^2$ constant: from Eq.~\eqref{deltamHuLL} this monotonically decreasing function is
\begin{equation}
\MS(x)|_{\delta m_{H_u}^2} = \Lambda\, \exp\!\left(\tfrac{1}{2}\, W_{-1}\!\left( \frac{-16\pi^2\,\delta m_{H_u}^2}{(2+x)\Lambda^2} \right)\,   \right)
\end{equation}
where $W_{-1}(\ldots)$ is the lower branch of the Lambert $W$ function\footnote{the multivalued function $W(z)$ satisfying $z = We^W$, with the lower branch $W_{-1}$ defined in the interval $(-\infty,-1/e]$.}.
The one-parameter function $\delta m_h^2(x,\,\MS(x)|_{\delta m_{H_u}^2})$ then gives the range of Higgs masses possible for a given $\delta m_{H_u}^2$; the {\it maximum} occurs at $x_{\text{natural}}$.

\begin{figure}
\centering
\subfigure{
\includegraphics[width=0.47\linewidth]{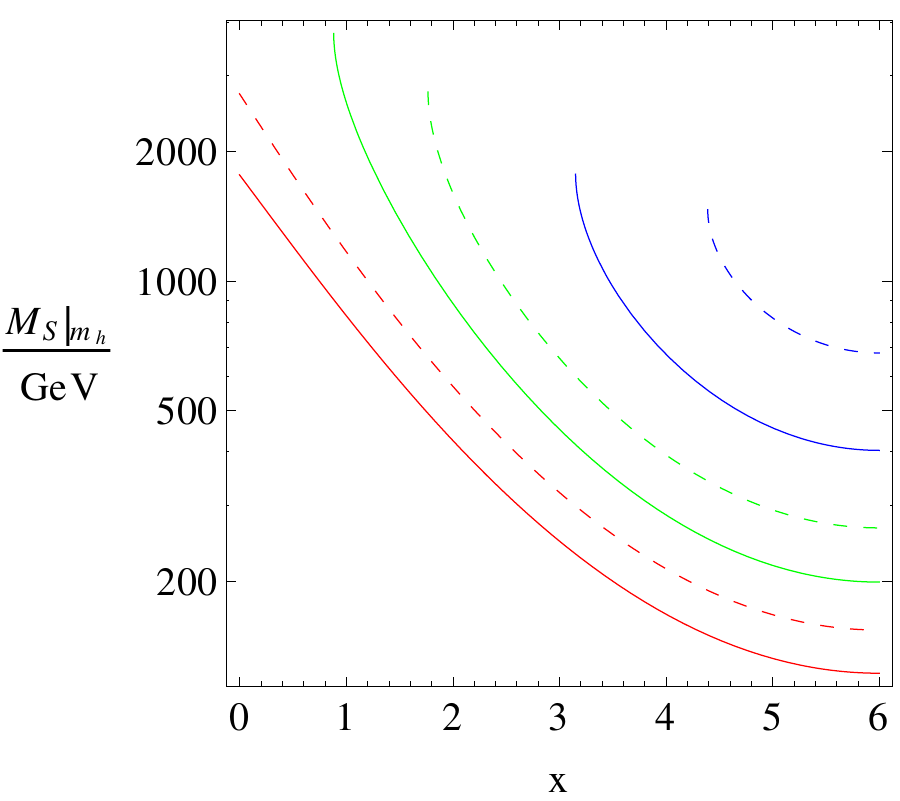}
}
\hspace*{5mm}
\subfigure{
\includegraphics[width=0.43\linewidth]{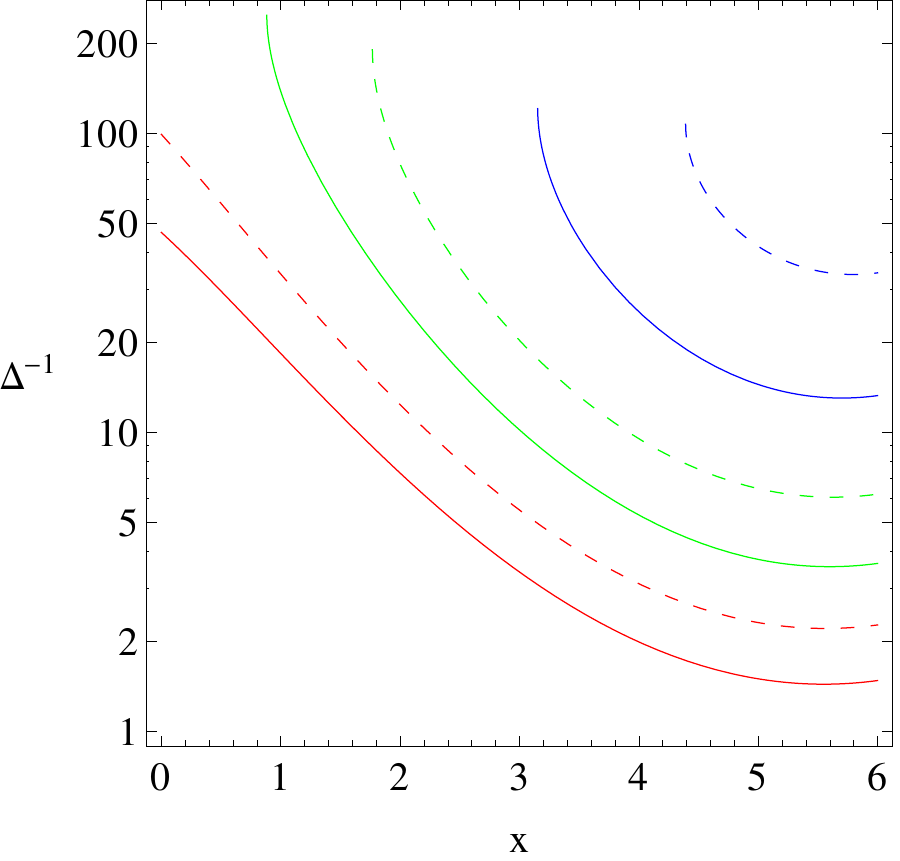}
}
\caption{$x$ axis: $x\equiv(A_t^2/\MS^2)$.
The left panel shows the average stop mass $\MS$ required for constant Higgs mass $m_h$; the right panel shows the fine-tuning $\Delta^{-1}\equiv|\delta m_{H_u}^2| / (\tfrac{1}{2}M_Z^2)$ that results. The mediation scale $\Lambda$ is taken to be $10^5$~GeV ($10^{16}$~GeV would increase the fine-tuning by a factor $\sim\!6$).
Red curves (the lowest two) have $m_h=115$~GeV, green curves (the middle two) $m_h=119$~GeV, and blue curves (highest) $m_h=123$~GeV. Dashed (solid) lines have $\tan\beta=8$ ($30$).
I take $M_t = 173.1$~GeV.}
\label{fig:ConstHiggsMasses}
\end{figure}

Secondly, one could invert the $\delta m_{h}^2$ expression to find the function $\MS(x)|_{\delta m_{h}^2}$ for how the stop mass varies as a function of $x$ for a constant Higgs mass.
This function is easily obtained from Eq.~\eqref{CarenaWagner} which is a quadratic equation in $\log(\MS^2/M_t^2)$; I plot it in the left panel of Fig.~\ref{fig:ConstHiggsMasses}.
The one-parameter function $\delta m_{H_u}^2(x,\,\MS(x)|_{\delta m_{h}^2})$ then gives the range of $\delta m_{H_u}^2$ possible for a given Higgs mass, depending on the amount of stop mixing ($x$) one uses to achieve that Higgs mass.
The {\it minimum} occurs at $x_{\text{natural}}$.
I plot this in the right panel of Fig.~\ref{fig:ConstHiggsMasses}, normalised to $\tfrac{1}{2}M_Z^2$ for a transparent indication of fine-tuning.

The different colours (line styles) in Fig.~\ref{fig:ConstHiggsMasses} correspond to different $m_h$ ($\tan\beta$), see the caption.
We see that the greater the $m_h$ we require (and the lower $\tan\beta$ is), the larger $x$ must be to even find a solution: no-mixing scenarios are more limited in the Higgs mass they can reach before the $m_h$ expression~\eqref{CarenaWagner} breaks down.
Indeed as was noted in~\cite{Draper:2011aa}, even using the program {\tt FeynHiggs}~\cite{Frank:2006yh,Degrassi:2002fi,Heinemeyer:1998np,Heinemeyer:1998yj} for a higher-order calculation, in the no-mixing $x=0$ scenario breakdown occurs before one can reach $m_h\approx126$~GeV and one must resort to a matching of the MSSM onto the Standard Model (with RG evolution from $\MS$ to $M_t$ resumming the large logs of this ratio of scales).

The left panel of Fig.~\ref{fig:ConstHiggsMasses} illustrates the obvious fact that the smallest stop mass for a given Higgs mass occurs at exactly maximal mixing $x=6$.
Close inspection of the right panel shows the more subtle point that the lowest fine-tuning occurs at {\it almost} maximal mixing.
We see from the flatness of the curve for $x\in[5,6]$, however, that the difference between the two is essentially nil.

\subsection{Working Backwards -- Finding {\it The} Stop Mass?}

\begin{figure}
\centering
\subfigure{
\includegraphics[width=0.47\linewidth]{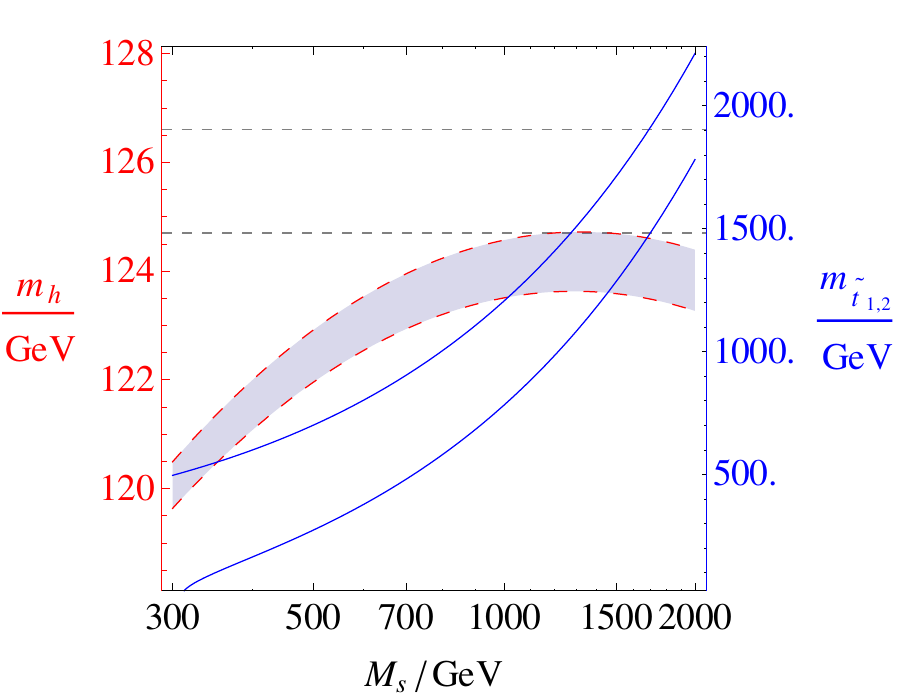}
}
\subfigure{
\includegraphics[width=0.47\linewidth]{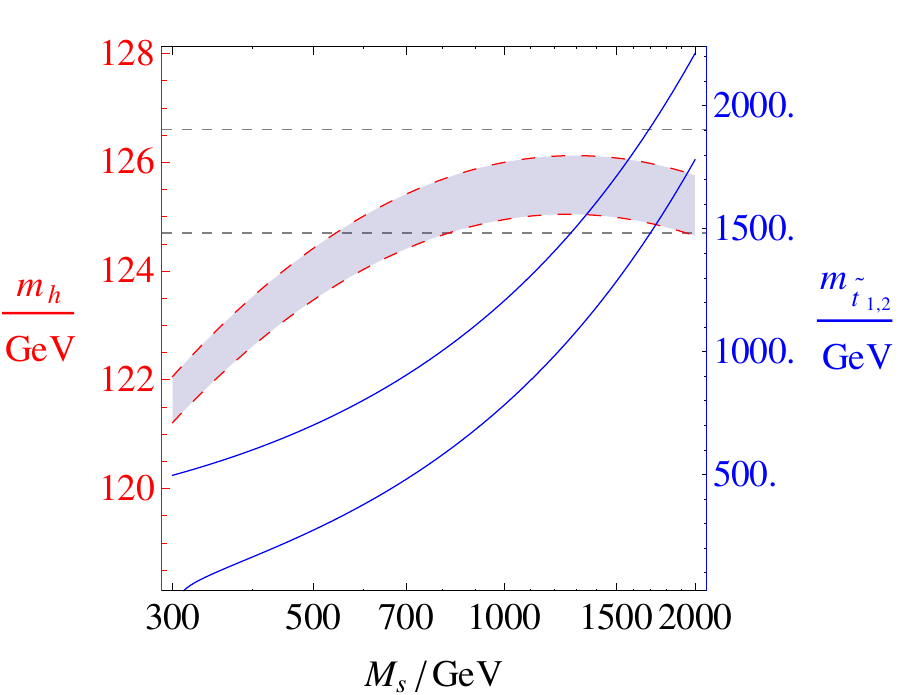}
}
\\
\subfigure{
\includegraphics[width=0.47\linewidth]{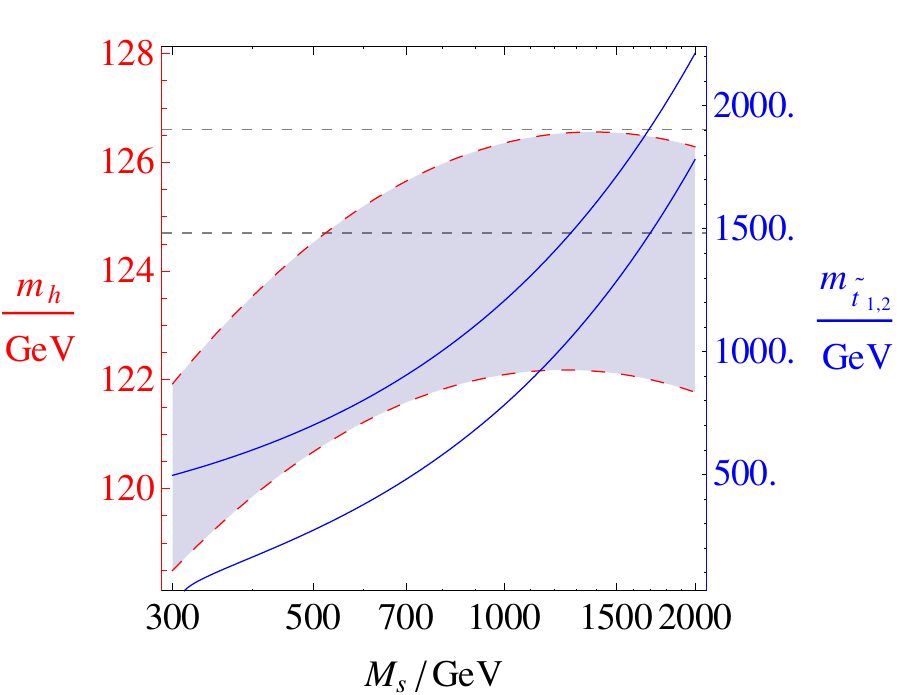}
}
\subfigure{
\includegraphics[width=0.47\linewidth]{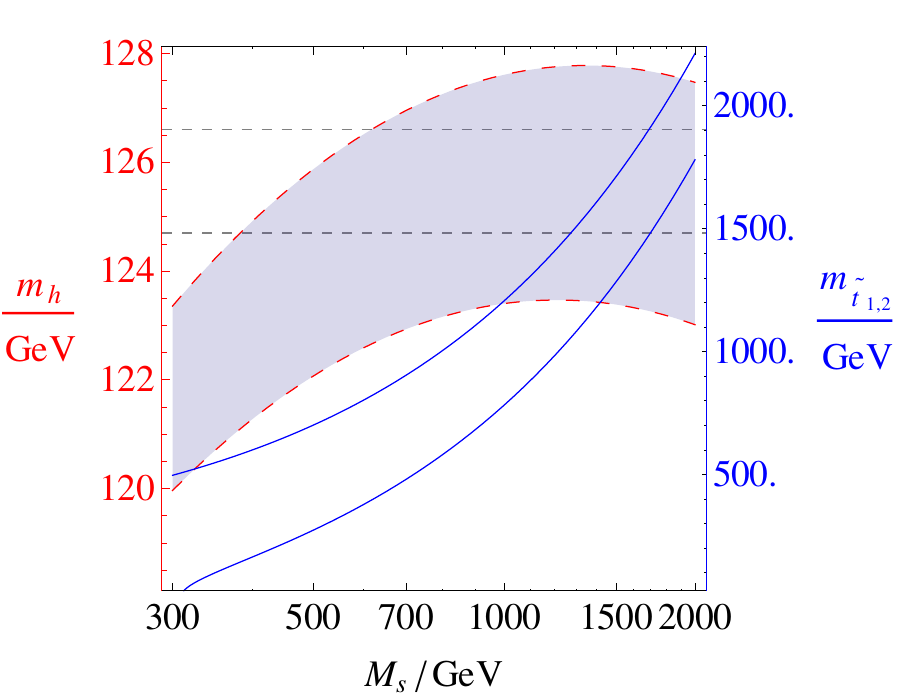}
}
\caption{Blue solid lines, right-hand $y$ axis: the tree-level stop mass eigenvalues $m_{\tilde{t}_{1,2}}$, assuming $m_{\tilde{Q}_3}^2 = m_{\tilde{u}_3}^2$.
Red dashed lines, left-hand $y$ axis: the two-loop expression of~\cite{Carena:1995bx} for the mass of the lightest CP-even Higgs boson $m_{h}$, valid for $850~\text{GeV} \lesssim \MS \lesssim 1500~\text{GeV}$.
Further (grey) dashed lines indicate the lowest $m_h$ compatible with the ATLAS Collaboration's measurement $m_h = (125.5 \pm 0.2\,(\text{stat})\: _{-0.6}^{+0.5}\,(\text{syst}))$~GeV \cite{ATLAS-CONF-2013-014} and the highest $m_h$ compatible with the CMS Collaboration's measurement $m_h = (125.8 \pm 0.4\,(\text{stat})\, \pm 0.4(\text{syst}))$~GeV \cite{CMS-PAS-HIG-12-045}.
Curves are plotted as a function of the average stop mass $\MS$, with the ratio $A_t^2/\MS^2$ taking its {\it most natural} value as defined in Eq.~\eqref{MostNaturalRatio2loop} and plotted in Fig.~\ref{fig:RatioNLO}.
Grey shading shows the Higgs mass uncertainty due to the top quark mass uncertainty.
Upper panels take the top quark pole mass as measured by the ATLAS and CMS Collaborations and the Tevatron: $M_t = (173.1 \pm 0.7) $~GeV  \cite{Degrassi:2012ry}; lower panels take $M_t = (173.3 \pm 2.8)$ as extracted from the Tevatron's $\sigma(pp\rightarrow t\bar{t}+X)$ measurement~\cite{Alekhin:2012py}.
The left (right) panels are for $\tan\beta=9$ ($30$).}
\label{fig:HiggsMassesNLO}
\end{figure}

Varying $\MS$ while keeping $x=x_{\text{natural}}$ fixed traces out the Higgs mass that results in this most natural setting.
Of course to go to the full Higgs mass from only the stop radiative corrections one must either neglect the corrections from other sparticles (and so certainly steer clear of the large bottom-Yukawa regime at $\tan\beta \gtrsim \frac{m_t}{m_b}$), or else pick some `representative' value for all other sparticle masses and calculate their fixed contribution.
I do the former in Fig.~\ref{fig:HiggsMassesNLO}.
I will first explain the range of validity of Fig.~\ref{fig:HiggsMassesNLO} before discussing the uncertainty arising from the top quark mass, shown with grey bands.

The Higgs mass expression~\eqref{CarenaWagner} arises from the effective theory in which both stops have been integrated out at a single scale $\MS$, thus requiring \mbox{$m_{\tilde{t}_1} \gtrsim \tfrac{3}{5}\,m_{\tilde{t}_2}$} \cite{Carena:1995bx} which is tantamount to a lower bound on $\MS$ for validity of the expression.
The lower bound is minimal when $m_{\tilde{Q}_3}^2 - m_{\tilde{u}_3}^2$ (which I have argued can be freely chosen) vanishes; I plot the resulting stop mass eigenvalues also in Fig.~\ref{fig:HiggsMassesNLO}.
The bound $m_{\tilde{t}_1} \gtrsim \tfrac{3}{5}\,m_{\tilde{t}_2}$ can be seen to imply $\MS\gtrsim 850$~GeV.
Eq.~\eqref{CarenaWagner} also does not contain higher-order logs $\mathcal{O}(\log^3(\MS^2/M_t^2))$, giving a corresponding upper bound for its validity.
Its accuracy is $\sim\! 2$~GeV for $\MS \lesssim 1.5$~TeV \cite{Carena:1995bx}.

Notice in Fig.~\ref{fig:RatioNLO} that at $\MS\sim 1.3$~TeV, $x_{\text{natural}}$ becomes as high as $6$ (and takes higher values still for $\MS \gtrsim 1.3$ TeV).
This signals a breakdown in my procedure, since the Higgs mass expression is a symmetric function of $x$ about the value $6$, but naturalness always favours lower values to minimise $A_t$.
From Fig.~\ref{fig:HiggsMassesNLO}, we see that at $\MS\sim 1.3$~TeV the derivative of the Higgs mass with respect to $\MS$ vanishes\footnote{
Note that the derivative of interest is $m_h$ with respect to $\MS$, with $x$ held constant; in Fig.~\ref{fig:HiggsMassesNLO} the latter is {\it not} constant.
However it is varying sufficiently slowly that when we instead hold it exactly constant, the relevant derivative still vanishes at the same point $\MS\sim 1.3$~TeV.
}, which is purely an artefact of the truncated expression.
The Lagrange constrained optimisation, Eq.~\eqref{LagrangeMin}, is then solved by the Higgs mass {\it alone} maximised with respect to both of its arguments, with the Lagrange multiplier $\lambda$ vanishing i.e. the naturalness consideration decouples.
Hence the solution is pushed onto exactly maximal mixing.
Even higher order terms in the Higgs mass expression would be needed to push this breakdown point out to higher stop masses.

The authors of~\cite{Degrassi:2012ry}, following a similar analysis to~\cite{Bezrukov:2012sa}, take the top quark mass measurement relevant for calculation of the (Standard Model) Higgs mass to be a combined measurement of the pole mass from the ATLAS and CMS Collaborations and the Tevatron: $M_t = (173.1 \pm 0.7) $~GeV.
In~\cite{Alekhin:2012py} it was argued that direct experimental measurement of the top quark pole mass gives a theoretically ill-defined quantity, and that a more theoretically rigorous approach is to extract the running mass from measurement of the top pair production cross-section, and thence obtain $M_t = (173.3 \pm 2.8)$~GeV.
I show both cases in Fig.~\ref{fig:HiggsMassesNLO}; the choice of error in $M_t$ has a striking effect on the Higgs mass uncertainty.

An initial hope for this work was to see whether a given Higgs mass could give an indication of the average stop mass, using the principle of optimal naturalness to reduce the function\footnote{
Necessarily for a choice of $\tan\beta$; large but less than $\frac{m_t}{m_b}\sim 40$ gives us a maximal Higgs mass and without a fine-tuning penalty, which is clearly optimal.
}
\begin{equation} \label{1DHiggs}
m_h\; \approx \;m_h(\MS,x) \quad \rightarrow \quad\left.m_h(\MS)\right|_{x=x_{\text{nat}}}
\end{equation}
The latter is shown in Fig.~\ref{fig:HiggsMassesNLO} with its uncertainty arising from the top quark mass uncertainty $\Delta M_t$; where it intersects with the observed Higgs mass, which has its own error $\Delta m_h$, we see which stop masses are possible.
Fig.~\ref{fig:HiggsMassesNLO} shows that even with the simplification of Eq.~\eqref{1DHiggs}, the uncertainties $\Delta M_t$ and $\Delta m_h$ alone make any inference of $\MS$ from $m_h$ very difficult.
This is compounded by a theoretical uncertainty in the calculation of $m_h$, widely taken to be \mbox{$\sim\!\!3$~GeV} \cite{Allanach:2004rh} (or perhaps larger still for heavy or highly non-degenerate stops), and the smaller contributions from the particles besides the stops.
The smallest $\MS$ compatible with the observed $m_h$, found for large $\tan\beta$ and conservative $\Delta M_t$, can be read off as $\sim\!\! 350$~GeV; however the stop quark mass eigenvalues are then too split to trust a calculation based on integrating them both out at once (as discussed earlier).
Fig.~\ref{fig:HiggsMassesNLO} also makes clear that even higher order terms than the two-loop corrections to $m_h$ are necessary to constrain $\MS$ from above, as the monotonic increase of $m_h$ with $\MS$ needs to be captured.
(An upper limit on $\MS$ {\it without} naturalness is given in~\cite{Giudice:2011cg} -- a few $10^{8}$~GeV for split Susy and unconstrained for high-scale Susy; MSSM-to-SM matching is needed to calculate $m_h$ with stops far beyond the weak scale.)

I consider RG improvement to go beyond a leading log expression for $\delta m_{H_u}^2$, but relegate this to Appendix~\ref{OptimalNaturalnessBeyondLeadingLogdeltamHu2} as the discussion is more involved though ultimately gives the same $x_{\text{natural}}$.

\subsection{Implications}

This study being analytic throughout, it is complementary to the many numerical investigations of the Higgs in Susy performed recently, illustrating more clearly the Higgs-stop-naturalness connection.
I have shown that {\it almost maximal mixing}, with $x\equiv A_t^2 / \MS^2$ slightly lower than $6$, is optimal; though I have also shown that the distinction between this case and maximal mixing $x=6$ is academic.
In other words to achieve a given Higgs mass $m_h$, balancing $A_t$ and $\MS$ to optimise naturalness gives almost the same result as simply trying to minimise $\MS$.
(Note that maximal mixing is not `needed' to achieve $m_h\approx126$~GeV, as has been reported for example in~\cite{Haisch:2012re} -- Fig. 6 of~\cite{Draper:2011aa} for example shows that $\MS=\mathcal{O}(5~\text{TeV})$ is sufficient with {\it no} mixing -- it is merely a less tuned method of doing so.)

However conversely, even remaining in the MSSM, comparing how easily different models accommodate a $126$~GeV Higgs (a major focus of recent Susy phenomenology) based on {\it how light the stops are} is misleading.
A maximal-mixing scenario will certainly have larger $m_h$ than a no-mixing scenario at the same $\MS$.
But note that the reasoning of the previous paragraph applies to a fixed mediation scale $\Lambda$.
If the maximal-mixing scenario has much larger $\Lambda$ than the no-mixing scenario -- e.g. if we take the former to represent supergravity and the latter low-scale gauge-mediated Susy breaking (GMSB) -- then it will be more unnatural not only due to the large $A_t$ but also due to large amount of RG running, i.e. the large logarithm in Eq.~\eqref{deltamHuLL}.
Taking $\Lambda = 10^{16}$~GeV and $10^{5}$~GeV as representative of these two cases, the former will have an unnatural $\delta m_{H_u}^2$ term $\sim\!25$ times larger; the two should thus compare their Higgs masses with GMSB having stops (roughly $\surd 25$ times) heavier than supergravity for similar fine-tuning, changing perhaps qualitatively the result of a comparison at fixed $\MS$, c.f.~\cite{Arbey:2011ab,Arbey:2012dq}.
The optimal situation in the MSSM is clearly (nearly) maximal mixing with low-scale mediation: ~\cite{Kang:2012ra} and~\cite{Craig:2012xp} realised this with the introduction of large $A_t$ terms into GMSB via Higgs-messenger superpotential couplings.

Investigating whether minimal stop masses coincide with optimal naturalness, as done here for the MSSM, is particularly important for extensions of the MSSM.
Introducing a new particle which couples strongly to the Higgs, in order to boost the latter's mass without heavy stops, is naively good for naturalness.
However pushing this new coupling as far as it will go can easily be imagined to introduce a new source of tuning somewhere in the theory (analogous to the effect of large $A_t$ on $\delta m_{H_u}^2$ considered here).
Exactly this effect in the Next-to-Minimal Supersymmetric Standard Model (NMSSM) was considered in~\cite{Agashe:2012zq}: the stop masses and mixing were kept small, and the naturalness implications of NMSSM-specific contributions to $m_h$ calculated.
These were found to be a tuning of the lighter scalar's couplings to hide it from current collider constraints, in the case where the $126$~GeV Higgs is the second-lightest scalar; and worse tuning still when it is the lightest scalar, to undo the push-down effect of level-repulsion between mass eigenvalues.
Nevertheless these tunings are at the level of $1$ part in $5$, making the NMSSM more natural when compared to the inescapable tuning of $1$ part in $\sim\!\!100$ in the MSSM to obtain $m_h\approx126$~GeV \cite{Hall:2011aa}.

\newpage
\section{The Higgs In The NMSSM} \label{NMSSM}
{\it This section is based on my work~\cite{PhysRevD.86.035023} done in collaboration with Daniel Albornoz Vasquez, Genevi\`{e}ve B\'{e}langer, C\'{e}line B\oe{}hm, Jonathan Da Silva and Peter Richardson; the text has been mostly re-written.}\\

\subsection{Introducing The NMSSM} \label{NMSSMintro}

The NMSSM (see~\cite{Ellwanger:2009dp} for a review and original references) contains the same field content as the MSSM -- Table~\ref{MSSMcontent} -- with the addition of a gauge-singlet superfield $S$.
The latter contains a neutral fermion -- the singlino $\tilde{S}$, which mixes with the four MSSM-like neutralinos to give $\chi_{i=1\ldots5}^0$; together with a neutral scalar and pseudo-scalar, which mix with the doublet-like/MSSM-like $h$ and $H$ to give $H_{i=1,2,3}$, and with the MSSM-like $A$ to give $A_{1,2}$, respectively.

A major motivation for this extension is to solve the $\mu$-{\it problem} of the MSSM.
From Eq.s~\eqref{HiggsMassMatrix} and~\eqref{REWSB1} it is clear that $|\mu|^2$ cannot be considerably greater than the Higgs soft mass-squareds (particularly $m_{H_u}^2$), or else electroweak symmetry will not be broken.
However this is precisely what we would expect in the MSSM as $\mu$ is supersymmetric (appearing in the superpotential) and hence its natural scale is the cutoff of the supersymmetric theory, which we are imagining might be the GUT scale or beyond.
To tie the size of $\mu$ to the size as the soft terms one could appeal to the anthropic principle, but since the latter may address the small size of the Standard Model Higgs mass {\it without} Susy, a primary motivation is lost.
Instead one can have an effective $\mu$ term arise as the VEV of a new superfield, which only acquires a VEV due to Susy breaking; in the NMSSM that new superfield is the gauge singlet $S$.

Generally of course the explicit $\mu$ term is also present, and by introducing $S$ there are two more dimensionful supersymmetric terms possible -- a mass and a tadpole for $S$ -- which also need to be suppressed well below values comparable to the cutoff scale in order to allow EWSB.
\begin{equation}
\begin{split}
W_{\rm NMSSM} = &\: W_{\rm MSSM,\:Yukawas} + W_{\rm Higgs-only}\\
\text{with} \quad W_{\rm Higgs-only} = &\: \lambda S H_u H_d + \mu H_u H_d + \chi_F S + \tfrac{1}{2} \mu' S^2 + \tfrac{1}{3} \kappa S^3
\end{split}
\end{equation}
The three dimensionful supersymmetric terms -- $\mu H_u H_d$, $\chi_F S$ and $\tfrac{1}{2} \mu' S^2$ -- can be forbidden with a $\mathbb{Z}_3$ symmetry, giving a scale-less superpotential only containing terms with three superfields.
If this discrete symmetry (or any other) is exact down to the electroweak scale where it is spontaneously broken, cosmologically unacceptable domain walls would exist between regions of space with different charges under the symmetry -- {\it bubbles}~\cite{Vilenkin:1984ib}.
It is therefore desirable to suppress such terms with only an approximate symmetry (see the discussion in~\cite{Ellwanger:2009dp}).

The $\mathbb{Z}_3$ symmetry forbidding dimensionful supersymmetric terms also forbids three Susy-breaking terms: $\mathcal{L} \slashed{\supset} -b H_u H_d + \tfrac{1}{2}m_S'^2 S^2 + \chi_S S + \text{c.c.}$.
The soft terms that are present and contribute to the Higgs potential at tree-level are
\begin{multline}
 -\mathcal{L}_{\rm soft,\:Higgs} = m_{H_u}^2 |H_u|^2 + m_{H_d}^2 |H_d|^2 + m_{S}^2 |S|^2\\
+ (\lambda A_\lambda H_u H_d S + \tfrac{1}{3}\kappa A_\kappa S^3 + \text{c.c.})
\end{multline}
The coefficients of these five terms, together with those of the two remaining $\mathbb{Z}_3$-symmetric terms in $W_{\rm Higgs-only}$ (that is, $\lambda S H_u H_d$ and $\tfrac{1}{3} \kappa S^3$) define the seven parameters in the $\mathbb{Z}_3$-symmetric NMSSM Higgs sector: $m_{H_u}^2$, $m_{H_d}^2$, $m_{S}^2$, $A_\lambda$, $A_\kappa$, $\lambda$ and $\kappa$.
The three scalar mass-squareds can be traded for the observed $M_Z$, $\tb$, and the effective $\mu$ term $\mu_{\rm eff} \equiv \lambda\langle S \rangle$. 

\begin{figure}	
\centering
\includegraphics[width=0.5\textwidth]{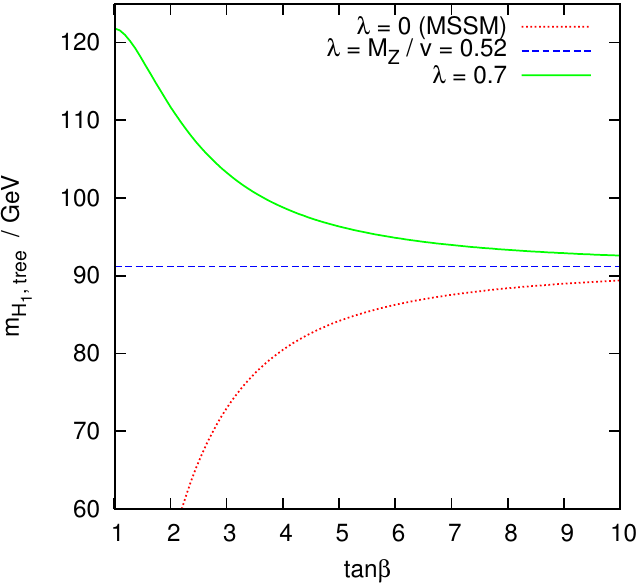}
\caption{The maximum value (achieved in the decoupling limit) of the tree-level mass of a purely doublet-like $H_1$ in the NMSSM: $m_{h,\text{tree}} = \surd(M_Z^2\cos^2 2\beta + \lambda^2 v^2 \sin^2 2\beta)$.
A singlet-doublet coupling of strength $\lambda\approx0.7$ is the maximum allowed by perturbativity / absence of Landau poles for a range of $\tb$, however as $\tb$ approaches $1$ a much smaller $\lambda$ is required.
}
\label{fig:NMSSMHiggsmass}
\end{figure}

The physical Higgs mass is boosted at tree-level, compared to the MSSM, by the coupling to the singlet: a purely doublet-like $H_1$ has $m_{H_1,\text{tree}}^2 \leq$\linebreak$M_Z^2\cos^2 2\beta + \lambda^2 v^2 \sin^2 2\beta$ with equality in the decoupling limit.
Vanishing $\lambda$ is the MSSM case, for which $m_h$ plummets as $\tan\beta\rightarrow 0$; for $\lambda=M_Z/v \approx 0.52$, $m_h=M_Z$ independent of $\tan\beta$; and $\lambda>0.52$ is when we start seeing a boost beyond what is possible in the MSSM: see Fig.~\ref{fig:NMSSMHiggsmass}.
There is however an upper limit on the size of $\lambda$ at the low scale in order to have it not encounter a Landau pole before the GUT scale -- this upper limit is $\sim\!0.7$ but becomes much smaller at low $\tan\beta$.
Pushing into the low-$\tb$ large-$\lambda$ region, which is desirable for an enhanced tree-level $m_h$, therefore pushes the viability of perturbation theory.
According to~\cite{Barbieri:2006bg,Hall:2011aa} minimal tuning is achieved for maximally non-perturbative $\lambda$ (namely $\lambda\approx2$), where a Landau pole is encountered and at the lowest scale permitted by precision measurements $\sim\!\!10$~TeV.
This was challenged in~\cite{Gherghetta:2012gb} where, with a slightly different measure of fine-tuning, $\lambda\sim1$ was found to be preferred.

\subsection{Scanning Parameter Space} \label{SettingUpNMSSMscans}

In~\cite{PhysRevD.86.035023} my collaborators and I analysed the Higgs and collider phenomenology of two scans of NMSSM parameter space which had been performed in~\cite{Vasquez:2010ru,AlbornozVasquez:2011js,AlbornozVasquez:2012px} and explored in a predominantly astrophysical context.
Details of how the scans were set up can be found in the original works, however I will summarise here.
\begin{itemize}
\item The key difference between the two scans was that the earlier one demanded a lightest-neutralino mass $m_{\chi_1^0} < 15$~GeV, motivated by hints of a signal in direct detection experiments~\cite{Aalseth:2010vx,Bernabei:2010mq}.
Since these did not materialise into more concrete observation this requirement was dropped for the second study, which contains points with arbitrary $m_{\chi_1^0}$.
In our work~\cite{PhysRevD.86.035023} we analysed the two cases separately, as $m_{\chi_1^0} < 15$~GeV requires a delicate fine-tuning of parameters in order to be viable, and exhibits some special features.
The arbitrary $m_{\chi_1^0}$ case was extended with an additional exploration of the $\lambda>0.5,\tb<5$ region in order to study in greater detail well-mixed light Higgs bosons with large singlet component, which may enhance the {\it signal strength} (see the glossary) in the Higgs to diphoton channel -- see Section~\ref{NMSSMheavyNeut}.
\item A large number of independent `model' parameters were defined at the low scale, i.e. with no particular UV completion in mind.
These parameters were taken to be the gaugino masses $M_1$, $M_2$ and $M_3$; Higgs sector parameters $\mu_{\rm eff}$, $\tan\beta$, $\lambda$, $\kappa$, $A_\lambda$ and $A_\kappa$ (as discussed in the previous section); flavour-blind soft masses for the left- and right-handed sleptons $m_{\tilde{L}}$ and $m_{\tilde{\overline{e}}}$; common soft masses for the squarks of the first and second generation ($M_{\tilde q_{1,2}}$) and third generation ($M_{\tilde q_3}$), and a single non-zero trilinear coupling, $A_t$.
For the $ m_{\chi_1^0} < 15$~GeV study three further restrictions were imposed: a common soft mass was taken for both `chiralities' of sleptons ($m_{\tilde{L}} = m_{\tilde{\overline{e}}}$), squark masses were taken to be flavour blind ($M_{\tilde q_3} = M_{\tilde q_{1,2}}$), and the gaugino mass-unification relation $M_2=\frac{1}{3}M_3$ was assumed.
The latter was taken to reduce the number of free parameters knowing that the gluino does not play an important role in dark matter observables for light neutralinos.  For the later analysis {\it without} the $ m_{\chi_1^0} < 15$~GeV requirement these three conditions were relaxed, increasing the parameter space dimensionality by three.
The spectrum and observables were calculated from the model parameters using {\tt micrOMEGAs~2.4} \cite{Belanger:2010gh} and {\tt NMSSMTools~2}~\cite{Ellwanger:2006rn,Ellwanger:2004xm,Ellwanger:2005dv}. 
\item Some experimental constraints were applied as simple pass/fail criteria, while for others a likelihood was calculated, giving a goodness of fit for the model for that particular observable.
A total likelihood (as the product of all of the separate likelihoods) was calculated and used to help steer the Markov Chain Monte Carlo's parameter-space exploration towards areas in better agreement with the full set of observables.
Contributing factors to this likelihood were:
\begin{itemize}
 \item A comparison of the LSP relic density to the Wilkinson Microwave Anisotropy Probe (WMAP) observed value $\Omega_{\rm WMAP} h^2=0.1131\pm 0.0034$~\cite{Komatsu:2008hk}, providing a constraint on the neutralino pair annihilation cross-section in the primordial Universe.
 The relic density was calculated assuming Friedman-Robertson-Walker cosmology, standard thermodynamics and the freeze-out mechanism.
 A relic density deficit $\Omega<\Omega_{\rm WMAP}$ calls for another type of particle to (partially) solve the dark matter problem, or else a modification of gravity (e.g.~\cite{Skordis:2005xk}).
  \item Dark matter direct detection limits from XENON100~\cite{Aprile:2011hi}, gamma rays from dwarf spheroidal (dSph) galaxies probed by Fermi-LAT \cite{Abdo:2010ex,Strigari:2006rd} and the radio emission in the Milky Way and in galaxy clusters~\cite{Boehm:2002yz,Boehm:2010kg}.
 \item The anomalous magnetic moment of the muon: $(g-2)_\mu$.
 \item Constraints included within {\tt NMSSMTools} from $b$ physics, and LEP and Tevatron Higgs and Susy searches (including invisible decays of the $Z$).
 \item Theoretical constraints, namely the absence of Landau poles and unphysical global minima of the scalar potential.
\end{itemize}
\end{itemize}

In~\cite{PhysRevD.86.035023} we studied the effect of applying further constraints to these scans.
LHC searches for Susy were considered, in the form of the ATLAS Collaboration's $1.04~\text{fb}^{-1}$ jets$+\slashed{E}_T$ search which I implemented and validated in~\cite{Grellscheid:2011ij}.
For the reader unfamiliar with such searches, the points in this paragraph may become clearer after the discussions in Section~\ref{1/fbSearches}.
As the standard tool for the calculation of Next-to-Leading Order (NLO) cross-sections in Susy -- {\tt Prospino}\cite{Beenakker:1996ed,Beenakker:1996ch,Beenakker:1999xh,Spira:2002rd,Plehn:2004rp} -- is restricted to the MSSM, constraints were derived at Leading Order (LO) unlike in Section~\ref{1/fbSearches}.
Without the {\it K-factor} (see the glossary), which is $\mathcal{O}(1-3)$ in the MSSM, these constraints are expected to be slightly conservative.
To derive them, {\it events} (see the glossary) were generated with {\tt Herwig++~2.5.1}~\cite{Bahr:2008pv,Gieseke:2011na} and analysed with {\tt RIVET~1.5.2} \cite{Buckley:2010ar}. 
These limits are capable of excluding first and second generation squarks lighter than $0.6-1$~TeV and gluinos lighter than $\sim\!0.5$~TeV; however they rely on large branching ratios of these particles into jets and a stable neutral particle such as a neutralino.
In the NMSSM with a {\it singlino}-like neutralino LSP, the squarks and gluinos cannot decay to this LSP directly but must do via an intermediate particle, frequently the second-lightest (MSSM-like) neutralino.
As noted in~\cite{Das:2012rr} this reduces the {\it acceptance} (see the glossary) into jets$+\slashed{E}_T$ search channels, as the extra step reduces the $\slashed{E}_T$ and may result in leptons\footnote{
Susy searches with leptons would then have greater sensitivity, but this typically does not compensate for the loss of sensitivity in the 0-lepton search~\cite{Das:2012rr}; at first glance a strange result since leptons are more striking objects over the quantum chromodynamics (QCD) background, but when large $\slashed{E}_T$ is replaced by modest $+\slashed{E}_T$ and a modestly hard lepton, backgrounds with $W$ bosons become relevant.
}.
A third factor which I observed being of equal importance is that even if the intermediate state (between squark/gluino and LSP) decays into the LSP with a jet rather than a lepton, there will be greater alignment between the $\slashed{\mathbf{p}}_T$ and one of the jets, causing greater difficulty in meeting the angular separation trigger $\Delta\phi(\text{jet},\slashed{\mathbf{p}}_T)$.
For the $m_{\chi_1^0} < 15$~GeV scan, whenever the coloured sparticles were light enough to be within reach of the LHC they were associated with a singlino-like $\tilde\chi_1^0$.
The jets+$\slashed{E}_T$ search has highly reduced sensitivity for the aforementioned reasons and excluded very few of the points.
For the arbitrary $m_{\chi_1^0}$ scan, however, the singlet sector particles were generally much heavier, so that the LSP was not singlino-like.
The MSSM-like $\tilde{q}\rightarrow q \tilde\chi_1^0$ and $\tilde{g}\rightarrow qq \tilde\chi_1^0$ decays then take place, resulting in the familiar limits $m_{\tilde{q}}\gtrsim0.6-1$~TeV, $m_{\tilde{g}}\gtrsim0.5$~TeV.

\begin{figure}	
\centering
\includegraphics[width=0.7\textwidth]{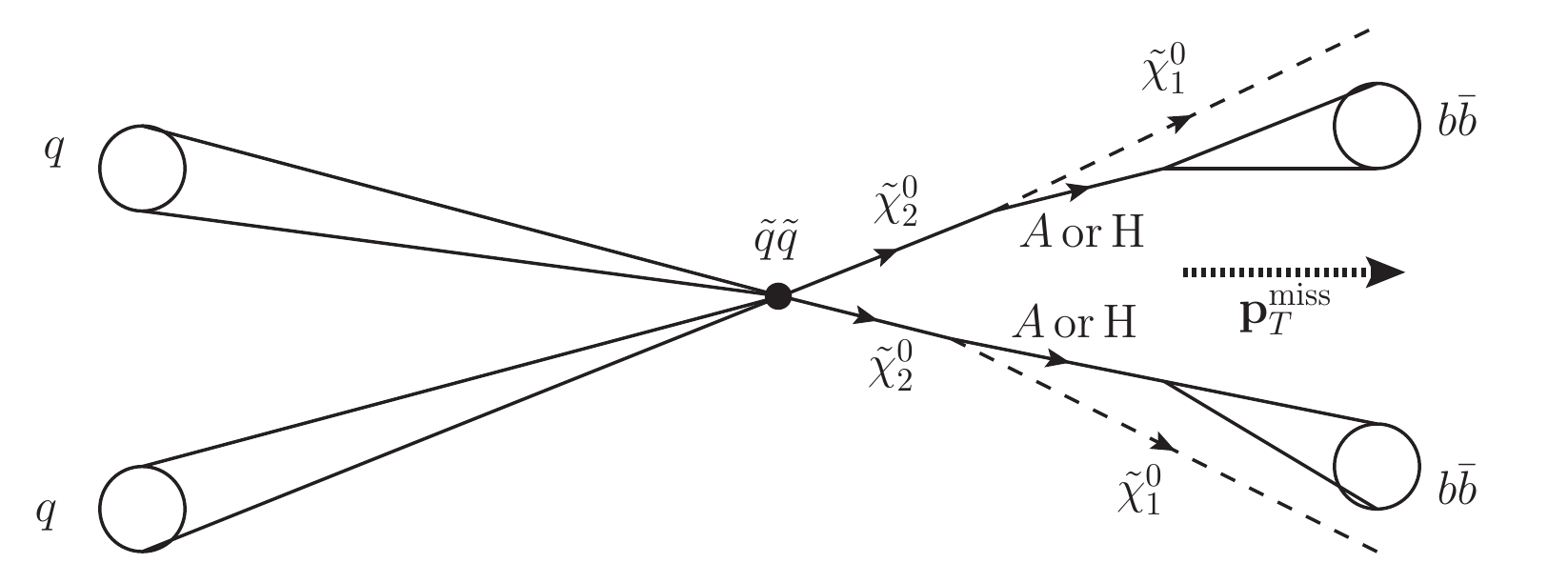}
\caption{The jets$+\slashed{E}_T$ signal arising in the NMSSM with a singlino-like LSP, and a $\chi_2^0$ which is much lighter than the squarks and which decays to $\chi_1^0$ with a Higgs (`$A$ or $H$').
}
\label{fig:fatjets}
\end{figure}

The observation of increased alignment between $\slashed{\mathbf{p}}_T$ and a jet prompted me to note in~\cite{PhysRevD.86.035023} the signal shown in Fig.~\ref{fig:fatjets}.
This arises in the case of a singlino-like $\chi_1^0$ (forcing squarks to decay instead via $\chi_2^0$) and a Higgsino-like $\chi_2^0$ which is (a) able to decay to $\chi_1^0$ with a Higgs (the latter going to $b\bar{b}$) and (b) much lighter than the squarks (making it boosted).
The result is a $\slashed{\mathbf{p}}_T$ vector which is sandwiched in between two jets, each consisting of two $b$ quarks.
In~\cite{PhysRevD.87.074004} (Section~\ref{MakingMostMET} of this thesis) I showed the relevance of this topology to simpler final states and many more classes of models, and how it can be exploited to reconstruct mass peaks.

The main constraint we were interested in investigating in~\cite{PhysRevD.86.035023} was compatibility of the NMSSM with the first observation of the $126$~GeV Higgs.
At the time this had also been studied in~\cite{Hall:2011aa,Kang:2012sy,Ellwanger:2011aa,King:2012is,Gunion:2012zd} (and our preliminary results had appeared in~\cite{Brooijmans:2012yi}).
Following the suggestion of~\cite{Heinemeyer:2011aa} we considered $122 < m_h / \text{GeV} < 128$ as the favoured range, within experimental and theoretical errors.
Note that the observed Higgs $h$ can correspond to $H_1$, $H_2$ or both at once (also in principle $H_3$ is possible, though we did not find a viable example).
As well as having the correct mass, a candidate for the observed signal must also have the correct signal strengths.
These quantities are measured in the experiments, normalised to a Standard Model-like Higgs and denoted by $\mu$, as the factor by which the expected signal should be multiplied in order for the signal plus background to best agree with the observed events, as a function of $m_h$ (see my discussion of {\it blue-band plots} in Appendix~\ref{BrazilBandPlots}).

Proton collisions can produce a Higgs in different ways, however the mechanism with the largest total cross-section (by a factor $\sim\!\!10$) for $m_h \sim 126$~GeV is the fusion of two gluons.
If {\it cuts} (see the glossary) are imposed, they can have different acceptances for the different production mechanisms and thus preferentially select some more than others, possibly reducing the dominance of gluon fusion.
This is not the case for the cuts used in the three most sensitive channels $h\rightarrow \gamma\gamma,WW,ZZ$, and so for the signal strength in these channels a good approximation is to consider the normalised production cross-section as being given by the normalised coupling to gluons (squared):
\begin{gather}
\sigma_{\rm prod}/\sigma_{\rm prod,\,SM} \approx g_{hgg}^2/g_{hgg,\,\text{SM}}^2,\\
R_{ggXX} \equiv  \frac{g_{hgg}^2 \,BR(h\rightarrow X X)}{g_{hgg,\,\text{SM}}^2\, BR(h\rightarrow X X)_\text{SM}},
\end{gather}
and we take $R_{ggXX}$ as the signal strength in the $XX$ channel, $XX=\gamma\gamma,WW,ZZ$.
As a loose criterion for sufficient visibility to correspond to the observed signal, I separated out points with $R_{gg\gamma\gamma} \geq 0.4$ for $2\sigma$ agreement with the most significant measurement -- $h\rightarrow \gamma\gamma$ as seen by the ATLAS Collaboration.

To supplement the Higgs constraints hard-coded into {\tt NMSSMTools}, I interfaced the latter to {\tt HiggsBounds}~\cite{Bechtle:2008jh,Bechtle:2011sb} for a further thorough check of existing Higgs-sector limits (see also Section~\ref{DerivationOfExclusionContours}).
The code for doing this is available at~\cite{Me}.

\subsection{Higgs Signals For $m_{\chi_1^0} < 15$~GeV}

\begin{figure}	
\centering
\includegraphics[width=0.47\textwidth]{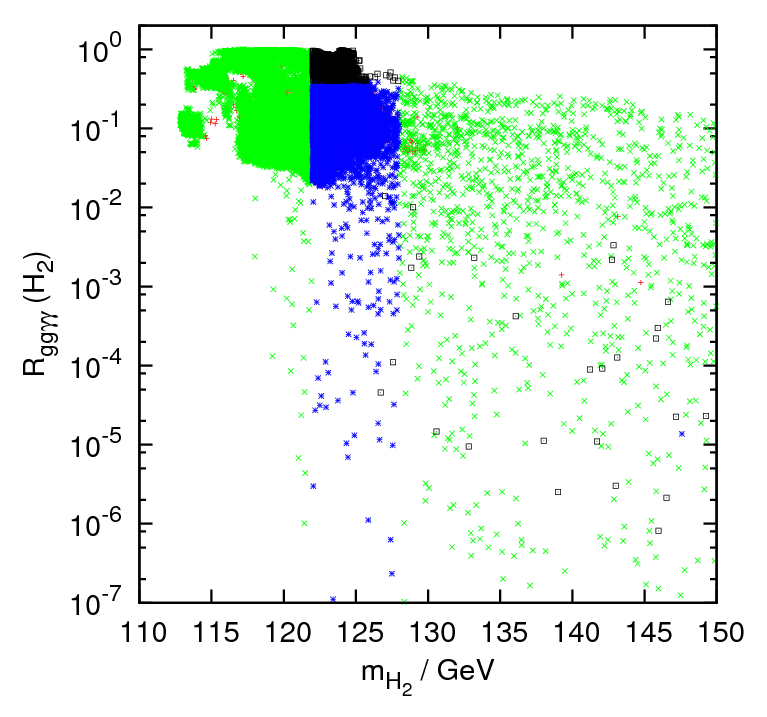}
\caption{Diphoton signal strength $R_{gg\gamma\gamma}$ as a function of the mass of $H_2$ in the scan with $m_{\chi_1^0}<15$~GeV.
Red points are ruled out either by jets$+\slashed{E}_T$ constraints (only three points) or by {\tt HiggsBounds~3.6.1}; other colours pass these constraints.
Green points have no scalar with mass $\in[122,128]$~GeV, blue points do have such a scalar ($H_1$ and/or $H_2$) but with $R_{gg\gamma\gamma}<0.4$, and black points have such a scalar with $R_{gg\gamma\gamma}\geq0.4$.
}
\label{fig:gggg_nmssm}
\end{figure}

Firstly we considered the broader case of $\chi_1^0$ as only a partial contributor to the observed relic density: $10\% \Omega_{\rm WMAP} h^2 < \Omega_{\chi_1^0} h^2 \leq \Omega_{\rm WMAP} h^2$ (with the lower bound admittedly somewhat arbitrary).
In Fig.~\ref{fig:gggg_nmssm} I plot the distribution of masses and diphoton signal strengths for $H_2$.
In our scan $H_2$ is most often doublet-like, i.e. with couplings like $h$ in the MSSM, and thus values of $m_{H_2}$ extend roughly down to the LEP limit and up to $130-150$~GeV.
The four colours are explained in the caption; black points are good candidate points -- at least one Higgs has mass $\in [122,128]$~GeV with $R_{gg\gamma\gamma} \geq 0.4$.
Fig.~\ref{fig:gggg_nmssm} shows that this Higgs is almost always $H_2$ in our scan, and for a handful of points it is $H_1$ (visible as black points with $m_{H_2}\notin [122,128]$~GeV).

\begin{figure}
\centering
\includegraphics[width=0.47\textwidth]{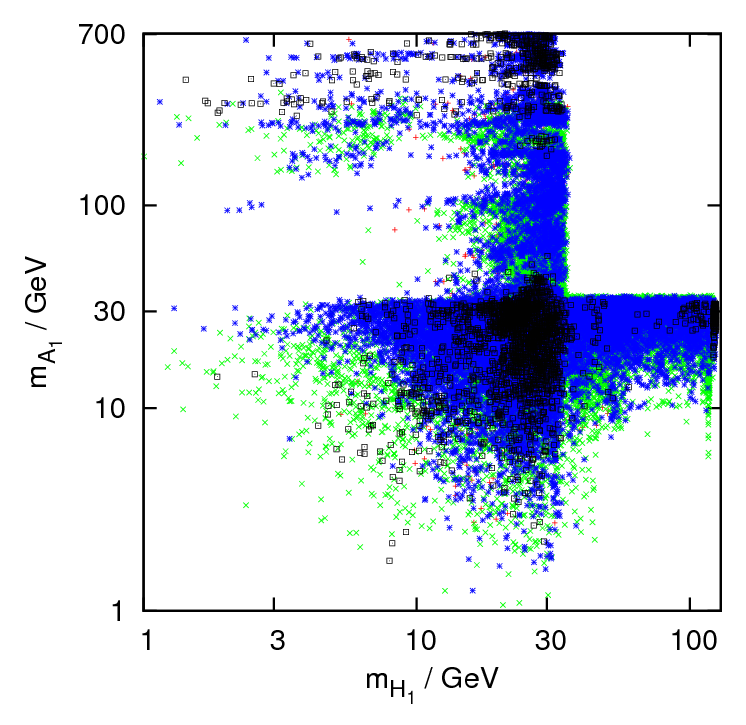}
\caption{Masses of the lightest scalar $H_1$ and pseudoscalar $A_1$, in the scan with $m_{\chi_1^0}<15$~GeV.
The colour coding is as in Fig.~\ref{fig:gggg_nmssm}.
}
\label{fig:mh12}
\end{figure}

The (normalised) diphoton signal strength can be seen to extend to much lower values than the Standard Model value of $1$.
This is chiefly because with $m_{\chi_1^0} < 15$~GeV, at least one extra decay mode is open to the candidate Higgs of mass $\sim\!\!126$~GeV: $h\rightarrow\chi_1^0\chi_1^0$.
Furthermore, sufficiently efficient annihilation of such light neutralinos for acceptable relic density is not found to be possible without a resonant $s$-channel scalar $A_1$ or pseudo-scalar $H_1$~\cite{Vasquez:2010ru}.
Hence either $A_1$ or $H_1$ has mass $\sim\!\!2m_{\chi_1^0}\lesssim30$~GeV, as shown in Fig.~\ref{fig:mh12}, and a Higgs-to-Higgs decay is also open.

\begin{figure}
\centering
\includegraphics[width=0.47\textwidth]{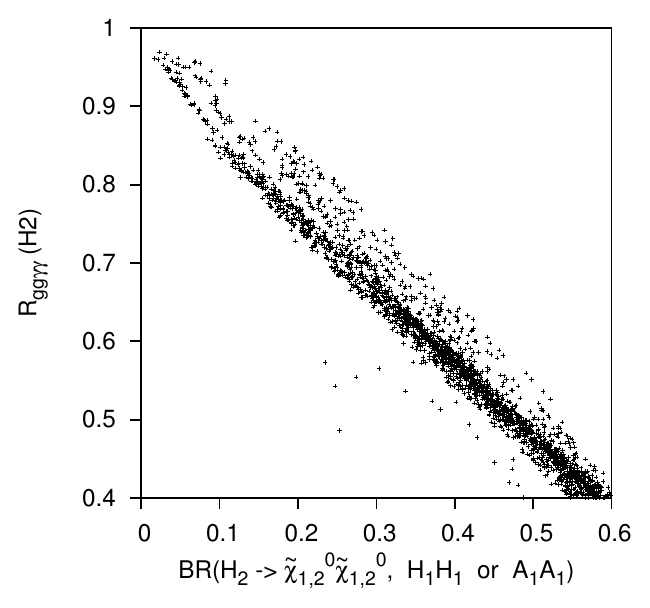}
\caption{Showing the reduction in diphoton signal strength from $H_2\rightarrow\gamma\gamma$ competing with new BSM decays.
Good candidate points (those with a scalar of mass $\in[122,128]$~GeV and $R_{gg\gamma\gamma}\geq0.4$) in the scan with $m_{\chi_1^0}<15$~GeV are shown.}
\label{fig:SignalStrength_V_BSMdecays}
\end{figure}

\begin{figure}
\centering
\includegraphics[width=0.4\textwidth]{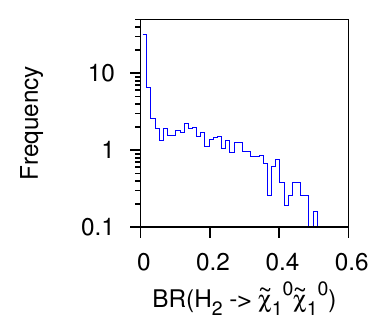}
\includegraphics[width=0.4\textwidth]{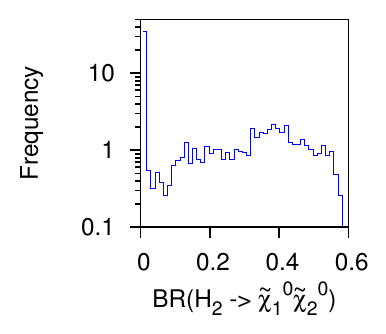}
\includegraphics[width=0.4\textwidth]{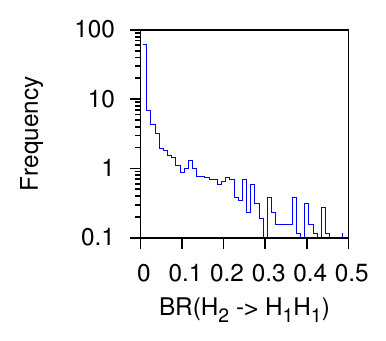} 
\includegraphics[width=0.4\textwidth]{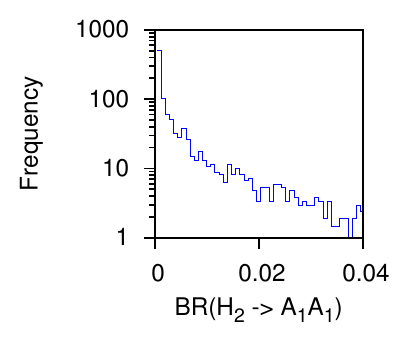}
\caption{Unit-normalised distributions of the different BSM decay channels for good candidate points (those with a scalar of mass $\in[122,128]$~GeV and $R_{gg\gamma\gamma}\geq0.4$), in the scan with $m_{\chi_1^0}<15$~GeV.}
\label{fig:distrib}
\end{figure}

In Fig.~\ref{fig:SignalStrength_V_BSMdecays} I plot only the (previously black) good candidate points, and show the effect on the diphoton signal strength of competing beyond the Standard Model (BSM) decays $h\rightarrow\chi_1^0\chi_1^0,\chi_1^0\chi_2^0,A_1A_1,H_1H_1$.
One sees $R_{gg\gamma\gamma} \approx 1 - BR(h\rightarrow \text{BSM})$, showing that the presence or absence of these decays is the dominant factor controlling signal strength.
In Fig.~\ref{fig:distrib} I show how the branching ratios for these decays are distributed amongst the good candidate points.

As an aside it is interesting to note that a light scalar or pseudo-scalar $x$, with reduced couplings to standard model particles (to explain current non-observation) but into which the Higgs can decay, is seen here in the NMSSM but is a widely motivated phenomenon. 
For example a symmetry of the Higgs potential explicitly broken by a small term in the Lagrangian will give a light degree of freedom strongly coupled to the physical Higgs (see the more detailed discussion in~\cite{Lisanti:2009uy}).
The decay of $x$ following $h\rightarrow2x$ is model dependent, and many possibilities have been considered.
One example is a decay to gluons which suffers from a huge QCD background~\cite{Bellazzini:2009xt}, however a boosted $h\rightarrow 2x\rightarrow 4g$ fat jet has characteristic (and intuitive) jet substructure~\cite{Falkowski:2010hi,Chen:2010wk}: the two hardest subjets are likely to have small masses (equal to $m_x$), they are likely to have similar masses, and they are likely to be much harder than the third hardest subjet.
Other possible decays include those to taus~\cite{Forshaw:2007ra,Englert:2011iz}, taus and muons~\cite{Lisanti:2009uy}, charm quarks~\cite{Lewis:2012pf}, and generic combinations of hadronic and missing energy~\cite{Englert:2012wf}.

A lighter Higgs could of course be produced directly, as well as via the decay of the $126$~GeV Higgs.
The substantial singlet component necessary to escape existing constraints suppresses the coupling to Standard Model particles.
In the NMSSM (and other type II two-Higgs-doublet models~\cite{Hall:1981bc}) couplings to leptons and down-type quarks are enhanced by $\tb$, possibly enhancing the associated production of the lighter Higgs with a $b$ quark, followed by a decay to $2\tau$.
In this scan however we found an $H_1 bb$ coupling equal to the Standard Model value at most.

Finally we checked the effect of requiring that points in the scan have a relic density within $1\sigma$ of the WMAP observed value (rather than merely not exceeding it).
This depleted the density of points surviving all constraints apparently uniformly, without causing any further noticeable correlations in the Higgs signals.

\subsection{Higgs Signals For Arbitrary $m_{\chi_1^0}$} \label{NMSSMheavyNeut}

The condition $m_{\chi_1^0}<15$~GeV required a fine tuning of parameters to be viable, as I have mentioned, and forced the lightest scalar or pseudo-scalar to have mass $\sim\!\!2m_{\chi_1^0}$ (and thus necessarily be singlet-like).
The scan without this condition typically gave rise to $\chi_1^0$ considerably heavier than $15$~GeV, which can annihilate efficiently through exchange of a $Z$ or light slepton, and therefore do not require (part of) the singlet sector to be light for acceptable relic density.
Indeed in this scan the singlet sector is generally heavier than in the previous one, and $H_1$ rather than $H_2$ is usually the candidate with \mbox{mass $\in [122,128]$~GeV}.

\begin{figure}	
\centering
\includegraphics[width=0.47\textwidth]{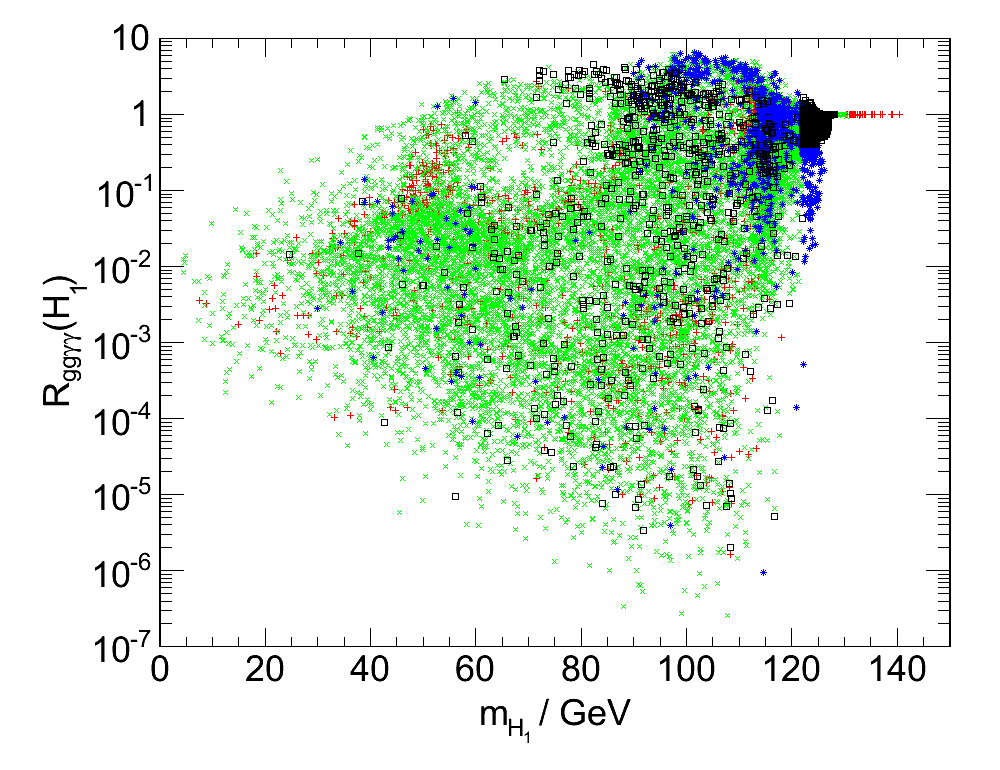}
\includegraphics[width=0.47\textwidth]{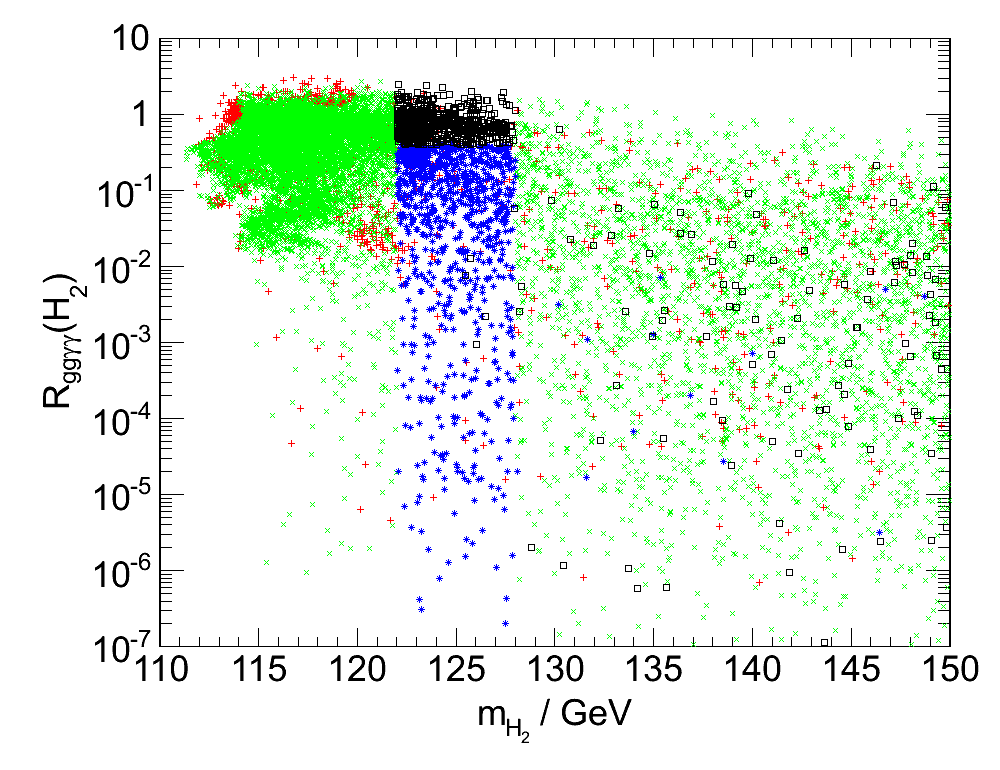}
\caption{Diphoton signal strength $R_{gg\gamma\gamma}$ as a function of the mass of $H_1$ (left panel) and of $H_2$ (right panel), in the scan with arbitrary $m_{\chi_1^0}$.
In the left panel most points are on top of one another at high mass and high $R_{gg\gamma\gamma}$.
The colour coding is as in Fig.~\ref{fig:gggg_nmssm} except jets$+\slashed{E}_T$ constraints are not checked.
Plots by Daniel Albornoz Vasquez.}
\label{fig:gggg_all}
\end{figure}

Again we started by considering $\chi_1^0$ as simply a contributor to the observed relic density, now without a lower bound: $\Omega_{\chi_1^0} h^2 \leq \Omega_{\rm WMAP} h^2$.
In Fig.~\ref{fig:gggg_all} I show plots made by my collaborator Daniel Albornoz Vasquez illustrating the distribution of diphoton signal with mass, for all such points.
Constraints from {\tt HiggsBounds~3.6.1} have been checked, those from jets+$\slashed{E}_T$ searches have not (yet).
Very strong enhancement of the $H_1$ diphoton signal strength is seen, though mostly for masses below $\sim\!\!126$~GeV.
For both $H_1$ and $H_2$ a modest enhancement $\mathcal{O}(2)$ is seen in the favoured mass range.
This is possible when the singlet scalar eigenstate is close in mass to the lightest doublet-like scalar, such that the singlet nature of the former partially mixes into the latter, suppressing the largest Higgs decay width $\Gamma_{h\rightarrow b\bar{b}}$ and permitting an increased branching ratio to two photons as explained in~\cite{Ellwanger:2010nf,Ellwanger:2011aa}.

Such a mechanism for enhancing the diphoton signal strength is weakly motivated by (statistically insignificant) excesses observed particularly by the ATLAS Collaboration -- currently at the level $2.4\sigma$~\cite{ATLAS-CONF-2013-012,Marumi} -- and previously by the CMS Collaboration (no more --~\cite{CMS-PAS-HIG-13-001}).
Let us understand it in greater detail.
With the matrix $S_{ig}$ relating the gauge eigenstate scalars $g=H_u^0, H_d^0, S$ to the mass eigenstates $H_{i=1,2,3}$, the modified couplings are at tree-level:
\begin{gather}
 \frac{g_{H_i bb}}{g_{H_i bb,\,\text{SM}}} = \frac{S_{id}}{\cos\beta},\label{HbbCoupling}\\
 \frac{g_{H_i VV}}{g_{H_i VV,\,\text{SM}}} = S_{id}\cos\beta + S_{iu}\sin\beta \label{HVVCoupling}
\end{gather}
The latter is relevant because the decay of a Higgs to two (massless) photons is purely loop induced, dominantly through a $W$ loop, thus strongly correlating the coupling to $\gamma\gamma$ with the coupling to $WW$.
If the coupling to $WW$ is decreased exactly in line with the coupling to $bb$, the diphoton signal strength will remain Standard Model-like (unless new light charged particles enhance the coupling of the Higgs to photons). 
Comparing Eq.s~\eqref{HbbCoupling} and~\eqref{HVVCoupling} we see that the diphoton signal strength is not enhanced when $\cos\beta=1$, which means the bottom-type Yukawas are exactly Standard Model-like; mixing a singlet with the doublet state then depletes all couplings uniformly, as would happen in the Standard Model, preventing modification of branching ratios.
We see that non-trivial mixing {\it between} the two doublet states ($H_u$ and $H_d$) is required in addition to mixing with the singlet.

As for the scan with $m_{\chi_1^0} < 15$~GeV, we checked the effect of requiring $\chi_1^0$ to make up all of the WMAP observed relic density (within $1\sigma$).
This eliminated all of the points in our scan where the diphoton signal strength was appreciably enhanced, $1 \lesssim R_{gg\gamma\gamma} \lesssim 2$ (leaving a new maximum $R_{gg\gamma\gamma}$ of $1.06$).
These points had a light singlet scalar, $m_S^2=(\kappa\mu_{\rm eff}/\lambda) (A_\kappa +4 \kappa\mu_{\rm eff}/\lambda)$~\cite{Belanger:2008nt,Ellwanger:2009dp}, by virtue of a small $\mu_{\rm eff} \lesssim 200$~GeV.
This leads to some amount of Higgsino component in $\chi_1^0$, allowing it to annihilate more efficiently and reducing the relic density (in our case always below $\Omega_{\rm WMAP} h^2$).
Nevertheless in~\cite{Cao:2012fz} an enhanced diphoton rate was found in the NMSSM with a WMAP-compatible relic density; note that small $\mu_{\rm eff}$ helps to make the singlet scalar lighter, but is not the only way of doing so.

\begin{figure}
\centering
\includegraphics[width=0.4\textwidth]{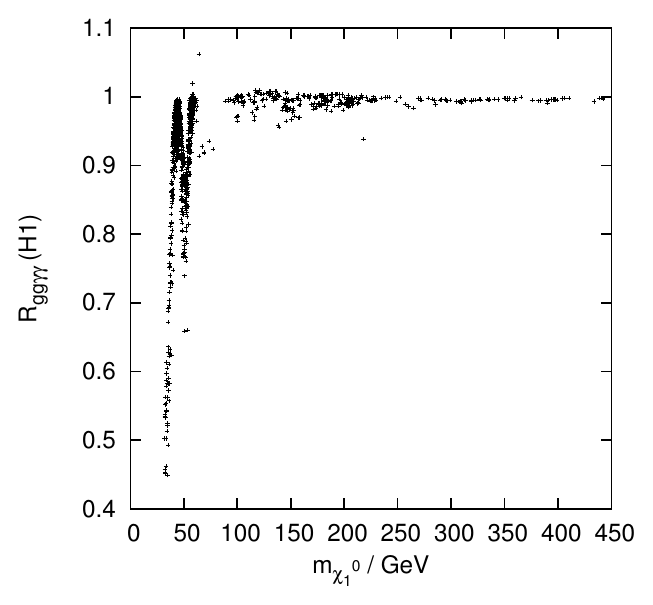}
\includegraphics[width=0.4\textwidth]{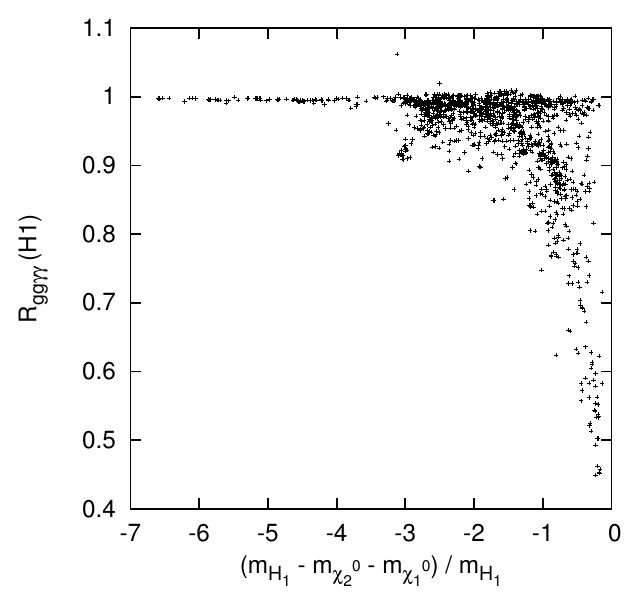}
\caption{Diphoton signal strength plotted against $m_{\chi_1^0}$ and against a dimensionless measure of the extent to which the $h\rightarrow\chi_2^0\chi_1^0$ decay is off shell, namely $(m_h - m_{\chi_2^0} - m_{\chi_1^0})/m_h$.
Points from the scan with arbitrary $m_{\chi_1^0}$, passing all considered constraints and with $122 < m_h / \text{GeV} < 128$ are shown.}
\label{fig:heavyneut}
\end{figure}

For these WMAP-saturating points, and further specialising to those with $122 < m_h / \text{GeV} < 128$, I checked jets$+\slashed{E}_T$ constraints\footnote{
A restricted application of LHC Susy search limits was chosen as they are computationally very expensive to check, requiring the generation of at least $10^4$ hadron-level events for each model point.
Doing this for exponentially large numbers of model points should be discouraged for environmental reasons, especially when there is questionable physical insight to be gained by doing so.
Examples include model points which can be discarded as physically uninteresting for some other reason, and model points with similar branching ratios and kinematics to the simplified or constrained models for which the cross-section limits have already been interpreted as mass limits.
} as mentioned in Section~\ref{SettingUpNMSSMscans}.
This eliminated all points with $m_{\tilde{q}}\lesssim0.6-1$~TeV, $m_{\tilde{g}}\lesssim0.5$~TeV, without further correlating other observables related to the Higgs signals.
In Fig.~\ref{fig:heavyneut} I plot the diphoton signal strength for remaining points against $m_{\chi_1^0}$, and against a dimensionless measure of the extent to which the $h\rightarrow\chi_2^0\chi_1^0$ decay is off shell: $(m_h - m_{\chi_2^0} - m_{\chi_1^0})/m_h$.
The cause of $R_{gg\gamma\gamma} < 1$ is apparent.

\newpage
\part{Gauge-Mediated Susy Breaking} \label{GGM}
\section{Introduction} \label{GGMintro}

As mentioned in Section~\ref{MSSM}, Susy breaking must occur in a hidden sector and be mediated to the visible sector (containing the MSSM), with the three main candidates for the mediator being gravitational interactions, extra dimensions, and gauge interactions.
In this part of the thesis I will discuss some phenomenological aspects of the last of these cases -- gauge-mediated Susy breaking (GMSB).

\subsection{$R$~Symmetry And Susy Breaking}

One of the relations of the Susy algebra is that the Hamiltonian is a positive semi-definite sum of the fermionic generators.
This means that, for a Susy theory, the vacuum $|0\rangle$ has unbroken Susy if and only if it has vanishing energy; Susy is broken if and only if the vacuum energy is strictly greater than zero.
To describe our universe with Susy (which is necessarily broken) we then must ensure either that the RHS of Eq.~\eqref{Fterms} is greater than zero ($F$-{\it term breaking}) or that the RHS of Eq.~\eqref{Dterms} is greater than zero ($D$-{\it term breaking}).
Phenomenological problems with the latter have concentrated most efforts on the former, with non-vanishing $F$-terms widely considered to be necessary (not just sufficient).

$R$~symmetry is a symmetry of the Lagrangian but not the superpotential: it is the continuous symmetry
\begin{equation}
\theta,\; d\theta,\; W(\Phi),\; \mathcal{L}\quad \rightarrow e^{i\alpha}\theta,\; e^{-i\alpha}d\theta,\;  e^{2i\alpha}W(\Phi),\;  \mathcal{L}
\end{equation}
Under this symmetry the superpotential has charge $2$: we write \mbox{$R(W)=2$}.
In~\cite{Nelson:1993nf} a deep connection between $R$~symmetry and dynamical $F$-term Susy breaking was established for generic, calculable theories.
{\it Dynamical} Susy breaking models are those in which none of the mass scales associated with Susy breaking are put in by hand -- all are generated by dimensional transmutation, being of the form $M_P e^{-a/g^2}$, with $a=\mathcal{O}(4\pi^2)$.
{\it Calculable} theories are those with a limit in which (asymptotically free) gauge dynamics can be integrated out at some scale, leaving an effective theory with only chiral superfields and no gauge fields (or gauge fields which do not couple to the remaining light chiral fields).
These chiral superfields $\Phi_{i=1\ldots N}$ can then be charged under {\it global} symmetries, and there are exactly three possibilities.
\begin{itemize}
 \item There are no global symmetries.
 The superpotential $W$ is a function of $N$ variables; the vanishing of the RHS of Eq.~\eqref{Fterms} amounts to $N$ constraints for the existence of a supersymmetric vacuum.
 \item There is no $R$~symmetry but there is a non-$R$~symmetry, with $l$ generators.
 $W$ must be a sum of terms with zero charge, and so must be a function on $N-l$ variables; a vanishing RHS of Eq.~\eqref{Fterms} thus gives $N-l$ constraints (with $l$ constraints trivially satisfied).
 \item There is an $R$~symmetry.
\end{itemize}
If the superpotential is {\it generic} -- that is, all terms not forbidden by symmetries are non-zero -- then in the first two cases a solution can be found and the vacuum is supersymmetric.
An $R$~symmetry is thus necessary for Susy breaking in the true vacuum.
This does not yet show, however, that Susy breaking is possible even then.
Now consider that the $R$~symmetry is spontaneously broken, with $\Phi_1$ (say) getting a VEV, and $R(\Phi_1)=q_1$.
Since $R(W)=2$ we can write $W = \Phi_1^{2/q_1}f(\Phi'_{i=2,\ldots N})$, where $\Phi'_i = \Phi_i / \Phi_1^{q_i/q_1}$ and $f$ is an arbitrary function of these $N-1$ uncharged ratios of fields.
The vanishing of the RHS of Eq.~\eqref{Fterms} now amounts to $N$ constraints on $N-1$ variables, which for a generic superpotential cannot be solved; therefore the vacuum energy is greater than zero everywhere in field space and Susy is broken in the vacuum.
A spontaneously broken $R$~symmetry is thus sufficient for Susy breaking.
These separate {\it necessary} and {\it sufficient} conditions are together called the Nelson-Seiberg Theorem.

Closely related to the Nelson-Seiberg Theorem is the theorem of Shih~\cite{Shih:2007av} for O'Raifeartaigh models of Susy breaking~\cite{O'Raifeartaigh:1975pr} with a single pseudo-modulus\footnote{
A {\it modulus} is a direction in field space which is classically massless; a {\it pseudo-modulus} is a modulus which becomes massive in the quantum field theory due to radiative corrections.
}.
This states that a sufficient condition for Susy breaking is an $R$~symmetry and a least one field with an $R$~charge not equal to either 0 or 2, under all consistent charge assignments (including e.g. mixing of the $U(1)_R$ with another global $U(1)$).
Recently a stronger version of the Nelson-Seiberg Theorem has been proposed~\cite{Kang:2012fn}: for a calculable theory with a generic superpotential, a necessary {\it and} sufficient condition is that there are more fields with $R$~charge 2 than with $R$ charge 0 under all consistent charge assignments.

An important ramification of the Nelson-Seiberg Theorem is that it pushes us towards {\it metastable} Susy breaking (see e.g.~\cite{Jaeckel200983c}).
If the $R$~symmetry is broken spontaneously, there is an exactly massless NGB -- the $R$~axion -- which is ruled out by astrophysical and experimental bounds~\cite{Jaeckel:2010ni,Beringer:1900zz}.
If the $R$~symmetry is broken explicitly, the $R$~axion becomes a permissible pNGB; but since the Lagrangian is not $R$~symmetric there is a supersymmetric vacuum.
As the explicit breaking tends to zero, however, the supersymmetric vacuum `moves out to infinity' away from a local minimum of the potential at finite field strengths (which should exist for metastability), making the latter arbitrarily long-lived.
The {\it ISS model} of~\cite{Intriligator:2006dd} is the prototypical model of this sort.

\subsection{The Gauge Mediation Parameter Space}\label{PartGGMSectionParamSpaceSubSectionSimpleToComplex}

{\it General Gauge Mediation} (GGM), defined in~\cite{Meade:2008wd}, is the collection of all models where the Susy-breaking hidden sector decouples from the supersymmetric MSSM as the Standard Model gauge interactions are turned off.
GGM allows up to six parameters $\Lambda_{G,r}$, $\Lambda_{S,r}$ ($r=1,2,3$) for specifying the gaugino and sfermion masses at some high scale characterising the hidden sector:
\begin{subequations} \label{eq:GGM}
\begin{eqnarray}
\label{gauginosoft}
M_{\lambda_r} &=& \,\frac{\alpha_r}{4\pi}\,\,\Lambda_{G,r} \, , \\
\label{scalarsoft}
m_{\tilde{f}}^2  &=& 2 \sum\limits_{r=1}^3 C_2(f,r) \,\frac{\alpha_r^2}{(4\pi)^2}\,\, \Lambda_{S,r}^2\,,
\end{eqnarray}
\end{subequations}
where $C_2(f,r)$ is the quadratic Casimir of the matter representation $f$ under the gauge group $r$, and a GUT normalisation of $\alpha_1$ is used: $g_1=\surd(5/3)g'$.
The $\Lambda$ parameters are all-order correlators of the (possibly strongly coupled) hidden sector.
GGM was further developed in~\cite{Buican:2008ws}, in which the following remark is relevant: ``the fact that the gaugino masses are complex$\ldots$ implies that GGM does not solve the Susy CP problem.
So additional mechanisms (such as an R-symmetry as in~\cite{Cheung:2007es}, or having the hidden sector be CP invariant) must be invoked to explain why the gaugino masses are real$\ldots$ we will assume that such a mechanism is at work and only consider CP invariant theories, so that the parameter space of GGM spans $\mathbb{R}^6$.''
Also in~\cite{Buican:2008ws} is an existence proof for the possibility of fully spanning this six-dimensional model space.

The relations~\eqref{eq:GGM} allow exploration of GGM phenomenology in blissful ignorance of the hidden sector dynamics, by scanning through $\Lambda_{G,r}$, $\Lambda_{S,r}$.
It is instructive however to consider concrete models with more restricted parameter spaces, for which the $\Lambda$ parameters can be calculated.
The simplest GMSB superpotential consistent with gauge-coupling unification is $W =\lambda X \tilde{\Phi} \Phi$, with the $\tilde{\Phi} \Phi$ messenger pair in the vector-like $5 \oplus \bar{5}$ representation of the SU(5) gauge group.
All information about Susy breaking relevant to the visible sector is parameterised in the (gauge-singlet) spurion superfield $X$ which, due to some unknown but ideally dynamical Susy breaking in the hidden-sector, acquires a VEV $\langle X \rangle = M + \theta^2 F$.
The gaugino (sfermion) masses that result can be calculated either explicitly with one-loop (two-loop) diagrams~\cite{Dine:1993yw,Dine:1994vc,Dine:1995ag} or more simply using wavefunction renormalisation~\cite{Giudice:1997ni}; in the language of the $\Lambda$ parameters of Eq.~\eqref{eq:GGM} they are
\begin{equation} \label{eq:mGM}
\begin{split}
W = \lambda X \tilde{\Phi} \Phi \quad \implies \quad \Lambda_G = \Lambda_S = \frac{F}{M},
\end{split} 
\end{equation}
where a suppressed gauge-group index $r$ on a $\Lambda$ parameter implies a common value across the gauge groups.
Eq.~\eqref{eq:mGM} is corrected by terms of higher order in $F/M^2$; in practice we often consider this dimensionless parameter to be small, so that the effective theory below the scale $M$ has only soft Susy-breaking effects\footnote{
With the superpotential Eq.~\eqref{eq:mGM} the mass given to the messenger pair $\tilde{\Phi} \Phi$ is $M$ for the fermion and $\surd(M^2\pm F)$ for the two scalar degrees of freedom; note that this symmetric Susy breaking obeys Eq.~\eqref{eq:supertrace} as it is a tree-level effect.
Thus $F<M^2$ is necessary to avoid tachyonic messengers, and $F\ll M^2$ avoids a {\it hard} Susy breaking theory where one messenger scalar $m_{\phi_2}^2 = F+M^2$ has been integrated out but the other $m_{\phi_1}^2 = F-M^2$ has not.
Susy broken only softly is necessary for a wavefunction renormalisation calculation~\cite{Giudice:1997ni}.
}.

The simple superpotential~\eqref{eq:mGM} has all six $\Lambda$ parameters equal; their splitting and independency comes from more general superpotentials.
I will demonstrate this with just the fundamental representation of $SU(5)$.
Only up to four independent scales will be obtained: we always have $\Lambda_{G,1} = \frac{3}{5} \Lambda_{G,2} + \frac{2}{5} \Lambda_{G,3}$ and $\Lambda_{S,1}^2 = \tfrac{3}{5}\Lambda_{S,2}^2 + \tfrac{2}{5}\Lambda_{S,3}^2$ according to the relative hypercharges of the doublet $l$ and triplet $q$ components of the $SU(5)$ fundamental when decomposed\footnote{
In~\cite{Buican:2008ws} it is shown that decomposing the total messenger sector into separate irreps with respect to the Standard Model SU(3)$\times$SU(2)$\times$U(1), each irrep contributes \textit{additively} to $\Lambda_{G,r}$ and $\Lambda_{S,r}^2$, in proportion to its Dynkin index with respect to gauge group $r$.
Contributing linearly to $\Lambda_{G,r}$ but in quadrature to $\Lambda_{S,r}$ means a single irrep can result in up to two free parameters; six independent $\Lambda_{G,r},\Lambda_{S,r}$ therefore requires at least three different irreps, for example as in the $\mathbf{10}$ representation of $SU(5)$.
The fundamental has only two: $q$ and $l$.
} onto $SU(3)\times SU(2)\times U(1)$.

As discussed in many places, such as \cite{Dimopoulos:1996ig,Giudice:1998bp,Cheung:2007es,PhysRevD.79.035002,Carpenter:2008he} (this list is not selective), one can generalise~\eqref{eq:mGM} with (a) multiple $\mathbf{5} \oplus \bar{\mathbf{5}}$ messengers, (b) multiple spurions (or equivalently supersymmetric messenger mass terms), and (c) doublet-triplet splitting, i.e. independent Yukawa couplings for the doublet $l$ and triplet $q$ components of the $\mathbf{5}$.
I summarise these $2\times2\times2$ possibilities -- corresponding to using or not using each of these three generalisations -- in Tables~\ref{SU5whole} and~\ref{2/3split}.
The resulting soft terms follow from~\eqref{eq:mGM} with no further computation:
\begin{itemize}
 \item {\it Multiple messengers}.
 Each messenger appears in the one-loop gaugino mass diagrams and the two-loop sfermion mass-squared diagrams, and so contributes additively to $\Lambda_G$ and $\Lambda_S^2$.
 Therefore generalising $\Phi\rightarrow\Phi_{i=1\ldots N}$, with the Yukawa coupling $\lambda$ becoming a matrix in messenger flavour space $\lambda_{ij}$, has the effect
 \begin{equation}
 \Lambda_G = \Lambda_S = \frac{F}{M} \quad \xrightarrow{\Phi\rightarrow\Phi_{i=1\ldots N}} \quad \Lambda_G = \sum_{i=1}^N \frac{F}{M}, \; \Lambda_S^2 = \sum_{i=1}^N \left(\frac{F}{M}\right)^2,
 \end{equation}
 provided we can diagonalise $\lambda_{ij}$ giving a basis in which the messengers are independent (more in this later).
 \item {\it Multiple spurions}.
 In Eq.~\eqref{eq:mGM}, note that $F/M$ is really $\lambda F / \lambda M$, since these dimensionful parameters only occur in conjunction with the Yukawa couping in the superpotential.
 Thus generalising to multiple spurions, $\lambda X \rightarrow \lambda^a X_a$, replaces $\lambda F / \lambda M$ by $\lambda^a F_a/\lambda^b M_b$ in the soft masses.
 Unless $F_a/M_a$ (no sum over $a$) is independent of $a$, the Yukawa coupling no longer cancels in this ratio.
 \item {\it Doublet-triplet splitting}.
 Even if the Yukawa couplings of $l$ and $q$ to the spurion are equal at the GUT scale, as would be required by gauge invariance under the GUT group, the breaking of the GUT group will cause these couplings to run differently at lower scales~\cite{Carone:1995kp}.
 The Yukawa coupling $\lambda$ becomes $\lambda_2$ in the expressions for $\Lambda_{G,2}$ and $\Lambda_{S,2}$, and $\lambda_{3}$ in the expressions for $\Lambda_{G,3}$ and $\Lambda_{S,3}$.
\end{itemize}

\begin{table}[!ht]
\centering
\begin{threeparttable}
\caption{Manifestly SU(5) symmetric messenger $\tilde{\Phi} \Phi$}
\begin{tabular}{ r | c c c c }
 & $1 \times (\mathbf{5} \oplus \bar{\mathbf{5}})$ &&& $N \times (\mathbf{5} \oplus \bar{\mathbf{5}})$ \\
\hline
&&&&\\
\multirow{2}{*}{$1 \times X$} & $W = \lambda \, X \, \tilde{\Phi} \Phi$ &&& $W = \lambda_{ij} \, X \, \tilde{\Phi}_i \Phi_j$ \\
 & $\Lambda_G = \Lambda_S = \frac{F}{M}$ &&& $\Lambda_G = N\frac{F}{M},\quad \Lambda_S = \sqrt{N}\frac{F}{M}$ \\
&&&&\\
&&&&\\
\multirow{4}{*}{$N_X \times X$} & $W = \lambda^a \, X_a \, \tilde{\Phi} \Phi$ &&& $W = \lambda_{ij}^a \, X_a \, \tilde{\Phi}_i \Phi_j$ \\
& $\Lambda_G = \Lambda_S = \frac{\lambda^a F_a}{\lambda^b M_b}$ &&& $ \xrightarrow{?} \sum\limits_{i}\lambda_{i}^a  \, X_a \, \tilde{\Phi}_i \Phi_i$ \\
&&&& $\Lambda_G = \sum\limits_{i} \frac{\lambda_i^a \, F_a}{\lambda_i^b \, M_b},$ \\
&&&& $\Lambda_S^2 = \sum\limits_{i} \left(\frac{\lambda_i^a \, F_a}{\lambda_i^b \, M_b} \right)^2$
\end{tabular}
\label{SU5whole}
\end{threeparttable}\\

\vspace*{10mm}

\begin{threeparttable}
\caption{Doublet-triplet-split messenger $\tilde{l}l,\:\tilde{q}q$}
\begin{tabular}{ r | c c c c }
 & $1 \times (\mathbf{5} \oplus \bar{\mathbf{5}})$ &&& $N \times (\mathbf{5} \oplus \bar{\mathbf{5}})$ \\
\hline
&&&&\\
\multirow{2}{*}{$1 \times X$} & $W = \lambda_2 \, X \, \tilde{l} l + \lambda_3 \, X \, \tilde{q} q$ &&& $W = \lambda_{2ij} \, X \, \tilde{l}_i l_j + \lambda_{3ij} \, X \, \tilde{q}_i q_j$ \\
 & $\Lambda_G = \Lambda_S = \frac{F}{M}$ &&& $\Lambda_G = N\frac{F}{M},\quad \Lambda_S = \sqrt{N}\frac{F}{M}$ \\
&&&&\\
&&&&\\
\multirow{4}{*}{$N_X \times X$} & $W = \lambda^a_2 \, X_a \, \tilde{l} l + \lambda^a_3 \, X_a \, \tilde{q} q$ &&&$W = \lambda^a_{2ij} \, X_a \, \tilde{l}_i l_j + \lambda^a_{3ij} \, X_a \, \tilde{q}_i q_j$ \\
&$\Lambda_{G,2} = \Lambda_{S,2} = \frac{\lambda^a_2 F_a}{\lambda^b_2 M_b}$ &&& $ \xrightarrow{?} \sum\limits_{i}\lambda^a_{2i} \, X_a \, \tilde{l}_i l_i + \lambda^a_{3i} \, X_a \, \tilde{q}_i q_i$ \\
 &$\Lambda_{G,3} = \Lambda_{S,3} = \frac{\lambda^a_3 F_a}{\lambda^b_3 M_b}$ &&& $\Lambda_{G;r=2,3} = \sum\limits_{i} \frac{\lambda_{r;i}^a \, F_a}{\lambda_{r;i}^b \, M_b},$ \\
&&&& $\Lambda_{S;r=2,3}^2 = \sum\limits_{i} \left(\frac{\lambda_{r;i}^a \, F_a}{\lambda_{r;i}^b \, M_b} \right)^2$
\end{tabular}
\label{2/3split}
\end{threeparttable}
\end{table}

Any matrix can be diagonalised by a bi-unitary transformation; rotating $\tilde{\Phi}_i$ and $\Phi_j$ independently we can always have diagonal $\lambda_{ij}X = \lambda_{ij} (M+\theta^2F)$.
For multiple spurions $X_a$ with non-universal $F_a/M_a$ (no sum), $\lambda_{ij}^a M_a$ is not aligned in messenger flavour space with $\lambda_{ij}^a F_a$, and so each requires a different bi-unitary transformation.
Said differently, the required rotation in messenger flavour space must involve $\theta$ and thus not commute with Susy transformations.
Such a rotation can be performed but it will necessarily mix the scalar and fermionic components of the messengers differently in the K\"{a}hler potential, and so one cannot integrate out each messenger separately\footnote{
Thanks to Alberto Mariotti for pointing this out.
}.
This encourages restriction of the couplings $\lambda_{ij}^a$ to diagonal form $\lambda_{i}^a$, in which case the simple logic of the multiple messenger and multiple spurion bullet points preceding this paragraph suffices to calculate the tricky case of multiple messengers {\it and} multiple spurions.


Restriction to a diagonal messenger-spurion coupling is also necessary in general, as noted in~\cite{PhysRevD.79.035002}, to suppress potentially tachyonic hypercharge $D$-term contributions to sfermion masses.
This is because a diagonal form ensures the existence of a {\it messenger parity}, defined in~\cite{Dimopoulos:1996ig}, where the messenger sector is unchanged at tree level by $\Phi_i, \tilde{\Phi}_i, V \; \rightarrow \tilde{\Phi}_i, \Phi_i, -V$ (with $V$ a gauge field).
This parity forbids the troublesome hypercharge term.
It is an approximate symmetry only, necessarily broken by the visible sector couplings with the gauge fields, but this is enough to make the hypercharge term arise safely at three or more loops instead of at one loop.

A different approach to calculating the soft terms for multiple messengers and spurions is taken in~\cite{Cheung:2007es}, which I will summarise below.
One can rotate in the basis of spurions to give just one spurion that has an $F$-term, plus a supersymmetric mass matrix.
Explicitly:
\begin{equation}
\label{RotateSpurions}
\begin{split}
\lambda_{ij}^a X_a = & \: \lambda_{ij}^a (M_a + F_a \, \theta^2) \\
= & \: (\frac{F_a}{F_1} \,\lambda_{ij}^a)\:(M_1 + F_1\, \theta^2) + \lambda_{ij}^a(M_a - \frac{F_a}{F_1} M_1) \\
\equiv & \: \lambda_{ij}X + m_{ij}
\end{split}
\end{equation}
The explicit spurion-relabelling symmetry of the LHS is implicit in the RHS due to a freedom to shift an arbitrary amount of the supersymmetric part of the spurion into the supersymmetric mass matrix:
\begin{equation} \label{SpurionSplitting}
\lambda_{ij}M + m_{ij} = \lambda_{ij}(M+M') + (m_{ij} - \lambda_{ij}M').
\end{equation}

In the spurion basis defined by the rotation in Eq.~\eqref{RotateSpurions}, the authors of~\cite{Cheung:2007es} assume a non-trivial $R$~symmetry (i.e. $R(X)\neq0$).
This uniquely specifies the splitting in Eq.~\eqref{SpurionSplitting}, and picks out one spurion from amongst the $X_a$ as unique, since the charged and uncharged terms cannot mix; in this way the spurion re-labelling symmetry is seen to disappear from both the LHS and RHS of Eq.~\eqref{RotateSpurions}.
(Note that if the $R$~symmetry existed in the pre-rotation basis $X_a$, is might prevent rotation to the single-spurion form; an alternative to the idea of rotation is of course a set up where there genuinely is just one spurion and supersymmetric masses for the messengers.)
For any $\lambda_{ij}$ and $m_{ij}$ consistent with the $R$~symmetry (which enforces $\lambda_{ij}=0$ if $m_{ij}\neq0$ and $m_{ij}=0$ if $\lambda_{ij}\neq0$) the soft masses are then calculated as
\begin{subequations}
\begin{equation}
\label{EOGMLambdaG}
\begin{split}
\Lambda_G = \; & F \, \partial_X \log \det (\lambda_{ij}X + m_{ij})\,\arrowvert_{X=M} = \frac{nF}{M} \\
\text{with}\quad n = \; & \frac{1}{R(X)} \sum\limits_{i=1}^N (2-R(\Phi_i)-R(\tilde{\Phi}_i))),
\end{split}
\end{equation}
\begin{equation}
\label{EOGMLambdaS}
\Lambda_S^2 = \; \frac{1}{2} |F|^2 \frac{\partial^2}{\partial X \partial X^{*}} \sum\limits_{i=1}^N \left( \log |\mathcal{M}_i|^2 \right)^2 \,\arrowvert_{X=M} 
\end{equation}
\end{subequations}
where $\mathcal{M}_i$ is the $i$th eigenvalue of $\mathcal{M}_{ij}=\lambda_{ij}X + m_{ij}$.
I add to Eq.~\eqref{EOGMLambdaS} only the comment that it is necessary for the interest of the setup that the diagonalisation of $\mathcal{M}_{ij}$ does not respect the $R$~symmetry.
Otherwise, each eigenvector $\mathcal{M}_{i}$ is a state of definite $R$~charge, with an eigenvalue which is either a coupling to the spurion or a supersymmetric mass; there are $n$ of the former and $N-n$ of the latter kind of eigenvector.
Eq.~\eqref{EOGMLambdaS} then evaluates trivially to $\Lambda_S^2=n\left(\frac{F}{M}\right)^2$ here as it must, matching $\Lambda_G = n\frac{F}{M}$ to give nothing but minimal GMSB with $n$ messengers and one-spurion.

For doublet-triplet split EOGM, independent $\Lambda_{S,2}$ and $\Lambda_{S,3}$ are found: Eq.~\eqref{EOGMLambdaS} with $\mathcal{M}_i$ replaced by $\mathcal{M}_{2i}$ and $\mathcal{M}_{3i}$ respectively.
$\Lambda_{G,2}$ and $\Lambda_{G,3}$ however remain constrained to be equal based on the assumption of the doublets' $R$~charges matching those of the triplets -- they thus share a value for the quantity $n$ in~\eqref{EOGMLambdaG}.

\newpage
\section{The Role Of The Messenger Scale, Sum Rules And RG Invariants} \label{RoleMessengerScale}
{\it This section is based on my works~\cite{Jaeckel:2011ma,Jaeckel:2011qj} done in collaboration with J\"{o}rg J\"{a}ckel and Valya Khoze; the text has been mostly re-written.
Unlike those works, here I use a GUT normalisation for the $U(1)$ gauge coupling: $g_1=\surd(5/3)g'$.}\\

The GGM relations for the soft masses, Eq.~\eqref{eq:GGM}, hold at some scale characterising the hidden sector.
For models with a single (or several degenerate) explicit messenger(s) of supersymmetric mass $M_{\rm mess}$, coupled to a Susy-breaking $F$-term with $F\ll M_{\rm mess}^2$, that scale is unambiguously $M_{\rm mess}$: the scale at which all messenger degrees of freedom are integrated out.
Consider holding all high-scale parameters constant, but changing the scale $M_{\rm mess}$ at which they set the soft masses.
Since this changes the amount of running required to reach the low scale, and all MSSM soft-mass beta functions are non-zero, then trivially the low-scale spectrum will change.
From this point of view, $M_{\rm mess}$ is an additional parameter impacting upon the low-scale GGM spectrum together with the six $\Lambda$ parameters.

Let us ask a more subtle question instead.
If varying $M_{\rm mess}$ with constant $\Lambda$ parameters changes the low-scale spectrum, can this effect be entirely absorbed into appropriately varying $\Lambda$ parameters?
If this were the case, $M_{\rm mess}$ would merely parameterise a one-dimensional family of models, $\Lambda(M_{\rm mess})$, with identical low-scale spectra (neglecting the precise value of the sub-GeV gravitino).
The answer to this question lies in sum rules and RG invariants (RGIs).

In GGM, the five sfermion soft masses (set in a flavour-blind manner) are given in terms of three parameters, $\Lambda_{S;r=1,2,3}$, and thus there are two sum rules~\cite{Meade:2008wd}, exact at $M_{\rm mess}$, which can be expressed as
\begin{align}
 \text{Tr}[Ym^2] = &\: 0,\\
\text{Tr}[(B-L)m^2] = &\: 0
\end{align}
After running to the low scale, these are expected to hold for the first two generations but not the third, as they are de-tuned under RG evolution only by the Yukawas~\cite{Meade:2008wd}.
Let me say this a little differently.
Setting the Yukawas to zero, the running of the sfermion soft masses depends only on the three gaugino masses; there are thus two linear combinations of sfermion mass-squared beta functions (which are always the same in the MSSM) that vanish.
In GGM, the linear combinations of {\it beta functions} which vanish are the same as the linear combinations of {\it mass-squareds} which vanish at $M_{\rm mess}$, hence the preservation of the sum rules.

In~\cite{Jaeckel:2011ma} I calculated the values of the sum rules at low scale resulting from models of GGM (allowing for non-zero $b(M_{\rm mess})$) with their spectra calculated using {\tt SOFTSUSY~3.1.6}~\cite{Allanach:2001kg} supplemented with GGM boundary conditions.
I found them to hold to $\mathcal{O}(1\%)$ for the first two generations and for the third-generation hypercharge sum rule; only the third-generation $B-L$ sum rule was broken strongly, by $\sim10-60\%$ over the parameter space investigated.

The goodness of the first two generation sum rules at lower scales allows for the meaningful definition of $\Lambda$ parameters which {\it run}, as opposed to being defined at the high-scale only -- five running mass-squareds continue to be described by three running parameters -- $\Lambda_{S,r}(Q)$ at a scale $Q$.
These are defined up to arbitrary additions of the two vanishing sum rules; an intuitive choice can be arrived at in the following way.
Amongst the one-loop RGIs of the MSSM (setting the Yukawas of the first two generations to zero) listed in~\cite{Carena:2010gr} are the six shown in Table~\ref{RGIsTable}.

\begin{table}[!h]
\centering
\begin{tabular}{ c | c | c }
RGI &  Definition & GGM high-scale value  \\
\hline
&&\\
$I_{B_r}$&$M_r/g_r^2$&$\frac{\Lambda_{G,r}}{16\pi^2}$ \\
&&\\
$\;I_{M_r}\;$&$\;M_r^2 + \sum_{\tilde{f}} D(\tilde{f},r) m_{\tilde{f},1}^2\;$&
$\;\left(\frac{\alpha_{r}(M_{\rm mess})}{4 \pi}\right)^{2} \left( \Lambda_{G,r}^{2} + \kappa_{r} \Lambda^{2}_{S,r} \right)\;$
\end{tabular}
\caption{Six one-loop RGIs of the MSSM, setting the Yukawas of the first two generations to zero, with $r=1,2,3$ denoting a gauge group, and $m_{\tilde{f},1}^2$ a first-generation sfermion mass-squared, taken from~\cite{Carena:2010gr}.
Their values in GGM at a scale $M_{\rm mess}$ are shown.
$D(\tilde{f},r)$ and $\kappa_{r}$ are numerical coefficients whose values (suppressed here for clarity) can be found in that reference.}
\label{RGIsTable}
\end{table}

\begin{figure}[t]
\begin{center}
\includegraphics[width=0.65\linewidth]{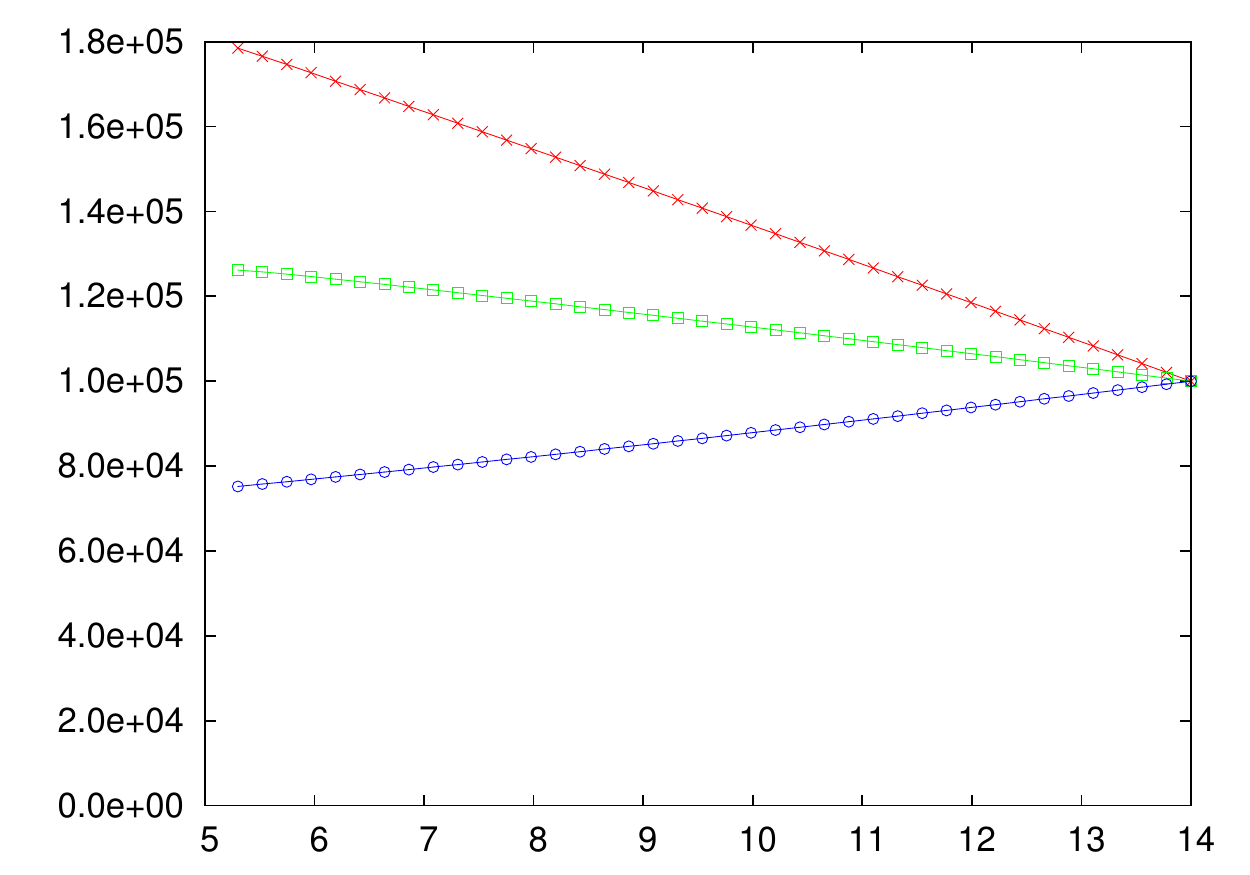}
\Text(-280,110)[c]{$\Lambda_{S,r}$ / GeV}
\Text(-110,-10)[c]{$\log_{10} \left(Q/{\rm GeV}\right)$}
\end{center}
\caption{Running $\Lambda_{S;r=1,2,3}(Q)$ as described in the text, for a model with unified $\Lambda_{S,r}=\Lambda_{G,r}=10^5~{\rm GeV}$, at the
messenger scale $M_{\rm mess}=10^{14}~{\rm GeV}$.
$\Lambda_{S,1}$ is shown with red crosses, $\Lambda_{S,2}$ with green squares and $\Lambda_{S,3}$ with blue circles.
}
\label{unifiedLambdas}
\end{figure}

The first three of these, $I_{B_r}$, show that the parameters $\Lambda_{G,r}$ are themselves RGIs; they can therefore trivially be defined at lower scales by $\Lambda_{G,r}(Q) = \Lambda_{G,r}$.
Then since the $I_{M_r}$ are constant, taking the expressions for their values in GGM at $M_{\rm mess}$ but allowing the $\alpha_{r}$ to run defines implicitly how $\Lambda_{S,r}(Q)$ must run to keep $I_{M_r}$ constant.
In other words:
\begin{equation} \label{runningLambdas}
I_{M_r} = \left(\frac{\alpha_{r}(Q)}{4 \pi}\right)^{2} \left( \Lambda_{G,r}^{2} + \kappa_{r} \Lambda^{2}_{S,r}(Q) \right)
\end{equation}
An example of how the $\Lambda_{S,r}(Q)$ run is shown in Fig.~\ref{unifiedLambdas}.

\begin{figure}[t]
\begin{center}
\includegraphics[width=0.4\linewidth]{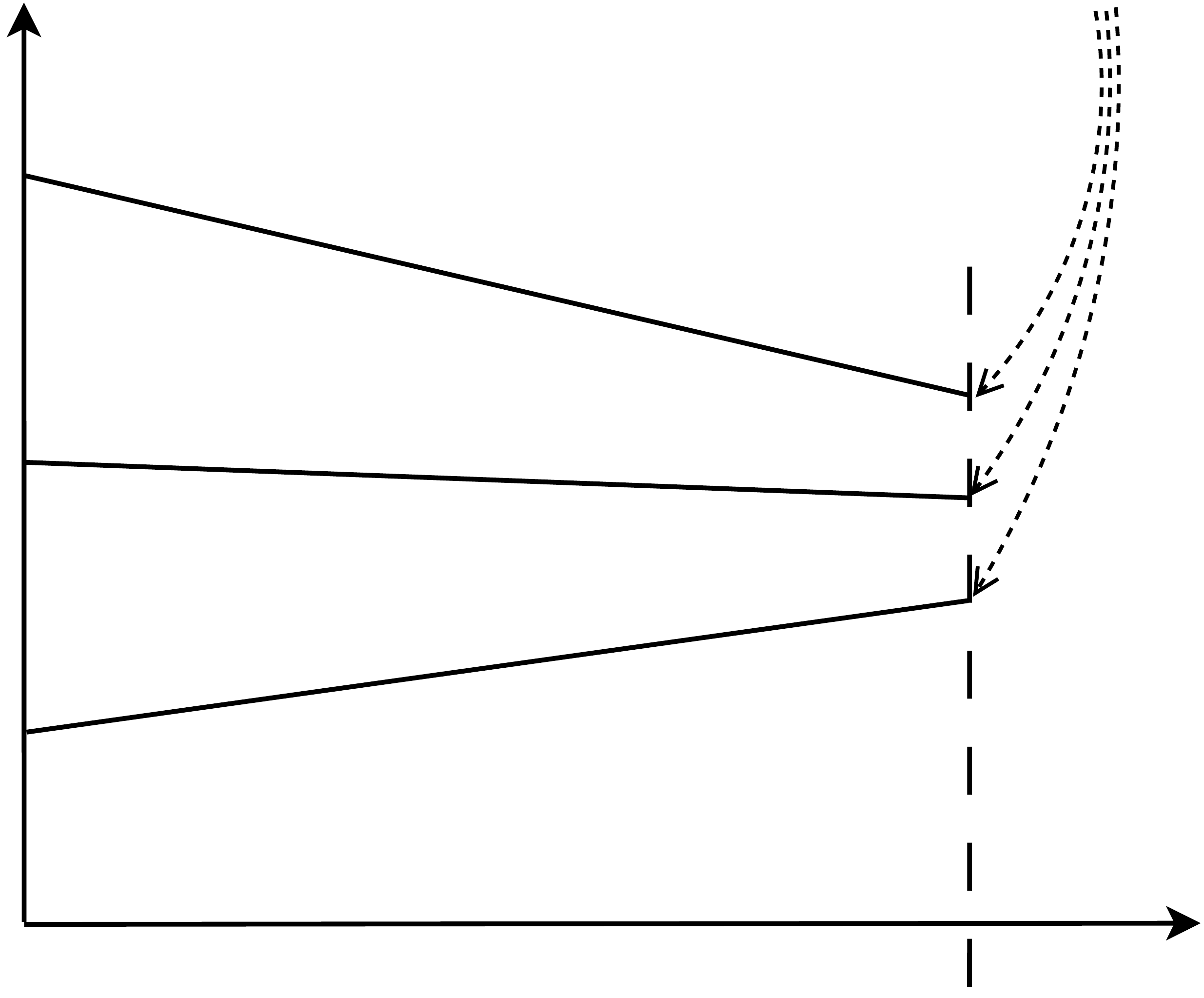}
\Text(-2,-2)[c]{$Q$}
\Text(-28,-10)[c]{$M_{\rm mess}$}
\Text(-20,140)[c]{$\Lambda_{S,r}(M_{\rm mess})$}
\rText(-178,145)[c][l]{$\Lambda_{S,r}$}
\hfill
\includegraphics[width=0.4\linewidth]{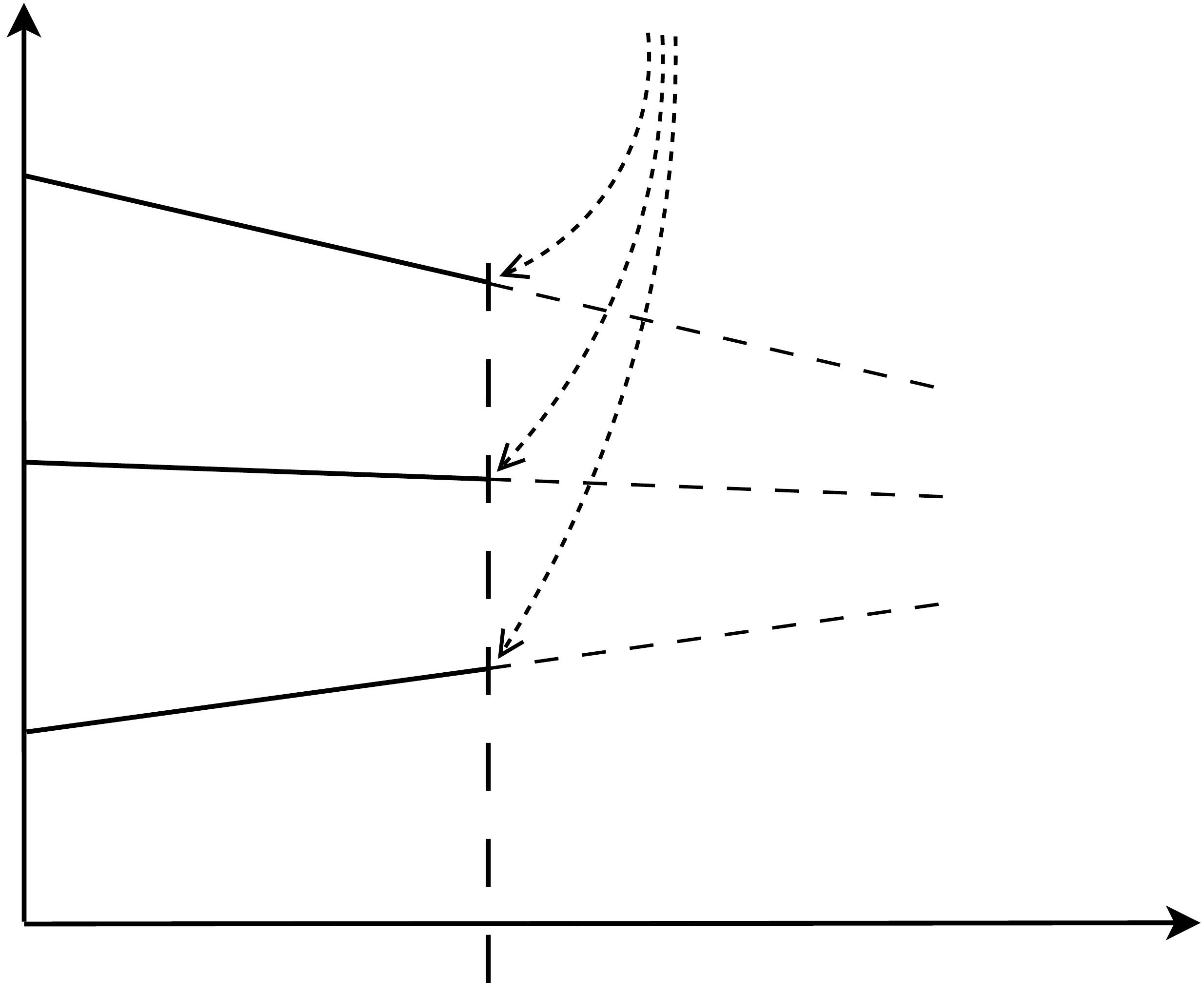}
\Text(-2,-2)[c]{$Q$}
\Text(-95,-8)[c]{$M_{\rm mess}'$}
\Text(-72,137)[c]{$\Lambda_{S,r}'(M_{\rm mess}')$}
\rText(-178,145)[c][l]{$\Lambda_{S,r}$}
\end{center}
\caption{The running $\Lambda_{S;r=1,2,3}(Q)$ for two models with different messenger scales and different values of $\Lambda_{S,r}$ at the messenger scale, but with matching $\Lambda_{S,r}(Q)$ below the smaller messenger scale $M_{\rm mess}'$.}
\label{MatchingLambdasCartoon}
\end{figure}

Where has defining running $\Lambda$ parameters got us?
Consider a taking a specific GGM model defined by $\Lambda_{S,r},\Lambda_{G,r}$ and $M_{\rm mess}$, and calculating the running of $\Lambda_{S,r},\Lambda_{G,r}$ down to a lower scale $Q$.
Now define another GGM model whose parameters are $M_{\rm mess}$ equal to that scale $Q$, and $\Lambda_{S,r},\Lambda_{G,r}$ equal to the running values just calculated.
By construction, these two models will have the same running $\Lambda$ parameters at all scales below the lower messenger mass, and thus be indistinguishable; the effect of larger $M_{\rm mess}$ / more RG running can be absorbed into the $\Lambda$ parameters (I am still neglecting Yukawa couplings -- this statement will shortly be revised).
I illustrate this with the cartoon Fig.~\ref{MatchingLambdasCartoon}.
Similarly Fig.~\ref{unifiedLambdas} can be reinterpreted, instead of showing the running $\Lambda_{S,r}(Q)$ for a single model, as showing $\Lambda_{S,r}(M_{\rm mess})$ for a series of equivalent models with different $M_{\rm mess}$.

\begin{figure}
\begin{center}
\hfill
\includegraphics[width=0.8\linewidth]{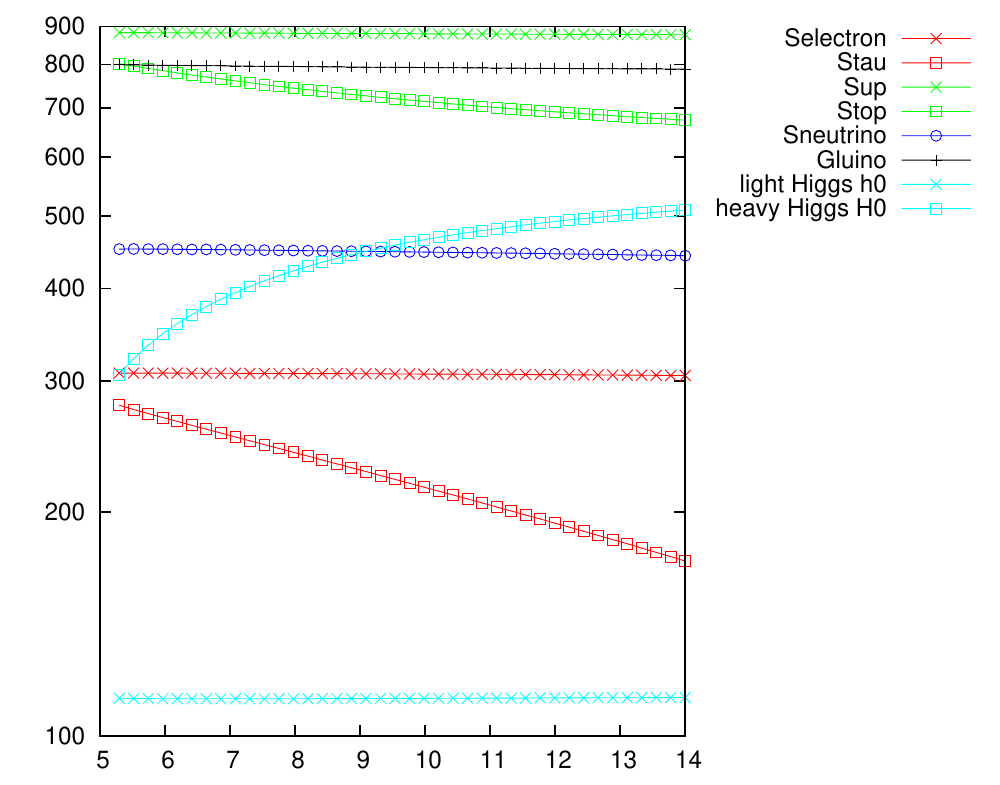}
\Text(-340,190)[c]{Mass / GeV}
\Text(-185,-10)[c]{$\log_{10} \left(M_{\rm mess} \,/\, \text{GeV}\right)$}
\end{center}
\caption{The masses of selected particles for GGM models where the $\Lambda_{S,r}$ are varied with $M_{\rm mess}$ as shown in Fig.~\ref{unifiedLambdas}, with $\Lambda_{G,r}$ held constant at $10^5~{\rm GeV}$.
`Sneutrino' refers to $\tilde{\nu}_e$.}
\label{fig:VaryingThirdGenMmess}
\end{figure}

The third generation sfermions (and Higgs scalars) however, have non-negligible Yukawa couplings.
This is why their high-scale sum rules do not hold at the low scale, and why the $I_{M_r}$ in Table~\ref{RGIsTable} are RGIs for the first (or second) generation only, not the third.
While the effect of gauge-group induced RG running can be absorbed into the $\Lambda$ parameters as discussed, the Yukawa-induced splitting of the third generation from the first two cannot (likewise for the splitting of $m_{H_u^2}$ and $m_{H_d^2}$ from $m_{\tilde{L}}^2$), as GMSB is by construction flavour-blind.
Therefore an inescapable role of the messenger scale is to control the amount of this Yukawa-induced splitting.
To illustrate this I chose one model to have unified $\Lambda_{S,r}=\Lambda_{G_r}=10^5~{\rm GeV}$ at a
messenger scale $M_{\rm mess}=10^{14}~{\rm GeV}$, and defined a series of models with lower $M_{\rm mess}$ and $\Lambda$ parameters chosen to make the models maximally equivalent (i.e. equivalent up to unavoidable Yukawa-induced splitting, as discussed), namely with $\Lambda_{G_r}$ constant and $\Lambda_{S,r}(M_{\rm mess})$ defined as shown in Fig.~\ref{unifiedLambdas}.
I plot the resulting masses of selected particles in Fig.~\ref{fig:VaryingThirdGenMmess}.

First generation sparticle masses in Fig.~\ref{fig:VaryingThirdGenMmess} stay constant -- the effect of changing $M_{\rm mess}$ has been fully absorbed into the appropriately chosen $\Lambda_{S,r}(M_{\rm mess})$ as promised.
The gaugino masses also stay constant (only the gluino is shown; the bino and wino mix with the Higgsinos).
Third generation sparticle masses split from their first generation counterparts increasingly with higher $M_{\rm mess}$; high $\tan\beta$ is chosen to show this effect maximally for the stau (likewise for the sbottom, not shown), at small $\tan\beta$ the stau and sbottom masses follow those of the selectron and sdown respectively.
The Yukawa-sensitive heavy Higgs mass also varies.

The goodness of the sum rules for the first two generations, and hence the continued parameterisation of the five mass-squareds in terms of running $\Lambda_{S,r}(Q)$, allows for a potential signature for unified $\Lambda_{S,r}(M_{\rm mess}) = \Lambda_{S}(M_{\rm mess})$ models.
If first/second generation superpartners were discovered, and their masses measured and found to satisfy the sum rules, then the low-scale $\Lambda_{S,r}$ could be calculated.
Discovery and measurement of the masses of gauginos (complicated by the need to determine the mixing angles with Higgsinos) would then determine the running $\Lambda_{S,r}(Q)$ up to high scales.
If these three quantities were observed to unify at a single scale, this would be suggestive of a unified $\Lambda_{S}$ GGM model with a messenger mass at the observed unification scale.
Note that unification of running $\Lambda_{S,r}(Q)$ is a separate issue from gauge-coupling unification: one may happen without the other, and the unification scales are independent.
Explicitly, from the definition of the $I_{M_r}$ in Table~\ref{RGIsTable} and the definition of running $\Lambda_{S,r}(Q)$ in Eq.~\eqref{runningLambdas}, the values of $\Lambda_{S,r}$ at a low scale $Q_{\rm low}$ are determined by measurement as
\begin{equation}
\Lambda_{S,r}^2(Q_{\rm low}) =
\frac{16\pi^2}{\kappa_r \alpha_{r}^2(Q_{\rm low})} \sum_{\tilde{f}} D(\tilde{f},r) m_{\tilde{f},1}^2(Q_{\rm low})
\end{equation}
These can be extrapolated up to any higher scale $Q$ by replacing $Q_{\rm low}$ with $Q$ throughout, of course; however knowing the one-loop running of the sfermion mass-squareds and gauge couplings this gives explicitly
\begin{equation} \label{RunningsLambdasFromData}
\Lambda_{S,r}^2(Q) =
\frac{16\pi^2}{\kappa_r \alpha_{r}^2(Q_{\rm low})} \left[ \sum_{\tilde{f}} D(\tilde{f},r) m_{\tilde{f},1}^2(Q_{\rm low}) + M_r^2(Q_{\rm low})\left(1-\frac{\alpha_{r}^2(Q_{\rm low})}{\alpha_r^2(Q)}\right) \right]
\end{equation}

\begin{figure}[!ht]
\centering
\includegraphics[width=0.8\linewidth]{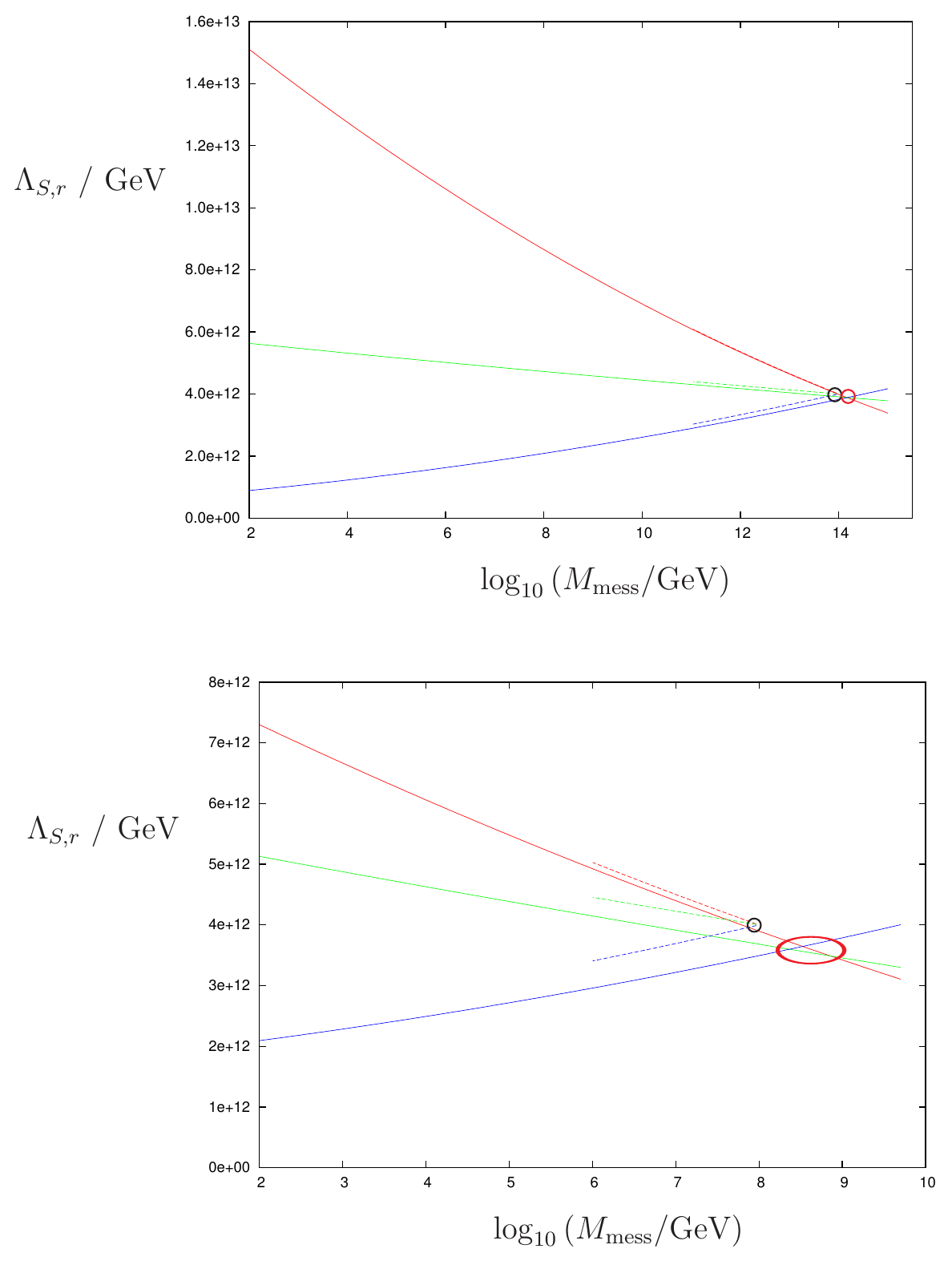}
\caption{Running $\Lambda_{S,r}$ in two models with the unification $\Lambda_{S,r}(M_{\rm mess}) = \Lambda_S$.
$\Lambda_{S,1}$ is shown in red, $\Lambda_{S,2}$ in green, and $\Lambda_{S,3}$ in blue.
Dashed lines show $\Lambda_{S,r}(Q)$ at and just below $M_{\rm mess}$ calculated with the high-scale $\Lambda_{S,r}(M_{\rm mess})$ parameters: these unify exactly, highlighted with a black circle.
Solid lines show $\Lambda_{S,r}(Q)$ calculated from the exact low-scale spectrum, with imperfect unification shown in the red ellipse.
The top panel has $\Lambda_G = 5 \times 10^{5}~{\rm GeV},\Lambda_S= 2 \times 10^{6}~{\rm GeV}, M_{\rm mess} = 10^{14}~{\rm GeV}$; the bottom panel $\Lambda_G = \Lambda_S= 2 \times 10^{6}~{\rm GeV}$ at $M_{\rm mess} = 10^{8}~{\rm GeV}$.
}
\label{fig:theor}
\end{figure}

Note that if the $\Lambda_{S,r}$ really are unified at a scale $M_{\rm mess}$, then even if the low-scale observables were measured perfectly there would still be some small error in the unification of the $\Lambda_{S,r}^2(Q)$ extrapolated from the low scale.
This is because the RGIs of Table~\ref{RGIsTable} are only RGIs at one loop; equivalently Eq.~\eqref{RunningsLambdasFromData} is obtained only by integrating the one-loop beta functions.
This can be considered a source of theoretical error in the calculated unification, and is illustrated in Fig.~\ref{fig:theor} for an optimistic and a pessimistic case: the true unification (obtained by using the model parameters) is compared to the result of a calculation from the exact low-scale spectrum.

\begin{figure}
\centering
\hspace*{1cm}
\subfigure{
\includegraphics[width=0.6\linewidth]{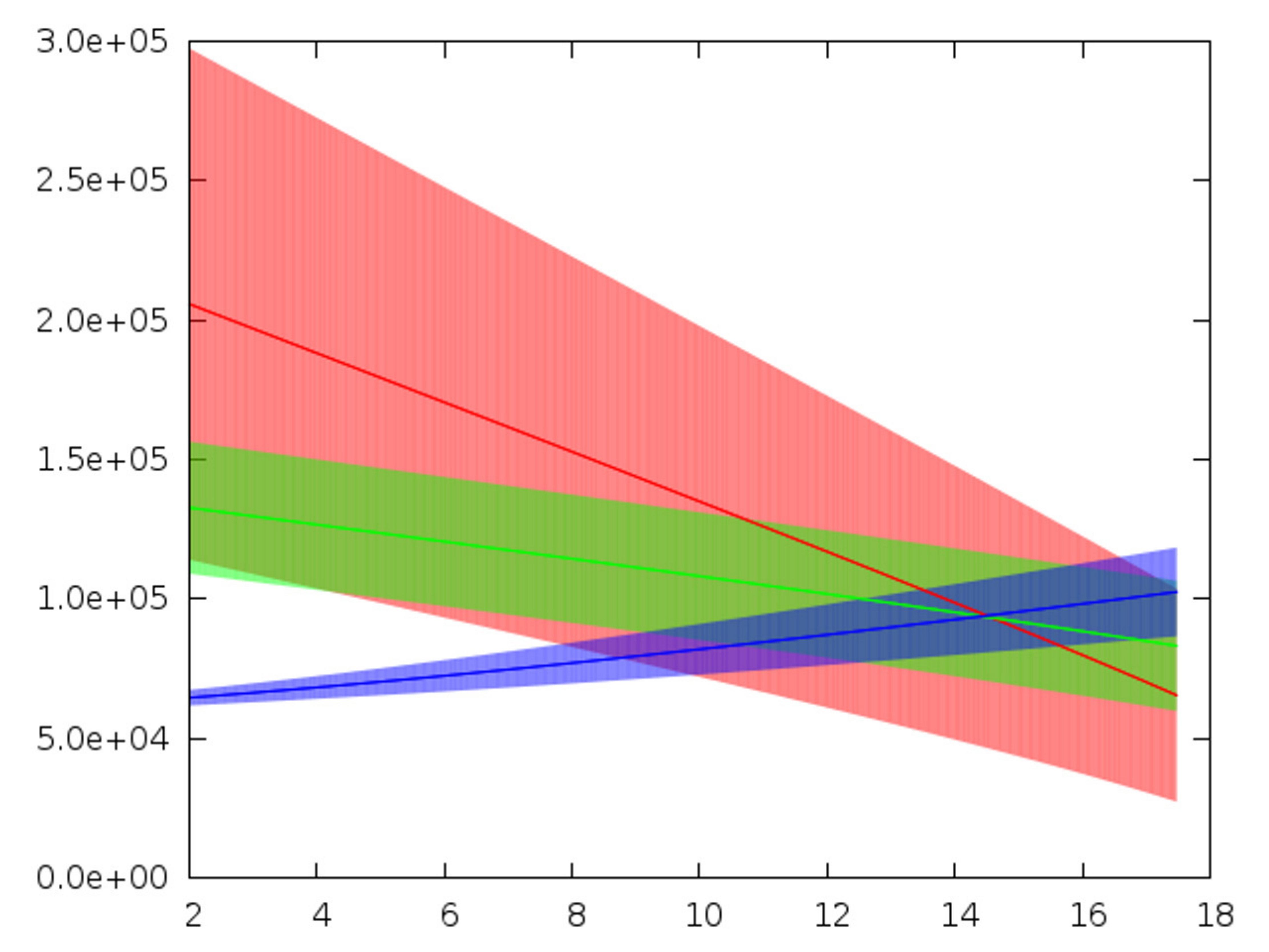}
\Text(-270,130)[c]{$\Lambda_{S,r}$ / GeV}
\Text(-90,-5)[c]{$\log_{10} \left(M_{\rm mess}/{\rm GeV}\right)$}
}\\\vspace*{3mm}
\hspace*{1cm}
\subfigure{
\includegraphics[width=0.6\linewidth]{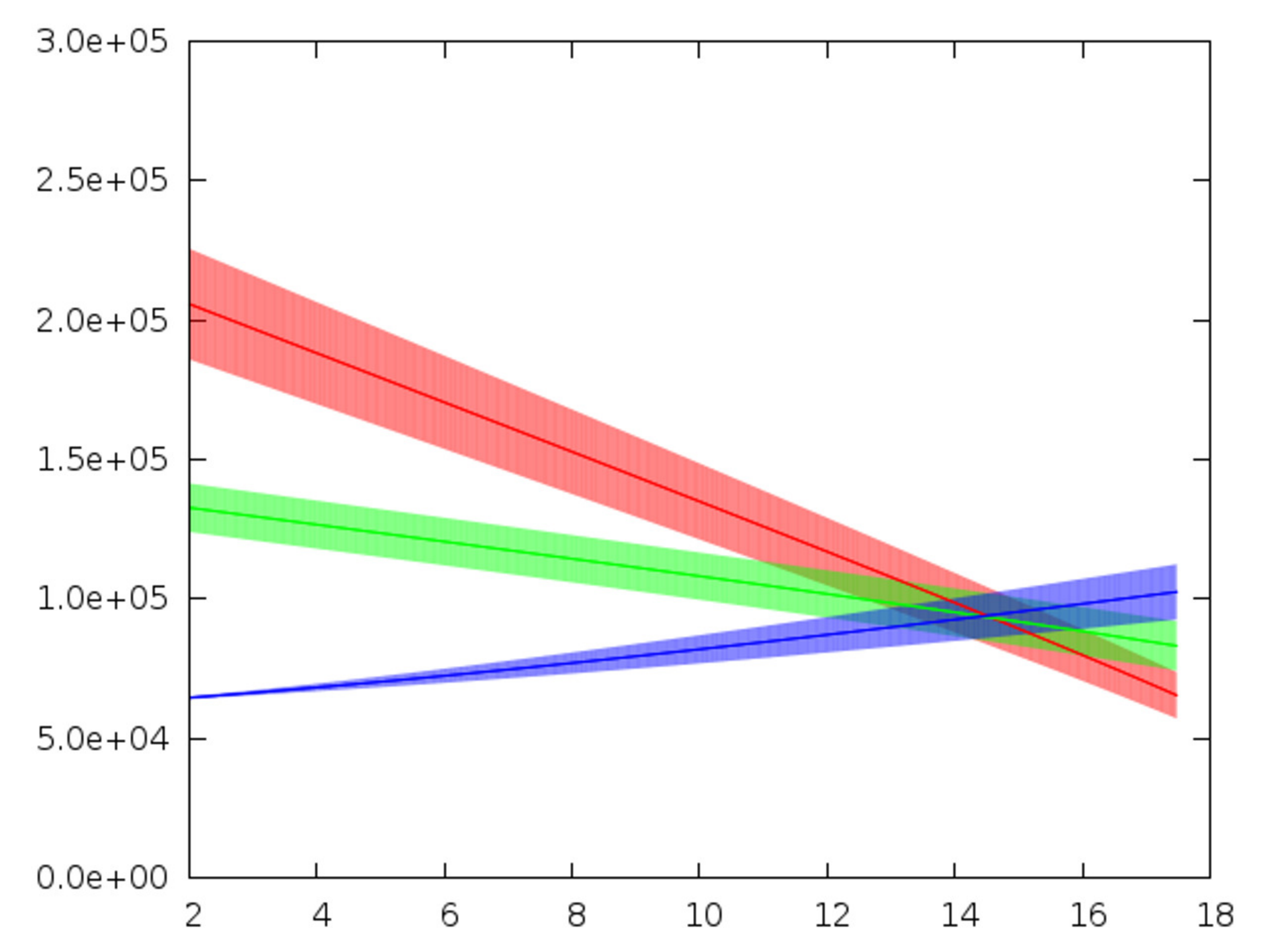}
\Text(-270,130)[c]{$\Lambda_{S,r}$ / GeV}
\Text(-90,-5)[c]{$\log_{10} \left(M_{\rm mess}/{\rm GeV}\right)$}
}
\caption{Reconstructed values of the running $\Lambda_{S,r}$ from the relevant soft masses -- those of the gauginos and the first generation spartners. 
In the top panel these are all taken to be measured to $5\%$; in the bottom panel an improved precision of $1\%$ is taken for $m_{\tilde{u}}$ and $m_{\tilde{d}}$.
The model has unified $\Lambda_{S,r}=\Lambda_{G,r}=10^5~{\rm GeV}$ at a messenger scale $M_{\rm mess}=10^{14}~{\rm GeV}$.
}
\label{fig:bands}
\end{figure}

In addition to the theoretical error we must consider the limited precision with which soft mass terms could be measured.
In Fig.~\ref{fig:bands} I illustrate the effect of this uncertainty on the observed unification.
Uncertainty in the running gauge couplings is neglected; uncertainty in the relevant soft masses -- those of the gauginos and the first generation spartners -- is first assumed to be $5\%$ (top panel of Fig.~\ref{fig:bands}), and then $1\%$ for $m_{\tilde{u}},m_{\tilde{d}}$ and $5\%$ for all the others (lower panel).
Squark mass measurements of $\mathcal{O}(1\%)$ are not implausible at the LHC with $300~\text{fb}^{-1}$~\cite{ATLAS:1999vwa}.
Unification in the former case is obscured by the errors whereas in the latter case it is more or less visible, due to the sensitivity to the splitting $m_{\tilde{u}}^2-m_{\tilde{d}}^2$, albeit with a few orders of magnitude uncertainty in the scale.
The experimental uncertainties dominate the theoretical uncertainties -- compare Fig.s~\ref{fig:theor} and~\ref{fig:bands}.

\newpage
\part{Susy Searches At The LHC}

\section{Introduction} \label{ColliderIntroduction}

\subsection{How To (Not) See Susy} \label{ColliderIntroHowToNotSeeSusy}

Proton collisions at the LHC may produce Susy particles {\it if} they exist with masses around the weak/TeV scale.
The final state(s) they decay into will then be enriched compared to the Standard Model\footnote{
This is is not a given in quantum field theory of course -- a cross-section is the square-modulus of a sum of amplitudes $\mathcal{A}$, and these may interfere destructively.
It is almost always the case however that the final states resulting from the decays of BSM particles are enhanced compared to the Standard Model alone, i.e. that $|\mathcal{A}_\text{signal}+\mathcal{A}_\text{background}|^2 > |\mathcal{A}_\text{background}|^2$.
}, or said another way, signal plus background is greater than background alone.
In order to have this effect be as visible as possible, we typically try to find the point in {\it phase space} (see the glossary) with the largest ratio of signal plus background to background, and focus on a region of phase space around that point which is small enough to retain this large ratio, but large enough for it to correspond to a statistically significant number of events $N$ when multiplied by our envisaged integrated luminosity $\mathcal{L}$:  $N = \mathcal{L} \times \sigma$, where sigma has been integrated over the chosen phase space.

Once one has a rough idea of the region of phase space to focus on, further discrimination of signal plus background from background alone may be possible if there is an observable (a quantity typically calculated from the final state four-momenta) which is distributed very differently in the two cases.
The classic example is the invariant mass of the sum of the four-momenta of all the particles into which a new particle decays: by definition this is a $\delta$-function for the signal, smeared out both by the finite width of the particle and by detector resolution, and some more broadly spread continuous distribution for the background.
It is thus highly desirable to calculate this observable not only to know the mass of any particles that might be discovered this way, but because it is essentially the ideal way to discover their existence in the first place, by concentrating all of the signal to a highly visible {\it bump} on a smooth background.

Mass reconstruction is generally not possible however when the particle we're searching for decays to neutral particles that are stable on a collider timescale, as such particles are not detected.
As mentioned in Section~\ref{SusyMotivation}, for $R$~parity conserving Susy the LSP is stable, and therefore it must be neutral for consistent cosmology.
Unless decays to the LSP are strongly suppressed, i.e the next-to-LSP (NLSP) or NNLSP etc. is quasi-stable, any sparticles produced at a collider will decay to an invisible LSP {\it inside} the detector.
Furthermore $R$~parity ensures sparticles can only be produced in pairs, and so all events will result in pairs of LSPs.
Mass peaks are therefore typically a pipe dream for $R$~parity conserving Susy, though see Section~\ref{MakingMostMET} in which the dream comes true in the specific case of boosted decays.

Without mass peaks we are often forced into {\it cut and count} style approaches: we identify the aforementioned roughly selected region of phase space which is promising, ignore/cut the rest of phase space, and simply count events.
If the number of events observed closely matches the number predicted by the background alone, then any new physics which predicts considerably more events than this can be excluded.
The problem with such analyses is that we need precise knowledge of the background normalisation -- the integral of the background differential cross-section over the selected phase space.
To see this note that a even a modest error in the background differential cross-section, when integrated over a large phase space, could result in an uncertainty that swallows the whole signal (which may live only in a certain part of the phase space though we don't know where).
This is to be contrasted with the presence or absence of a bump on top of a smooth function (the background), which is not sensitive to an overall scaling of the latter.

Cut and count analyses do however lead to exclusion limits which are straightforward to re-interpret in other models: they set an upper limit, at a certain confidence level which is usually chosen to be $95\%$, on the cross-section that any new physics can contribute to the final states and region of phase space that pass the cuts.
Said more precisely the upper limit is on the quantity $\sigma_{\rm prod} \times A \times \epsilon$, where $\sigma_{\rm prod}$ is the production cross-section for new physics, $A$ is the acceptance, and $\epsilon$ is the {\it efficiency} i.e. the ratio of the number of events that should pass the cuts to the number that actually do, due to imperfections of the detector.
Writing this product as $\sigma_{\text{signal}}$, new physics with $\sigma_{\rm limit}/\sigma_{\text{signal}}$ less than (greater than) one is excluded (allowed).
The boundary/boundaries between such regions are often illustrated with {\it Brazil-band plots}, which I explain in a manner suitable for novices in Appendix~\ref{BrazilBandPlots}.

\subsection{Complications At NLO}

To interpret a model-independent cross-section upper limit in a specific model, we need to know the model's acceptance, and so
we need the distributions in phase space resulting from production and decay of our new hypothesised particles.
This problem is almost always tackled by using Monte Carlo event generators to simulate large numbers of {\it events} (particle collisions in the manner of the chosen collider -- here the LHC).
In a nut-shell, an event generator simulates a single event by considering what could happen each time particles interact (a typical event involves a huge number of interactions), and choosing one of the possibilities randomly but weighted by the physical probability which has been calculated using QFT.
The user may force all of the events to go down a certain pathway on the branching tree of possibilities, which is sensible if only a subset of all possible outcomes are to be studied; a scaling of final cross-sections is then appropriate to account for this.
For example, in studies of decays of the Higgs to a given final state, the user would normally force the event generator to only decay the Higgs to that final state, and multiply any final cross-sections obtained by the branching ratio for such a decay.
A second example, more relevant here, is that to study decays of Susy particles, one should force the event generator to simulate only events where Susy particles are produced, rather than events where the colliding particles interact in any way possible!

Most Monte Carlo event generators work with LO {\it matrix elements} (see the glossary) for the production of Susy particles.
NLO corrections can be substantial however, with K-factors of $2$-$3$ common for total cross-sections for production of coloured sparticles.
A common approach is then to calculate $\sigma_{\text{signal}} = \sigma_{\rm prod} \times A$ (neglecting non-unit efficiencies $\epsilon\neq1$ for the moment) at LO using event generation, and then to multiply by the global K-factor defined for the total cross-section: $K = \sigma_{NLO} / \sigma_{LO}$.
There are two ways in which this can fail as an approximation for $\sigma_{\text{signal}}$ at truly NLO.
\begin{itemize}
 \item {\it High phase space dependence of $K$}.
 It is clear that to convert $\sigma_{\text{signal}}$ from LO to NLO we should multiply it by the K-factor associated with the region of phase space we are concerned with, {\it not} the K-factor associated with an integral over all phase space.
 The latter is only acceptable if the two K-factors are similar.
 (As an example of when this is not the case see the $\mathcal{O}(10-100)$ K-factor in the high-$p_T$ tail of light slepton Drell-Yan production shown in~\cite{FridmanRojas:2012yh}.)
 \item {\it High dependence of both $K$ and acceptance on particle-species}.
 $\sigma_{\text{signal}}$ may receive contributions from the production and subsequent decay of different new particles; let me say via different channels for brevity.
 If different channels have different K-factors {\it and} different acceptances, then multiplying the cross-section for the full signal (including all channels) by the total K-factor is not correct\footnote{
 Thanks to Peter Richardson for realising this in the context of our work, and to David Grellscheid for patiently explaining it to me.
 }, as the following toy example shows.
 Say our signal receives contributions from electroweak production (`EW') and coloured production (`C') and we use a search strategy geared more towards one than the other, say vetoing leptons, giving a higher acceptance for the latter than the former: $A_C > A_{EW}$.
 The strength of the strong interaction means that generically \mbox{$K_C > K_{EW}$}; so we have channels with different acceptances and K-factors.
 For concreteness let us take the cross-sections to be
 \begin{gather}
  \sigma_{C,\,LO} = \sigma_{EW,\,LO},\\
  K_C = 3, \:K_{EW} = 1\\
  \therefore \;K_{tot} = \sigma_{tot,\,NLO} \:/\: \sigma_{tot,\,LO} = 2,
 \end{gather}
where we assume interference diagrams to be negligible so that \linebreak\mbox{$\sigma_{tot} = \sigma_{C} + \sigma_{EW}$}.
Take the ({\it a priori} unknown) acceptances to be \mbox{$A_C = 1$}, $A_{EW}=0$.
Monte Carlo simulation (at LO) of both channels {\it simultaneously} will generate equal numbers of events for each (as \mbox{$\sigma_{C,\,LO} = \sigma_{EW,\,LO}$}), and since $A_C = 1, A_{EW}=0$ the total acceptance is determined to be $A_{tot} = 0.5$.
The LO signal cross-section is then \mbox{$\sigma_{int,\,LO} = \sigma_{tot,\,LO} \times A_{tot}$}; multiplying by the total K-factor one erroneously concludes
\begin{align}
 \sigma_{int,\,NLO} \stackrel{!}{=} &\:\sigma_{tot,\,LO} \times A_{tot} \times K_{tot} \quad \text{in general,}\label{wrongChannelDecompositionOfMC}\\
 = &\: 2\sigma_{C,\,LO} \quad \text{in this example.}
\end{align}
This should be compared with what is clearly the correct answer, obtained by simulating the two channels separately in order to determine their individual acceptances, then weighting each by its individual K-factor:
\begin{align}
 \sigma_{int,\,NLO} = &\: \sum_{i \,=\, C,EW} \sigma_{i,\,LO} \times A_{i} \times K_{i} \quad \text{in general,} \label{ChannelDecompositionOfMC}\\
 = &\: 3\sigma_{C,\,LO} \quad \text{in this example,}
\end{align}
where $i$ denotes the channel.
To see more clearly the difference between the two approaches, and when they coincide, I re-write Eq.
\eqref{wrongChannelDecompositionOfMC} explicitly denoting was is meant by the {\it total} acceptance and K-factor in terms of the channel-by-channel definition of these quantities:
\begin{equation} \label{wrongChannelDecompositionOfMC2}
 \begin{split}
 \sigma_{int,\,NLO} \stackrel{!}{=} &\:\sigma_{tot,\,LO} \times A_{tot} \times K_{tot}\\
 = &\: \left( \sum_{i} \sigma_{i,\,LO}  \right) \times \left( \frac{\sum_{i} \sigma_{i,\,LO} A_{i} }{\sum_{i} \sigma_{i,\,LO} } \right) \times \left( \frac{\sum_{i} \sigma_{i,\,LO} K_{i} }{\sum_{i} \sigma_{i,\,LO} } \right)\\
 = &\: \frac{ \left(\sum_{i} \sigma_{i,\,LO} A_{i}\right) \left( \sum_{i} \sigma_{i,\,LO} K_{i} \right)}{\sum_{i} \sigma_{i,\,LO} }
 \end{split}
\end{equation}
The incorrect combined channel approach, Eq.
\eqref{wrongChannelDecompositionOfMC2}, coincides with the correct channel-by-channel approach, Eq.
\eqref{ChannelDecompositionOfMC}, if there is a common acceptance across all channels $A_i = A$ or if there is a common K-factor across all channels $K_i = K$.
\end{itemize}
In the event generation of Section~\ref{1/fbSearches} I will not address the issue of phase space dependent K-factors.
I will however take into account the issue of non-universal K-factors and acceptances across different channels.

\subsection{The Importance Of The LSP}

Spontaneous breaking of a global supersymmetry results in a massless fermion called the Goldstino; this is just Goldstone's Theorem applied to a broken fermionic generator.
In the context of gravity Susy must be promoted to a local symmetry ({\it supergravity}) with the the spin-$\tfrac{3}{2}$ superpartner of the graviton -- the gravitino -- acting as the associated gauge field.
Upon spontaneous breaking of the local supersymmetry the massless gravitino eats the would-be Goldstino to obtain mass and longitudinal components: this is called the {\it Super-Higgs mechanism}, being a close parallel of the {\it Higgs-} or more justly the {\it Brout-Englert-Higgs mechanism}~\cite{Englert:1964et,Higgs:1964ia,Higgs:1964pj} in which a spin-one gauge boson obtains mass and a longitudinal component by eating a scalar.
The transverse components of the gravitino interact only gravitationally, and so its interactions are dominantly just those of its Goldstino component (and the distinction between gravitino and Goldstino is irrelevant, with both usually denoted $\tilde{G}$).
With Susy breaking coming from a single $F$-term VEV, $\langle F\rangle$, the resulting gravitino mass can be estimated from dimensional analysis~\cite{Deser:1977uq,Cremmer:1978iv} as
\begin{equation}
m_{3/2} \sim  \langle F\rangle/M_P,
\end{equation}
as it must vanish in the limit of decoupled gravity $M_P\rightarrow\infty$ or restored Susy $\langle F\rangle\rightarrow0$.
The other superpartners however receive masses $\sim\!\langle F\rangle/M_{\rm mess}$, suppressed only by the mass $M_{\rm mess}$ of the Susy-breaking mediator.
In GMSB this is much below $M_P$, and so the gravitino is always the LSP.

The decay of a sparticle to its Standard Model partner and the gravitino is suppressed, having decay width~\cite{Martin:1997ns}
\begin{equation}
 \Gamma(\tilde{X}\rightarrow X\tilde{G}) = \frac{m_{\tilde{X}}^5}{16\pi\langle F\rangle^2} (1-m_{X}^2/m_{\tilde{X}}^2)^4
\end{equation}
For GMSB with $m_{\tilde{X}}\sim\langle F\rangle/M_{\rm mess}$, the suppressing factor is $m_{\tilde{X}}^4/\langle F\rangle^2 \sim \langle F\rangle^2 / M_{\rm mess}^4$, which we typically take much less than one as discussed in Section~\ref{PartGGMSectionParamSpaceSubSectionSimpleToComplex}.
Therefore the dominant {\it decay cascades} (see the glossary) for $R$~parity conserving GMSB will be those where each pair-produced particle decays through a series of steps with gauge or Yukawa vertices until the NLSP is produced, and only then does the decay to the gravitino occur, as no other channel competes.
GMSB collider phenomenology then depends most importantly on the NLSP species (i.e. the identity of $\tilde{X}$, and thus $X$, in $\tilde{X}\rightarrow X\tilde{G}$) and whether or not this decay is sufficiently suppressed as to occur outside the detector.
In~\cite{Abel:2010vba} the associated decay length for $\Lambda_S,\Lambda_G$ models is given as
\begin{equation}
L_{\rm NLSP}\sim \frac{1}{k^2_{G}}\left(\frac{100~{\rm GeV}}{m_{\rm NLSP}}\right)^5 \left(\frac{\sqrt{\Lambda_{G} M}}{10^5~{\rm GeV}}\right)^4\, 10^{-4}\, {\rm m},
\end{equation}
where $k_{G}$ (the reciprocal of the parameter known as $C_{\rm grav}$) quantifies the coupling of messengers to the Susy-breaking sector, which may be $\mathcal{O}(1)$ or much smaller.
For $\Lambda_{G}\sim 10^{5}~{\rm GeV}$, messenger scales $M_{\rm mess}\gtrsim 10^{7}~{\rm GeV}$ lead to $L_{\rm decay}\gtrsim 10$~m and a decay outside of the detector.
When $k_{G}\ll 1$ even smaller values of the $M_{\rm mess}$ may suffice.
Detector-stable NLSPs are therefore a very realistic possibility.
The NLSP species in these models can be crudely characterised as a bino-like $\chi_1^0$ for $\Lambda_G\lesssim \Lambda_S$ and a charged slepton for $\Lambda_G \gtrsim \Lambda_S$ (specifically a stau for large $\tb$ and slepton-smuon-stau co-NLSP for small $\tb$).
Detector-stable charged particles have dedicated searches; a detector-stable $\chi_1^0$ however will manifest itself via missing energy.

\newpage
\section{A Historical Detour: Comparing Early LHC Direct Searches To LEP Higgs Bounds} \label{1/fbSearches}
{\it This section is based on my work~\cite{Grellscheid:2011ij} done in collaboration with David Grellscheid, J\"{o}rg J\"{a}ckel, Valya Khoze and Peter Richardson; the text has been mostly re-written.}\\

\subsection{Introduction And Main Results}

The final state in which we initially harboured the most hope for seeing Susy at the LHC was jets, $\slashed{E}_T$, and no leptons.
\begin{itemize}
 \item {\it Jets}.
 The particles with the largest production cross-sections for a given mass are coloured particles, since QCD is the strongest interaction at the weak/TeV scale and at the LHC we collide coloured objects.
 Decays of coloured particles must produce other coloured particles (as the colour quantum number is conserved), which will ultimately hadronise to produce jets.
 \item $\slashed{E}_T$.
 Susy is most often considered in the $R$~parity conserving scenario, so that the LSP is stable and constitutes a dark matter candidate.
 Provided the full decay cascade is {\it prompt} (takes place inside the detector), the escape of the two LSPs gives rise to $\slashed{E}_T$.
 To give large $\slashed{E}_T$, the LSPs must carry appreciable kinetic energy (and not be back-to-back in the transverse plane), which requires the initially produced particles to be appreciably more massive than the sum of the masses of all final state particles.
 \item {\it No leptons}.
 Backgrounds are either {\it reducible} or {\it irreducible} (see the glossary).
 The irreducible background for $\slashed{E}_T$ consists exclusively of $W$ and $Z$ bosons decaying to neutrinos.
 A neutrino from a decaying $W$ is accompanied by a lepton, unless it is $\nu_\tau$ and the associated $\tau$ decays hadronically.
 Thus for new physics decays that produce $\slashed{E}_T$ without leptons, such as the squark and gluino decays $\tilde{q}\rightarrow q\chi_1^0$ and $\tilde{g}\rightarrow q\bar{q}\chi_1^0$, a lepton veto will generally improve the ratio of signal to (reducible) background.
\end{itemize}

\begin{figure}
\vspace*{-3mm}
\begin{center}
\subfigure[
Showing models of GMSB with a messenger scale $M=10^7 (10^{14})$~GeV in the left (right) panel.
For $\tb=45$ there is a region where the stau becomes tachyonic which cuts off the Higgs constraint contour.
]{
\shortstack{
\vspace*{-5mm}
\includegraphics[width=0.47\linewidth]{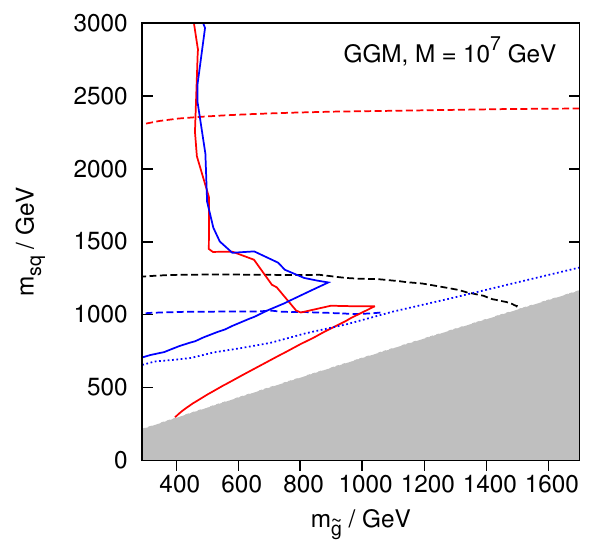}
\includegraphics[width=0.47\linewidth]{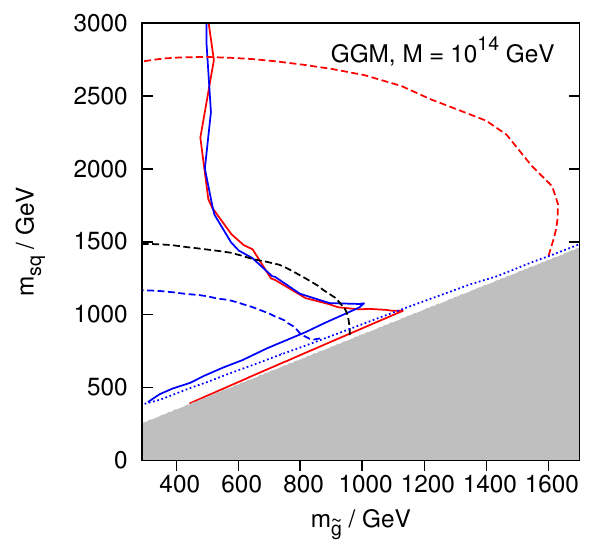}\\
\includegraphics[width=0.4\linewidth]{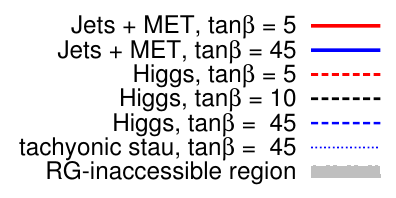}
}
\label{fig:GGM-sq-gl-plane}
}
\subfigure[Showing the CMSSM.
In the left panel I set $A_0=0$ and vary $\tb$; in the right panel I fix $\tb=10$ and vary $A_0$.
The jets$+\slashed{E}_T$ constraints are essentially independent of $\tb$ and $A_0$ in the CMSSM~\cite{Akula:2011zq}.
]{
\shortstack{
\vspace*{-3mm}\includegraphics[width=0.47\linewidth]{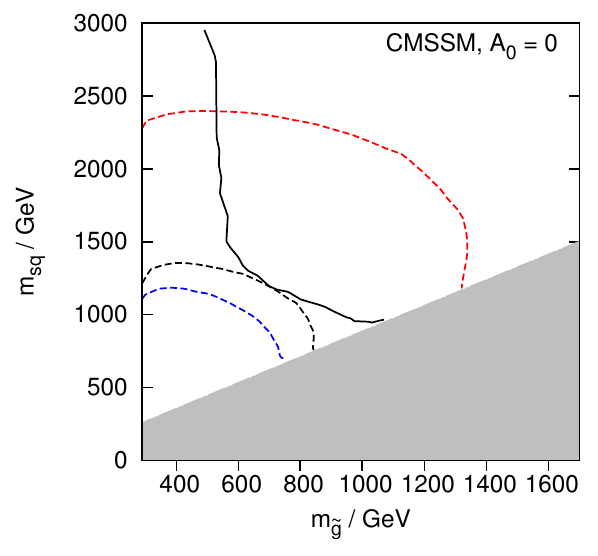}
\includegraphics[width=0.47\linewidth]{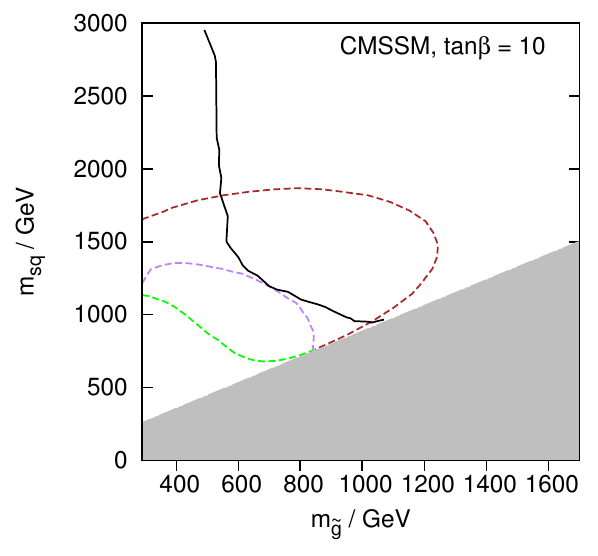}\\
\includegraphics[width=0.4\linewidth]{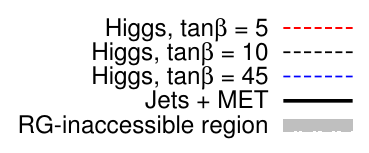} \hspace{0.1\linewidth}
\includegraphics[width=0.4\linewidth]{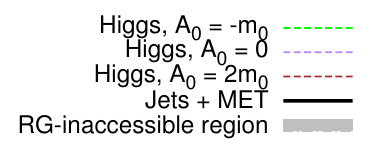}
 }
\label{fig:CMSSM-sq-gl-plane}
}
\caption{Bounds from the 2011 $1.04~{\rm fb}^{-1}$ search for Susy by the ATLAS Collaboration in the jets$+\slashed{E}_T$ final state (solid lines), and from the LEP Higgs bound $m_{h}> 114.4~{\rm GeV}$ (dashed lines), in the plane of gluino $\tilde{g}$ and squark masses.
Different colours show different values of $\tb$.
The grey area is theoretically inaccessible~\cite{Jaeckel:2011wp} (unless tachyonic squarks are allowed at high scales).
}
\label{1fbMainResults}
\end{center}
\end{figure}

Though discovery and exclusion do not have trivially connected statistics (e.g. the confidence level for discovery plus the confidence level for exclusion does not equal one!), a search strategy with strong potential for discovery will also be able to set strong exclusion limits (unless the hypothetical particle actually exists of course) and vice versa, since both of these come if and only if the strategy has sensitivity to the hunted particle's signal.
Therefore when data in the much anticipated jets, $\slashed{E}_T$ and no-lepton (henceforth jets$+\slashed{E}_T$) channels began coming in in earnest but {\it without} excesses over the background, strong exclusion bounds were set (and are continuing to be set of course).
In~\cite{Grellscheid:2011ij} I compared the strongest such bounds from 2011 -- those obtained in $1.04~{\rm fb}^{-1}$ of data taken in $7$~TeV collisions (and focussing on the ATLAS Collaboration's results~\cite{Aad:2011ib}) -- to the LEP limit on a Higgs with Standard Model-like couplings, $m_{h}> 114.4~{\rm GeV}$~\cite{Barate:2003sz}.
This was done for models of GMSB parameterised by $\Lambda_S, \Lambda_G, M_{\rm mess}$ and $\tb$ (see Part~\ref{GGM}) and for the Constrained Minimal Supersymmetric Standard Model (CMSSM, see e.g.~\cite{Martin:1997ns}).
The sign of the $\mu$ parameter was taken to be $+1$ throughout.
The results are shown in Fig.~\ref{1fbMainResults}, with the exclusion contours projected on the squark-gluino mass plane\footnote{
Thanks to J\"{o}rg J\"{a}ckel for providing the gradient of the RG-inaccessible region.
}.
Explanations and interpretations of these bounds follow.

For the GMSB case, as mentioned, the NLSP in these models is generally either a bino-like neutralino or a charged slepton, and may be quasi-stable or decay promptly.
Charged sleptons, when decaying promptly, produce a gravitino (seen as $\slashed{E}_T$) and a charged lepton; when stable a charged non-hadronic track is seen to leave the detector.
Promptly decaying bino-like neutralinos produce a gravitino usually with a photon ($\tilde{G}+Z$ becomes competitive for heavy neutralinos).
I focussed on a detector-stable neutralino as the NLSP, which gives large $\slashed{E}_T$ that is not shared with a hard and distinctive lepton or photon, and so is most appropriate for constraining by a jets$+\slashed{E}_T$ search.

\subsection{Derivation Of The Exclusion Contours} \label{DerivationOfExclusionContours}

\begin{figure}
\vspace*{-2mm}
\begin{center}
\vspace*{-2mm}
\includegraphics[width=0.49\linewidth]{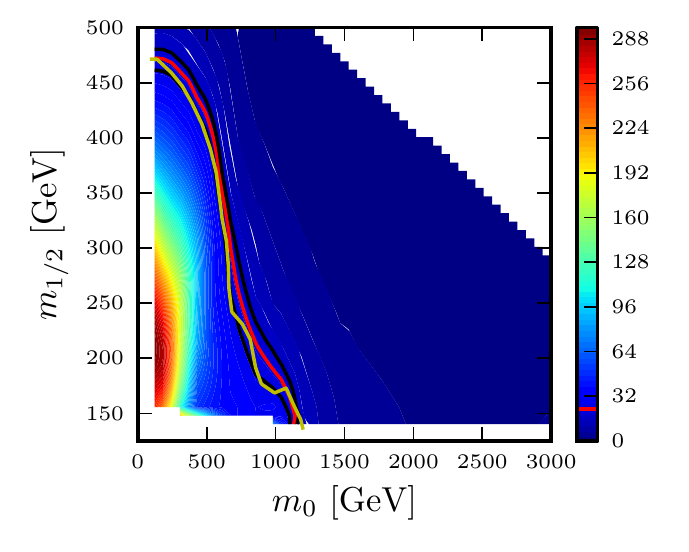}
\includegraphics[width=0.49\linewidth]{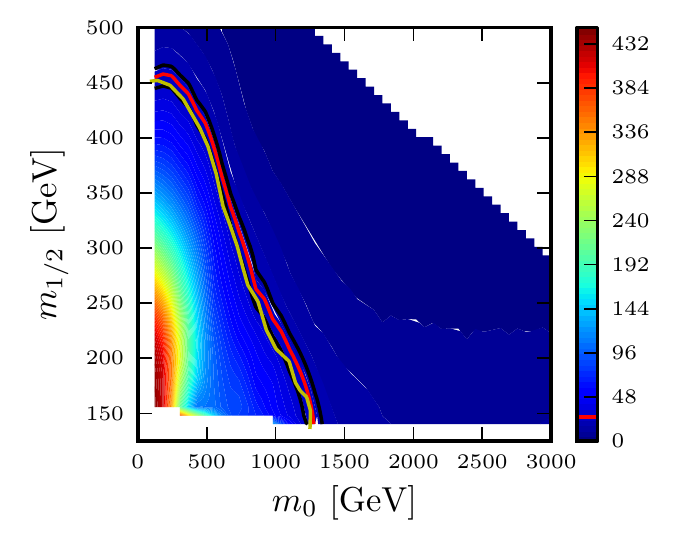}
\vspace*{-2mm}
\includegraphics[width=0.49\linewidth]{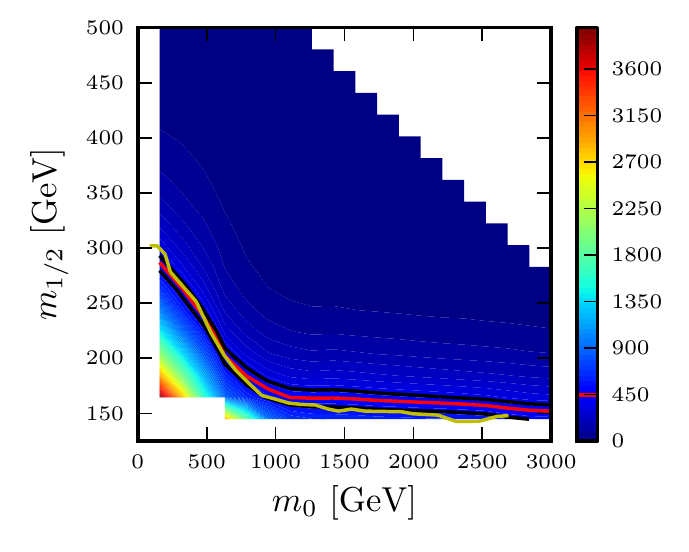}
\includegraphics[width=0.49\linewidth]{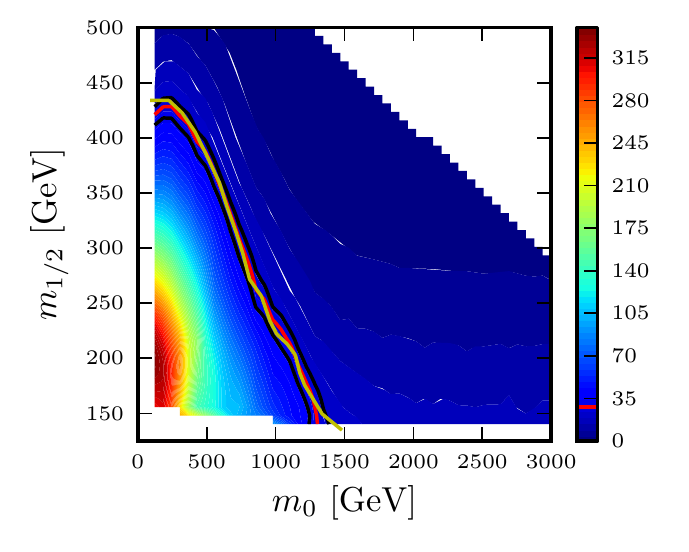}
\vspace*{-2mm}
\includegraphics[width=0.49\linewidth]{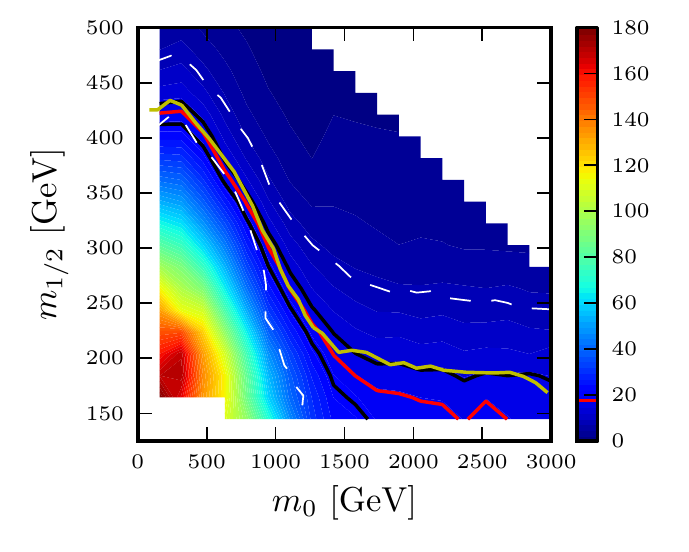}
\caption{
Validation of my reproduction of the five search channels defined in the $1.04~{\rm fb}^{-1}$ jets$+\slashed{E}_T$ search for Susy by the ATLAS Collaboration, shown in the CMSSM plane of $m_0$ and $m_{1/2}$ with $\tb=10,\,A_0=0$.
The colour `axis' shows $\sigma_{\text{signal}}$ in units of $1\text{fb}$; the red line is the contour $\sigma_{\text{signal}} = \sigma_{\text{limit}}$.
The black lines show the result of varying the renormalisation/factorisation scale by factors of $2^{\pm1}$.
The exclusion contour derived by the ATLAS Collaboration is shown in yellow.
The five search channels shown are, from left to right and top to bottom: $\ge$~2-jets, $\ge$~3-jets, $\ge$~4-jets with $m_{\rm eff}> 500$~GeV, $\ge$~4-jets with $m_{\rm eff}> 1000$~GeV, and {\it High Mass}.
In the last panel the dashed white lines show the $\pm 1 \sigma$ contours for the ATLAS Collaboration's expected exclusion.
}
\label{fig:validate}
\end{center}
\end{figure}

My route to obtaining the jets$+\slashed{E}_T$ constraints followed~\cite{Dolan:2011ie} closely\footnote{
I am grateful to David Grellscheid and Peter Richardson for providing the code necessary for {\tt Prospino} cross-section calculation and Susy event generation with {\tt Herwig++} and {\tt RIVET} on {\it The Grid}~\cite{TheGrid}.
}:
\begin{itemize}
 \item Low-scale Susy spectra were calculated from high-scale model input using {\tt SOFTSUSY~3.1.6}.
 \item Signal events were generated with {\tt Herwig++~2.5.1}, separated into samples of (a) pair-produced squarks and/or gluinos, (b) pair-produced sleptons, neutralinos and/or charginos, and (c) one squark or gluino produced with one slepton, neutralino or chargino.
(c) typically has negligible cross-section in these scenarios.
(b) may have comparable cross-section to (a) but is largely eliminated by a lepton veto; given this difference in acceptance (and the obvious difference in K-factors), the argument of Section~\ref{ColliderIntroduction} shows the importance in separating these production channels.
\item {\it High-level objects} (see the glossary) were defined from the final state particles using the {\tt RIVET~1.5.2} analysis framework and {\tt FastJet}~\cite{Cacciari:2011ma}.
The cuts for each search channel were also imposed with {\tt RIVET}; these cuts are detailed in Appendix~\ref{cutsAppendix}.
The fraction of events passing the cuts for each channel is the acceptance.
\item NLO production cross-sections were calculated with {\tt Prospino~2.1}; an indication of theoretical uncertainty was given by varying the renormalisation/factorisation scale by factors of $2^{\pm1}$. 
Multiplying the cross-section by the acceptance defined our $\sigma_{\text{signal}}$, which was compared to the experimentally determined $\sigma_{\text{limit}}$.
\item The previous steps were done for the same plane in Susy model space as used by the experimental collaboration to present its own results.
With a close matching of the original and reproduced exclusion contours, the reproduced search channels can be considered validated.
In the present case this agreement is shown in Fig.~\ref{fig:validate}, giving confidence in my reproduction of the jets$+\slashed{E}_T$ search channels (which we then made publicly available as part of {\tt RIVET} package: analysis \mbox{{\it ATLAS\_2011\_S9212183}}).
\item With the previous step validating those that came before, the same procedure can then be followed for other Susy models which were not considered by the experimental collaboration.
\end{itemize}

As discussed in Ref.~\cite{Aad:2011ib} a localised detector failure caused a loss of jet energy and ``a loss of signal acceptance which is smaller than $15\%$ for the models considered''.
This was taken into account here by including an efficiency factor $\epsilon = 0.85$ in $\sigma_{\text{signal}} \equiv \sigma_{\rm prod} \times A \times \epsilon$, for four of the five different channels -- all except the {\it High Mass} channel.
The latter places the strongest demands on jet {\it multiplicity} (see the glossary) and $p_T$; as can be seen from Fig.~\ref{fig:validate} the location of the resulting exclusion contour has large uncertainties\footnote{
This is because these demanding cuts mean the acceptance is a steeply rising function of sparticle mass in the part of the plane where $\sigma_{\text{signal}} \sim \sigma_{\text{limit}}$; the product of acceptance and production cross-section (with the latter always falling steeply with mass) is then fairly flat around $\sigma_{\rm prod}\times A \sim \sigma_{\text{limit}}$, and a systematic shift in this product (such as an efficiency factor or $\sigma_{\rm prod}$ scale uncertainties) can move this flat region from having $\sigma_{\rm prod}\times A \lesssim \sigma_{\text{limit}}$ to $\sigma_{\rm prod}\times A \gtrsim \sigma_{\text{limit}}$, causing a large shift in the exclusion contour.
}.
Including an efficiency factor in this way gave a minor improvement to my agreement with the ATLAS Collaboration's bounds in the same plane.

To check exclusion by Higgs-based searches I used {\tt FeynHiggs~2.8.5} to calculate Higgs-sector masses, couplings and cross-sections, and passed these to {\tt HiggsBounds 3.5.0beta} to compare with a comprehensive set of experimental limits from LEP, the Tevatron and the LHC.
{\tt HiggsBounds} returns $\sigma_{\text{signal}} / \sigma_{\text{limit}}$ for the search channel
with the highest statistical sensitivity: a ratio greater than $1$ indicates $95\%$ confidence-level exclusion by at least one search.
Note that this does not take into account a kind of {\it look elsewhere effect} that arises when one checks many different independent searches for the presence of a single one indicating exclusion.
A statistically thorough approach would calculate the full combined likelihood for all of the observed experimental results given the hypothesis of the model in question, and see whether this corresponds to a probability less than the critical value (nominally 0.05); this is beyond the scope of {\tt HiggsBounds}.

Higgs constraints were checked with {\tt HiggsBounds} because, in general, limits are fundamentally set in terms of cross-sections.
Their interpretation as mass limits is always a model-dependent process, requiring assumptions about couplings and branching ratios.
The LEP bound $m_{h}> 114.4~{\rm GeV}$ applies to a Standard Model-like Higgs, and carries across to a Susy Higgs only in the decoupling limit and when $h$ has no BSM decay channels open.
For the models investigated here this was the case: the {\tt HiggsBounds} result simplified to a comparison of $m_{h}$ with $114.4~{\rm GeV}$.
However away from the decoupling limit, and/or with $h\rightarrow \text{BSM}$ possible, and/or with the LEP constraint being superseded by LHC Higgs results (which are incorporated in updated versions of {\tt HiggsBounds}), this check will not in general be so trivial.
It is facilitated by my code available at~\cite{Me} which links {\tt SOFTSUSY} to {\tt FeynHiggs} and {\tt HiggsBounds} with automated plotting of the results.

\subsection{Interpretation Of The Exclusion Contours}

The most basic point to note is that large masses for squarks and gluinos make it easier to satisfy both jets$+\slashed{E}_T$ and $m_h$ constraints.
Coloured sparticles being heavy helps to avoid jets$+\slashed{E}_T$ constraints because this decreases signal production cross-section.
Heavy squarks boost $m_h$ because of the stop's positive contribution to $\delta m_h$, and the fact that both GMSB and the CMSSM are flavour-blind (at the high-scale at least -- this is slightly detuned by the Standard Model Yukawa couplings during RG running), tying the stop mass to the general squark mass.
Heavy gluinos may help boost $m_h$ because they contribute to the negative running of $A_t$ from high to low scale, thus increasing the latter's absolute value at the low scale (provided it is negative), and for constant stop mass, $m_h$ increases\footnote{
Unless of course $|A_t|$ is increased beyond $\surd6 \MS$.
In GGM however we start from $A_t=0$ at a high scale, and boosting its low scale value with a large gluino mass and/or many {\it decades} (see the glossary) of running will simultaneously boost the low scale stop mass which also runs strongly with $m_{\tilde{g}}$, generically forcing $A_t/\MS < \surd6$~\cite{Grajek:2013ola}.
This conclusion is escaped if we allow the stops to be tachyonic at the high scale, which unties the low-scale $\MS$ from $A_t$~\cite{Dermisek:2006ey}.
This implies the existence of other vacua beside our own in which charge and colour are broken, however cosmological constraints may allow such a situation~\cite{Ellis:2008mc}.
} with $|A_t|$.
This effect is clearly enhanced for more RG running, and indeed we see that for heavy gluinos the LEP bound is less constraining for high messenger scale than low messenger scale: in Fig.~\ref{fig:GGM-sq-gl-plane} the dashed lines are flatter in the left panel whereas they fall off for large $m_{\tilde{g}}$ in the right panel.

\begin{figure}
\begin{center}
\includegraphics[width=0.65\linewidth]{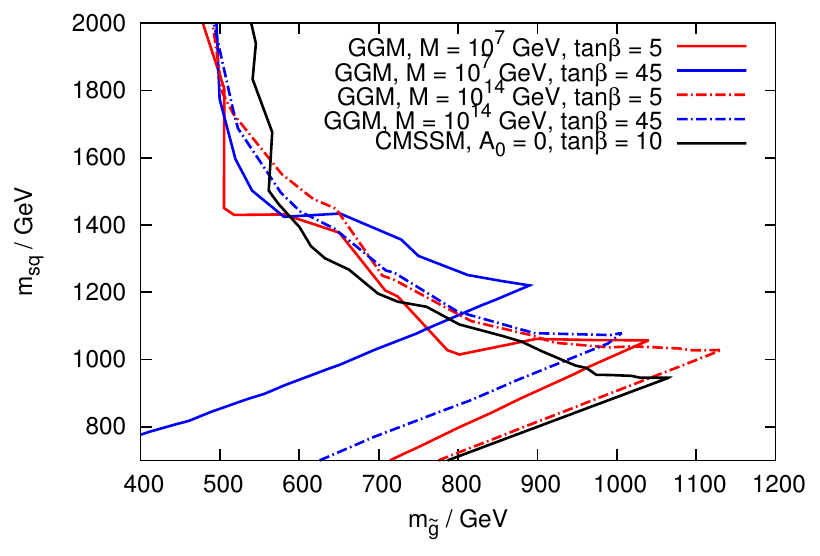}
\caption{Jets + $\slashed{E}_T$  exclusion for the CMSSM, and for GGM for different values of $\tan (\beta)$ and the messenger scale $M$.
The diagonals delimiting the excluded areas in each case arise from one of the following two effects.
Either the NLSP (in GMSB models) changes from a neutralino to a stau, for which jets$+\slashed{E}_T$ searches have no sensitivity (in the assumed context of a collider-stable NLSP), or we simply reach a region in the gluino-squark mass plane which is theoretically inaccessible.
}
\label{fig:CMSSM-GGM-comb}
\end{center}
\end{figure}

In addition to the squark and gluino masses, which are dominantly set by the mass parameters $m_0,m_{1/2}$ in the CMSSM and their counterparts $\Lambda_S,\Lambda_G$ in these GMSB models, there are two further free parameters in each case: $\tb,A_0$ in the CMSSM and $\tb,M_{\rm mess}$ for GMSB.
In the CMSSM case these extra parameters were shown in~\cite{Akula:2011zq} to have essentially no effect on the jets$+\slashed{E}_T$ constraints.
As can be seen in Fig.~\ref{fig:CMSSM-GGM-comb}, the parameters $\tb,M_{\rm mess}$ do not affect the jets$+\slashed{E}_T$ constraints for GMSB\footnote{
Note that for constant $\Lambda_S,\Lambda_G$, changing the messenger scale changes the low-scale mass of the squarks (though not the gluino), however as we have seen in Section~\ref{RoleMessengerScale} this effect can be entirely absorbed into a messenger-scale dependent $\Lambda_S$.
More precisely then, I mean that for constant $m_{\tilde{q}}$ with $\Lambda_S$ varying appropriately, varying the messenger scale has no effect on the constraints.
}; furthermore we see that for given $m_{\tilde{q}}$ and $m_{\tilde{g}}$, the constraints are virtually the same for GMSB as for the CMSSM.
All of these observations can be understood as follows.
The jets$+\slashed{E}_T$ signal in these models predominantly comes from the decays $\tilde{q}\rightarrow q+\chi_1^0$ and \mbox{$\tilde{g}\rightarrow (\bar{q}\tilde{q}^{(*)}\text{ or } q\bar{\tilde{q}}^{(*)})\rightarrow q\bar{q}\chi_1^0$}, which, together, depend on $m_{\tilde{q}}$, $m_{\tilde{g}}$ and $m_{\chi_1^0}$.
For the CMSSM and GMSB with a single $\Lambda_G$, gaugino masses unify at the GUT scale and so the mass of the gluino fixes the mass of the bino gauge eigenstate; assuming a bino-like $\chi_1^0$ this also fixes\footnote{
We have $m_{\tilde{B}} = (\alpha_1/\alpha_3)\,m_{\tilde{g}} \approx m_{\tilde{g}}/6$; if $\chi_1^0$ is instead Higgsino-like, this must be because $|\mu|<m_{\tilde{B}}$ and so the assumption that $m_{\chi_1^0} \approx m_{\tilde{B}}$ is an overestimate.
However, if $m_{\tilde{q}}$ is comparable to or greater than $m_{\tilde{g}}$, a change of $m_{\chi_1^0}$ from $m_{\tilde{g}}/6$ to something smaller is a negligible change in the kinematics of the decay of either $\tilde{q}$ or $\tilde{g}$.
If the squarks are somewhat lighter than the gluino, then $\chi_1^0$ actually being lighter than the estimate $m_{\tilde{g}}/6$ (due to a small $|\mu|$ making $\chi_1^0$ Higgsino-like) could potentially affect the squark decay -- increasing visible and missing energy, and thus $\sigma_{\text{signal}}$.
Note that with gaugino mass unification the gluino mass also fixes the wino mass, which controls the extent to which left-handed squarks decay via a wino-like chargino or neutralino instead of directly to the LSP. 
} $m_{\chi_1^0}$.
Therefore for given $m_{\tilde{q}}$ and $m_{\tilde{g}}$, the signal (and thus the constraints) looks the same regardless of other parameters or even whether we are in the CMSSM or GMSB.

Having said that $\tb$ does not affect the jets$+\slashed{E}_T$ constraints, it does however control the boundary between the qualitatively different regions where the neutralino is the NLSP and where a charged slepton(s) is the NLSP.
This is because large $\tan\beta$ enhances the tau Yukawa and the negative contribution it gives to running stau mass.
This particular search does not have sensitivity to a quasi-stable charged slepton NLSP and does not constrain the corresponding parameter space, hence the $\tb$-dependence of the straight diagonal edge to the GMSB jets$+\slashed{E}_T$ exclusion contours.

The aforementioned `extra' parameters -- $\tb,A_0$ in the CMSSM and $\tb,M_{\rm mess}$ for GMSB -- do however affect the Higgs constraints.
\begin{itemize}
 \item Larger $M_{\rm mess}$ is helpful if $m_{\tilde{g}}$ is also large, as discussed.
 \item Large $\tb$ increases $m_h$ at tree-level.
 \item $A_0$ is driven negative by the gluino during running, and so given the benefit of sizable $|A_t|/\MS$ at low scale to boost $m_h$, the CMSSM cases $A_0 = -m_0$, $A_0 = 0$ and $A_0 = 2m_0$ have increasing difficulty meeting the LEP bound.
\end{itemize}
Each of these effects can be seen in Fig.~\ref{1fbMainResults}.

\begin{figure}
\begin{center}
  \includegraphics[width=0.5\linewidth]{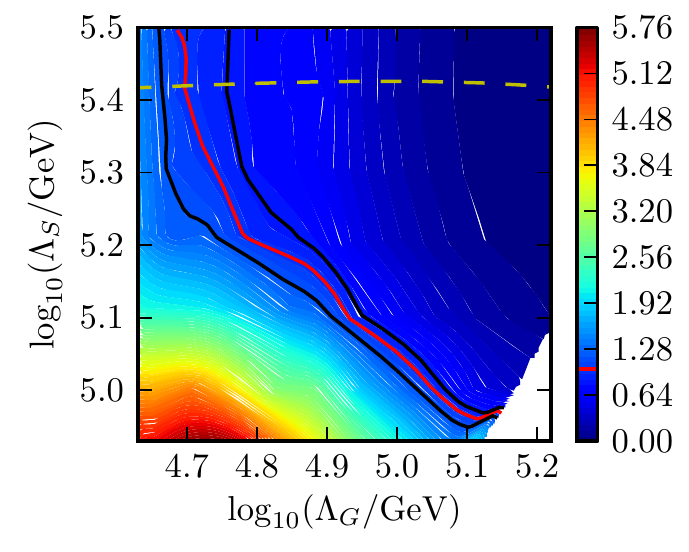}
\hspace*{-0.02\linewidth}
  \includegraphics[width=0.5\linewidth]{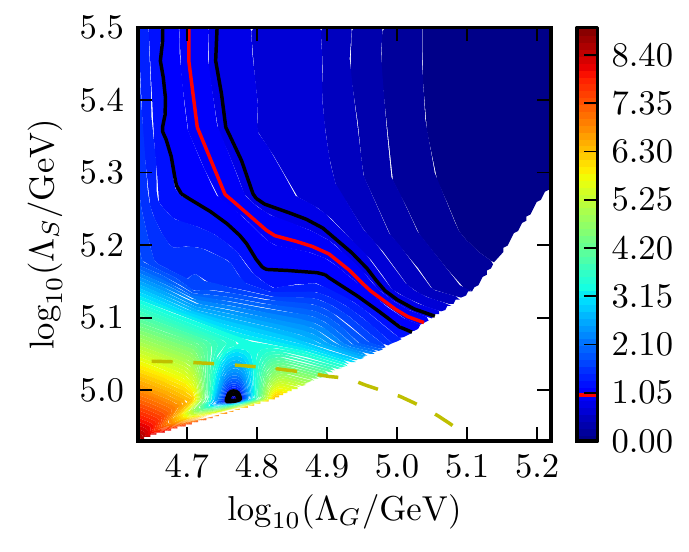}
  \includegraphics[width=0.5\linewidth]{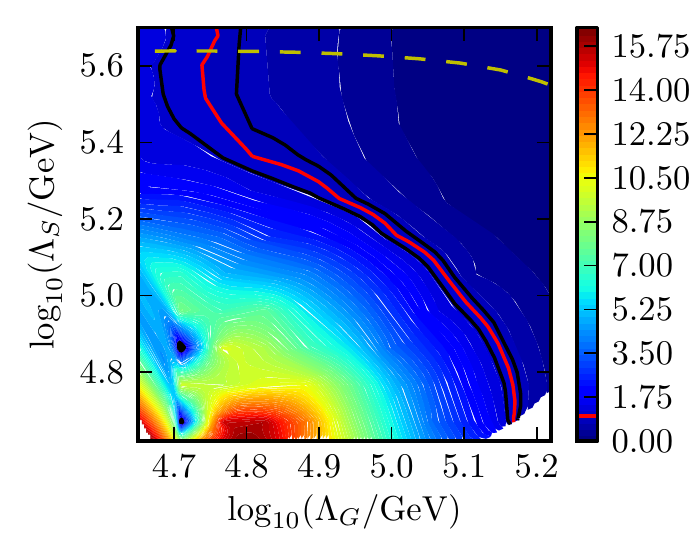}
\hspace*{-0.02\linewidth}
  \includegraphics[width=0.5\linewidth]{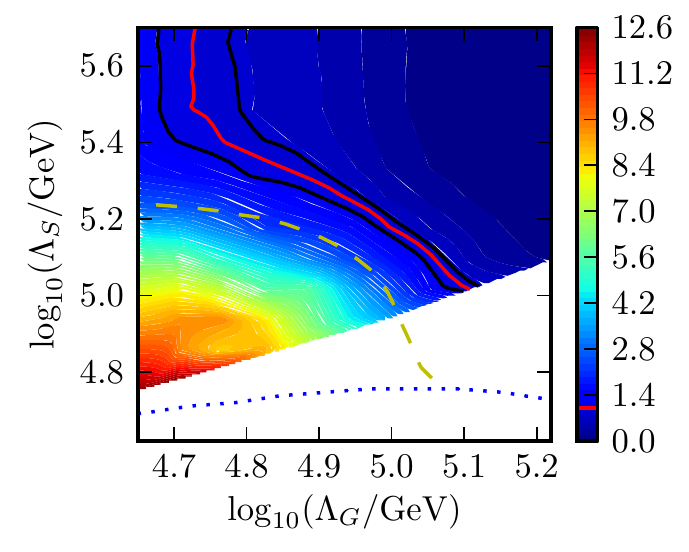}
\end{center}
\begin{center}
\caption{Exclusion contours for models of GMSB with a messenger scale $M=10^7$~GeV (top panels) or $M=10^{14}$~GeV (bottoms panels), and  $\tan\beta = 5$ (left panels) or $\tan\beta = 45$ (right panels).
The red solid line shows jets$+\slashed{E}_T$ constraints; the black solid lines show the result of varying the renormalisation/factorisation scale by factors of $2^{\pm1}$.
The yellow dashed line shows the LEP Higgs limit.
White regions have stau NLSP.
The blue dotted line in the bottom right panel delineates a tachyonic stau region.
The colour `axis' is $\sigma_{\rm signal} / \sigma_{\mathrm{limit}}$ for the most constraining search channel.}
\label{fig:GGM-LS-LG-7}
\end{center}
\end{figure}

Fig.~\ref{fig:GGM-LS-LG-7} shows the GMSB exclusion contours in terms of the original $\Lambda_G$ and $\Lambda_S$ model parameters.
The projection of these to the squark-gluino mass plane is Fig.~\ref{fig:GGM-sq-gl-plane}.

Though I have considered GMSB models with a single $\Lambda_S$ common to all three gauge groups, and likewise for $\Lambda_G$, the resulting exclusion contours have some broader applicability to six-dimensional GGM parameter space.
$m_h$ is mostly controlled by the mass of the squarks (truly the stop, but the squark masses are flavour-blind) and $A_t$.
In turn, the squark mass is controlled more by $\Lambda_{S,3}$ (at tree-level) and $\Lambda_{G,3}$ (at one-loop level) than by $\Lambda_{S;1,2}$ or $\Lambda_{G;1,2}$, and similarly $A_t$ more by $\Lambda_{G,3}$ than by $\Lambda_{G;1,2}$, in both cases because of the dominance of $\alpha_3$ over $\alpha_{1,2}$.
Therefore splittings of $\Lambda_{G;1,2}$ from $\Lambda_{G,3}$ and of $\Lambda_{S;1,2}$ from $\Lambda_{S,3}$ are subdominant parameters for setting the Higgs mass compared to $\Lambda_G$ and $\Lambda_S$, and for small splittings the same contours will hold.

Similarly the jets$+\slashed{E}_T$ signal depends most strongly on the squark and gluino masses, and thus on $\Lambda_G$ and $\Lambda_S$ (rather than splittings between the contributions for different gauge groups).
However (a) an independent $\Lambda_{G,1}$ may be tuned to give a bino arbitrarily close in mass to the squarks or gluino, {\it compressing} the spectrum and removing the visible energy; (b) with six fully independent $\Lambda$ parameters the NLSP has even more possible identities, many of which correspond to signals that are most strongly constrained by different searches, with the jets$+\slashed{E}_T$ search losing sensitivity.
The qualitative changes contained within the possibility (b) represent a strong dependence of the squark-gluino mass-plane contours set here on the decision to consider only a subset of the full GGM parameter space.

\newpage
\section{Making The Most Of MET: Mass Reconstruction From Collimated Decays} \label{MakingMostMET}
{\it This section is based on my work~\cite{PhysRevD.87.074004} done in collaboration with Michael Spannowsky; excepting the extended Introduction the text here follows it closely.
In this section neutralinos are denoted by $\tilde{N}_i$, with $\chi$ indicating a generic invisible particle.
}\\

\subsection{Introduction} \label{setup}
Missing transverse energy -- {\it MET} -- is of great importance at hadron colliders: it is our only way of inferring the presence of neutral (collider-)stable particles $\chi$, be they neutrinos or BSM particles.
However whenever {\it two} such particles are produced (which will always be the case if their stability is due to a $\mathbb{Z}_2$ symmetry, for example) our observation only of the vectorial sum of their transverse momenta $\slashed{\mathbf{p}}_T = \mathbf{p}_{a,T} + \mathbf{p}_{b,T}$ thwarts reconstruction of masses in the decay cascade\footnote{
$2\chi$ could also be directly produced, giving a final state with, at leading order, no large transverse energy (visible or invisible).
The universal possibility of hard initial state radiation allows essentially model-independent limits to be set on the direct production of new $\chi$ particles from monojet and monophoton searches.
Here I will focus only on production of $2\chi$ via a decay cascade.
}
ending with $2\chi$.
Popular methods for searching for heavy particles with partially invisible decays are transverse mass observables~\cite{Barger:1987re}, $M_{T2}$~\cite{Lester:1999tx}, razor analyses~\cite{Rogan:2010kb} and kinematic edges~\cite{Allanach:2000kt}.
I will introduce the first two of these.

\begin{figure}
\begin{center}
\subfigure[]{\includegraphics[width=0.4\linewidth]{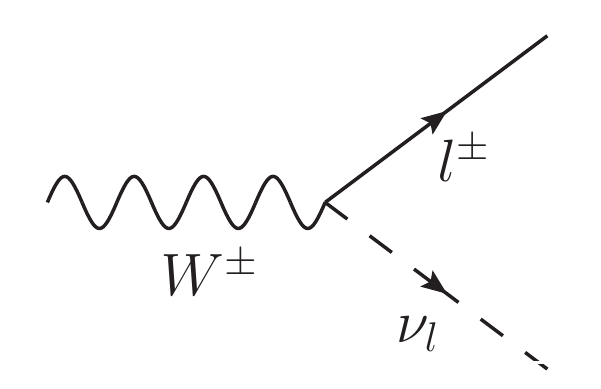}\label{Wdecay}} 
\subfigure[]{\includegraphics[width=0.4\linewidth]{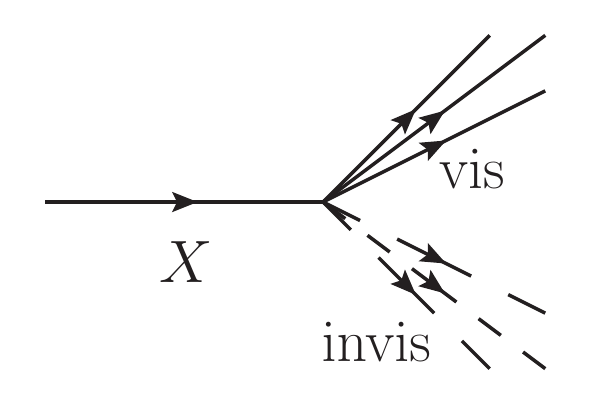} \label{GenVisInvisDecay}}
\end{center}
\vspace*{-6mm}
\caption{Cartoons of leptonic $W$ decay (left panel) and the decay of one particle $X$ to multiple visible and multiple invisible particles (right panel).}
\label{InvisDecays}
\end{figure}

Consider events with a single leptonically decaying $W$, Fig.~\ref{Wdecay}, together with any number of jets and photons (but not leptons: the sole lepton and $\slashed{\mathbf{p}}_T$ can then unambiguously identified with the lepton $l$ and neutrino $\nu$ from the $W$).
We have
\begin{equation}
\begin{split}\label{mW}
m_W^2 =&\; (E_{l} + E_{\nu})^2 - (\mathbf{p}_{l} + \mathbf{p}_{\nu})^2 \\
=&\;m_l^2+m_{\nu}^2+2(E_{T,l}E_{T,\nu}\cosh(\Delta y_{l\nu}) - \mathbf{p}_{T,l}\cdot\mathbf{p}_{T,\nu}),
\end{split}
\end{equation}
where $y$ is rapidity and $E_T$ is transverse energy -- $\surd(m^2 + p_T^2)$.
At hadron colliders we do not know the neutrino's momentum in the $z$ direction and hence $m_W$ is not calculable this way.
A quantity we can calculate is the {\it transverse mass} $m_T$:
\begin{equation}
\begin{split} \label{mT}
m_T^2\equiv & \; (E_{T,l} + E_{T,\nu})^2 - (\mathbf{p}_{T,l} + \mathbf{p}_{T,\nu})^2 \\
= & \;m_l^2+m_{\nu}^2+2(E_{T,l}E_{T,\nu}\, - \mathbf{p}_{T,l}\cdot\mathbf{p}_{T,\nu})
\end{split}
\end{equation}
As we have just one invisible particle, $\mathbf{p}_{T,\nu}=\slashed{\mathbf{p}}_T$; and since that it is massless, $E_{T,\nu}=|\slashed{\mathbf{p}}_T|$; hence the calculability of Eq.~\eqref{mT}.
Comparing it to Eq.~\eqref{mW} we see that $m_T \leq m_W$; an inequality which is violated by the finite width effects of the $W$ and by detector resolution effects.
If the inequality were typically far from being saturated it would not be helpful for determining $m_W$.
However the cross-section is enhanced at $m_T \approx m_W$~\cite{Han:2005mu}:
\begin{equation}
 \frac{d\sigma}{dm_{e\nu}^2 m_{e\nu,T}^2} \propto \frac{\Gamma_W m_W}{(m_{e\nu}^2-m_W^2)^2 + \Gamma_W^2 m_W^2} \frac{1}{m_{e\nu}\sqrt{m_{e\nu}^2-m_{e\nu,T}^2}},
\end{equation}
leading to what is called a {\it Jacobian peak} in the distribution right before the end point at $m_W$.

Now consider a single particle $X$ decaying into multiple visible and multiple invisible particles, Fig.~\ref{GenVisInvisDecay}.
By replacing the four-momenta of the lepton and neutrino in Eq.s~\eqref{mW} and~\eqref{mT} by the four-momenta summed over the visible and invisible particles respectively, the same arguments give a transverse mass with a Jacobian peak and end point at $m_X$.
There is one further subtlety here however, in that while $\slashed{\mathbf{p}}_T$ can still be identified with the summed $p_T$ of the invisible particles, the transverse energy of the invisible system is {\it not} given by $|\slashed{\mathbf{p}}_T|$ since the invariant mass of invisible system is in general not zero.
The latter vanishes only if all the invisible particles are massless and travelling in exactly the same direction.
In the absence of further information this must be assumed to be true; neglecting the invisible mass leads to an underestimate of the invisible transverse energy, thus smearing the Jacobian peak to values further below the end point at $m_X$.

\begin{figure}
\begin{center}
\includegraphics[width=0.4\linewidth]{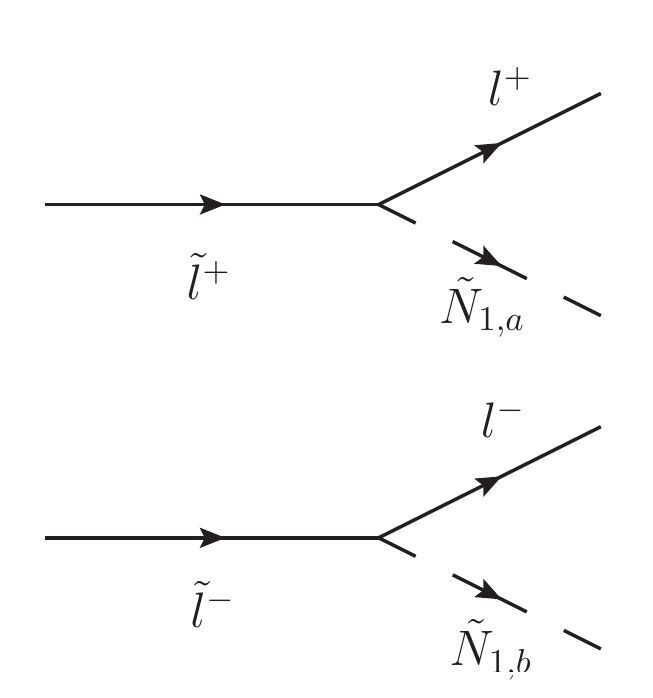} 
\end{center}
\caption{The decays of two sleptons to leptons and neutralinos -- an example signal for the $M_{T2}$ variable.}
\label{MT2eg}
\end{figure}

For two partially-invisibly decaying particles, with the example of sleptons $\tilde{l}$ decaying to leptons and neutralinos $\tilde{N}_1$ shown in Fig.~\ref{MT2eg}, the transverse mass as defined in the previous paragraph can be calculated in the same way.
However the result is now constrained to be less than or equal to the mass of the two-slepton system: if the sleptons are moving relative to each other, their combined mass is not an interesting quantity.
What we would like is a transverse mass associated with each separate slepton -- by definition these quantities would have end points at the slepton mass (for simplicity this example has a common mass for both decaying particles).
Their calculation would require knowledge of the decomposition of $\slashed{\mathbf{p}}_T$ into the two contributing components:
\begin{equation}
\slashed{\mathbf{p}}_T = \mathbf{p}_{T,\tilde{N}_{1,a}} + \mathbf{p}_{T,\tilde{N}_{1,a}},
\end{equation}
then we would have
\begin{equation} \label{mT2prelim}
\rm{max}\{ m_T^2(\mathbf{p}_{T,l^+},\mathbf{p}_{T,\tilde{N}_{1,a}}),m_T^2(\mathbf{p}_{T,l^-},\mathbf{p}_{T,\tilde{N}_{1,b}}) \} \leq m_{\tilde{l}}^2,
\end{equation}

Since we don't know the decomposition of $\slashed{\mathbf{p}}_T$, we cannot calculate the two transverse masses in the correct way.
If we were to use an incorrect decomposition and calculate the LHS of Eq.~\eqref{mT2prelim}, it would no longer be guaranteed to be smaller than $m_{\tilde{l}}^2$.
If we were to try every possible decomposition, trivially that would include the correct decomposition; the smallest value of the LHS of Eq.~\eqref{mT2prelim} obtained this way would therefore be equal to or smaller than the correct decomposition, which is equal to or smaller than $m_{\tilde{l}}^2$.
Hence
\begin{equation}
M_{T2}^2 \equiv \underset{\slashed{\mathbf{p}}_1 + \slashed{\mathbf{p}}_2 = \slashed{\mathbf{p}}_{T} }{\rm min}\left( \rm{max}\{ \vphantom{\frac{1}{1}} m_T^2(\mathbf{p}_{T,l^+},\slashed{\mathbf{p}}_1),
 m_T^2(\mathbf{p}_{T,l^-},\slashed{\mathbf{p}}_2) \} \right) \leq m_{\tilde{l}}^2
\end{equation}
In many examples $M_{T2}^2$ has been shown to nearly saturate the inequality, thus giving a visible end point which is useful for measuring a mass.

Here I wish to consider the circumstances under which we can do better than transverse masses and end points, by deducing the four-momentum associated with each of the two invisible $\chi$ particles separately and thence constructing mass peaks.
Clearly some feature of the rest of the event must suggest the correct decomposition of $\slashed{\mathbf{p}}_T$ into $\mathbf{p}_{a,T} + \mathbf{p}_{b,T}$.
If there are two well-localised visible objects that we expect, from some prior prejudice about the kinematics, to be parallel or antiparallel to the two unseen $\chi$ particles, then we have two directions in the transverse plane to give us $\mathbf{p}_{a,T}$ and $\mathbf{p}_{b,T}$.
Furthermore we can add longitudinal components to each of these two transverse vectors to make them \mbox{(anti)parallel} to their corresponding visible object in three dimensions, giving approximations for $\mathbf{p}_{\chi_{a,b}}$.
If $\chi$ is much lighter than the particle produced in the hard scattering, i.e. at the start of the decay cascade, we can promote $\mathbf{p}_{\chi_{a,b}}$ to massless four-vectors; I will show that combined with the four-vectors for the visible decay products, a strong mass peak for the initial particles can be reconstructed.

\subsection{Motivation}

Parallel or antiparallel visible and missing energy is not worth considering only for its ease: it can arise in many circumstances.
Spin correlations may make $\chi$ particles approximately \mbox{(anti)parallel} to other particles.
Two-body decays of particles $P$ nearly stationary in the lab frame are back-to-back: therefore in $2P\rightarrow2\chi+2\text{vis}$, each $\chi$ is nearly antiparallel to one of the `$\text{vis}$'.
However {\it antiparallelness}, unlike {\it parallelness}, is not preserved under $z$-boosts of the mother particle, and the $z$-boost is unknown.

A promising scenario is when each $\chi$ is produced together with visible energy from the decay of a {\it boosted} particle.
This will arise whenever (a) directly pair-produced particles are appreciably heavier than whatever they decay into in the first step of the cascade, and (b) $\chi$ are created following two or more steps.
Together these points imply that each of the two {\it sides} of the event (separated according to the mother particle) contains an intermediate particle which is boosted: the visible object(s) and $\chi$ it ultimately decays to will be collimated.

For some examples, consider the quintessential Susy decay of a pair-produced squark to a hard jet and {\it light} neutralino : $\tilde{q}\rightarrow q+\tilde{N}_1$.
There are many reasons why we might expect $\tilde{N}_1$ to be unstable, decaying to visible energy and a lighter, neutral, collider-stable particle -- the latter could be:
\begin{itemize}
\item a gravitino $\tilde{G}$, if Susy breaking is mediated at a low scale, i.e. some form of gauge mediation.
A low mediation scale is motivated by electroweak naturalness and an automatic solution of the Susy flavour problem.
See~\cite{Kats:2011qh} for a comprehensive list of possible collider signatures.
\item a pseudo-Goldstino $\tilde{G'}$, if more than one hidden sector independently breaks Susy and mediates it to the visible sector, as may occur in string theory or quiver gauge theories
\cite{Cheung:2010mc,Argurio:2011hs}.
See~\cite{Argurio:2011gu} for the collider phenomenology.
\item a singlino\footnote{
In this case the decay is not really $\tilde{q}\rightarrow q+\tilde{N}_1\rightarrow\ldots$ but $\tilde{q}\rightarrow q+\tilde{N}_2\rightarrow\tilde{N}_1+\ldots$, since new photinos/singlinos actually mix with the MSSM neutralinos. If $\tilde{N}_2$ is mostly `MSSM-like' (any mixture of Higgsino, wino and bino), and $\tilde{N}_1$ is mostly singlino or a new photino, then direct decay of $\tilde{q}$ to $\tilde{N}_1$ is suppressed relative to the two-step decay.
 } $\tilde{S}$ in the NMSSM.
See~\cite{Das:2012rr} for the modified collider signals.
\item a new photino\footnotemark[\value{footnote}] $\tilde{\gamma}'$, if the MSSM is extended with one or more extra $U(1)$ gauge symmetries, as is commonly expected to arise from the compactification of extra dimensions.
See~\cite{Baryakhtar:2012rz} for a discussion of collider prospects.
\end{itemize}
In the nomenclature of~\cite{Baryakhtar:2012rz}, $\tilde{N}_1$ here is the lightest {\it ordinary} supersymmetric particle (LOSP).
All of these examples have some other particle as the true LSP, and so cosmology allows a charged or coloured Susy particle to be lighter than $\tilde{N_1}$ and take its place as the LOSP in the cascade $\tilde{q}\rightarrow \text{vis}_1+(\text{LOSP})\rightarrow\text{vis}_1+(\text{vis}_2+\text{LSP})$, giving different visible energy.

\subsection{The Analysis} \label{sec:Analysis}
I will elaborate on the strategy outlined in Section~\ref{setup} in terms of a concrete example to allow clearer references to the particles involved in the signal: I consider the classic gauge-mediation decay\footnote{
A similar final state may arise from Universal Extra Dimensions~\cite{Macesanu:2002db}, though semi-invisibly decaying Kaluza-Klein photons from KK quark/gluon decays are not generally expected to be boosted; this will be important for my analysis.}
$2\tilde{q}\rightarrow 2q+2(\tilde{N}_1)\rightarrow 2q+2(\tilde{G}+\gamma)$.
The lightest neutralino is typically expected to be considerably lighter than the squarks in this scenario, as renormalisation-group evolution tends to drive squark masses up and the bino mass down, and the phenomenon of gaugino screening in the simplest models makes the gauginos much lighter than the scalars (see e.g.~\cite{Cohen:2011aa}).
This simple observation gives a powerful handle on the signal, as yet unexploited: the gravitinos and photons are normally collimated.
It is exploited as follows.
\begin{enumerate} \itemsep0pt \parskip0pt \parsep0pt
\vspace*{-1mm}
{\setlength\itemindent{-2pt}  \item Uniquely decompose $\slashed{\mathbf{p}}_T$ into $\mathbf{p}_{a,T} + \mathbf{p}_{b,T}$ which are defined to be parallel, in the transverse plane, to the two hardest isolated photons.
}
{\setlength\itemindent{-2pt}  \item Promote $\mathbf{p}_{a,b;T}$ to three-vectors $\mathbf{p}_{a,b}$ by adding the longitudinal components required to make them parallel to each of the photons in three dimensions.
}
{\setlength\itemindent{-2pt}  \item Promote $\mathbf{p}_{a,b}$ to massless four-vectors $p_{a,b}^\mu = (|\mathbf{p}_{a,b}|, \mathbf{p}_{a,b})$, giving approximations for the two gravitino four-vectors.
Adding each of these to the four-vector of the collinear photon gives massless approximations for the two neutralino four-vectors, $p_{\tilde{N}_{1;a,b}}^\mu$.
}
{\setlength\itemindent{-2pt}  \item If each neutralino $\tilde{N}_{1;a,b}$ can be paired with the `correct' jet in the event $j_{a,b}$, then taking the invariant mass of each pair reconstructs the mass of the initial squarks: $M_{\text{rec};a,b}^2 =  (p_{\tilde{N}_{1;a,b}}^\mu + p_{j_{a,b}}^\mu)^2$
}
\end{enumerate}

Steps 1-2 above reconstruct the three-momenta of the two neutralinos in the same way as is done for the two $\tau$ in $H\rightarrow2\tau\rightarrow e^{\pm}\mu^{\mp}\slashed{E}_T$ with the collinear approximation of~\cite{Plehn:1999xi}.
There, the two $\tau$ four-momenta are added together to get the mass of the single mother particle; here the four-momenta of the two neutralinos are separately added to those of other visible particles in the event to get the masses of two mother particles -- step 4.

Step 4 needs a criterion for the correct way to pair each reconstructed neutralino with one of the jets in the event.
The correct jet is considered to be the one most closely resembling the quark produced in the same $\tilde{q}\rightarrow \tilde{N}_1+q$ decay.
Keeping only the two hardest jets, there are two arrangements -- two ways of pairing each neutralino with a different jet.
More generally one can consider the $N$ hardest jets in the event, giving $N(N-1)$ arrangements to choose from.
Each squark is generally produced nearly at rest, therefore the neutralino and jet into which it decays are likely to be back-to-back; the jet is also expected to be hard, with an energy of roughly half the squark's mass.
Therefore one criterion is to pair the two neutralinos $\tilde{N}_{1;a,b}$ with jets $j_a$ and $j_b$ so as to make maximally negative the sum of dot products between the three-momenta of each neutralino and its jet:
\begin{equation*}
 \text{criterion } \alpha\!:\; -\!\left( \mathbf{p}_{\tilde{N}_{1,a}}.\mathbf{p}_{j_a} + \mathbf{p}_{\tilde{N}_{1,b}}.\mathbf{p}_{j_b} \right) \; \text{maximal}
\end{equation*}
If the pair-produced squarks are mass degenerate, this can also be exploited: the two reconstructed masses should coincide.
This gives the second possibility for finding the right jets:
\begin{equation*}
 \text{criterion } \beta\!:\; \left| (p_{\tilde{N}_{1,a}}^\mu + p_{j_a}^\mu)^2 - (p_{\tilde{N}_{1,b}}^\mu + p_{j_b}^\mu)^2 \right| \; \text{minimal}
\end{equation*}
Each criterion suggests the correct jets, defining two reconstructed masses $M_{\text{rec};a,b}^2 =  (p_{\tilde{N}_{1;a,b}}^\mu + p_{j_{a,b}}^\mu)^2$.
The maximisation/minimisation above is not differential but discrete -- the quantity is calculated once for each of the $N(N-1)$ arrangements of jets with neutralinos and only the largest/smallest is kept.
It thus takes negligible computational time (indeed $N=2$ is optimal in the example considered) and could potentially be incorporated into a search at the trigger level.
These two criteria are not specific to neutralinos and jets: they are relevant for final states where two objects need to be paired correctly with two other objects, both being the decay products of pair-produced particles (the second criterion also requires mass degeneracy of the two mother particles).
The solution chosen for this same problem in~\cite{Goncalves-Netto:2013nla}, for mass reconstruction of leptogluon pairs from $l_8 \bar{l}_8 \rightarrow llgg$, was to assign a hemisphere to each of the two hardest leptons and then pair with each lepton the hardest jet in the same hemisphere.

I considered a simplified model with squarks of the first two generations, a bino-like neutralino and a gravitino with masses $m_{\tilde{q}}= 1.2~\text{TeV}$, \mbox{$m_{\tilde{N}_1}=100~\text{GeV}$} and $m_{\tilde{G}}=1~\text{eV}$ respectively; this squark mass is at the edge of the strongest current constraints~\cite{CMSsearch}.
I calculated a full spectrum for this simplified model (all other superpartner masses are set $2$~TeV) with {\tt SOFTSUSY~3.3.4} and decay widths with
{\tt Herwig++~2.6.1}.
I then followed two routes to get to observable distributions.
In the first, {\tt MadGraph~5~1.5.5} \cite{Alwall:2011uj} supplied the matrix elements for disquark production; the subsequent decays, extra radiation, showering and hadronisation were performed by {\tt PYTHIA~6} \cite{Sjostrand:2006za}; and detector response was simulated with {\tt PGS~4}~\cite{PGS}.
In the second, {\tt Herwig++} was used to generate the complete event; jets were defined with {\tt FastJet~3.0.3}, and the final state objects analysed in the {\tt RIVET~1.8.1} framework.
My kinematical analysis -- steps 1-4 with criteria $\alpha$ and $\beta$ above -- was then applied.
Code for doing this, easily generalisable to other final states, can be found at~\cite{Me}.

We chose basic cuts for the analysis as follows:
\begin{itemize} \itemsep0pt \parskip0pt \parsep0pt
\vspace*{-1.5mm}
{\setlength\itemindent{-2pt}  \item At least two jets, clustered using the anti-kt algorithm~\cite{Cacciari:2008gp} with size parameter 0.4.
Jet candidates are required to have $p_T>30$~GeV and $|\eta|<4.5$.
}
{\setlength\itemindent{-2pt}  \item At least two isolated photons with $p_T>10$~GeV.
On the {\tt MadGraph}-{\tt PYTHIA} route, {\tt PGS} handles isolation.
On the {\tt Herwig} route, I considered a photon isolated based on total transverse energy deposited inside a surrounding cone (as in the relevant searches~\cite{CMSsearch,Aad:2012zza}), specifically $5$~GeV in a cone $\Delta R < 0.4$.
}
{\setlength\itemindent{-2pt}  \item A minimum and maximum azimuthal angular separation between the two hardest isolated photons \mbox{$\epsilon < \Delta \phi_{\gamma_1 \gamma_2} < \pi - \epsilon$} with $\epsilon = 0.01$, since photons which are exactly \mbox{(anti)}parallel in the transverse plane do not allow $\slashed{\mathbf{p}}_T$ decomposition.
}
{\setlength\itemindent{-2pt}  \item The missing energy vector $\slashed{\mathbf{p}}_T$ should lie in between the two photons in the transverse plane (i.e. inside the smaller of the two sectors delimited by the two photon directions).
This ensures that the event has $\slashed{\mathbf{p}}_T$ corresponding to the ansatz of both gravitinos being parallel (and not antiparallel) to their photons.
With this cut the kinematics are always in the `trivial zero' of the $M_{T2}$ observable (see~\cite{Lester:2011nj}).
}
\end{itemize}
Decomposition of $\slashed{\mathbf{p}}_T$ of course requires $\slashed{E}_T\neq0$; in practice this is always satisfied.
I did not cut on $\slashed{E}_T$ -- I analysed this particular signal not to optimise the associated cuts but simply as a demonstration of the mass reconstruction technique.
In the present example more than $90\%$ of events have $\slashed{E}_T > 100$~GeV and so a large requirement could be placed as in existing searches (likewise for the leading jet and photon which are typically hard in the signal).
Note that with a requirement for hard photons and $\slashed{E}_T$ there is typically very small background for new physics~\cite{Kribs:2009yh} and the priority is an observable that increases the visibility of the signal alone, ideally through a resonance.

\begin{center}
\begin{figure}[!ht] 
\centering
\includegraphics[width=0.35\linewidth]{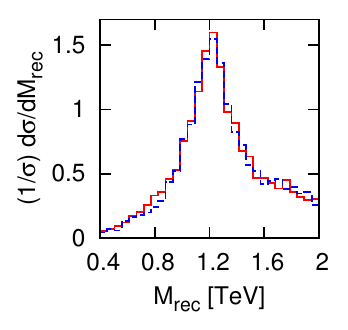}
\includegraphics[width=0.35\linewidth]{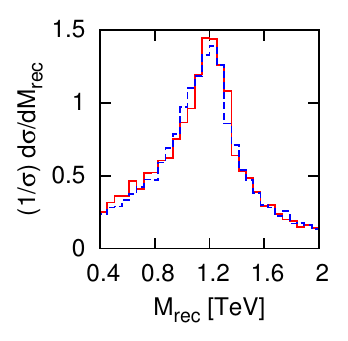}
\includegraphics[width=0.35\linewidth]{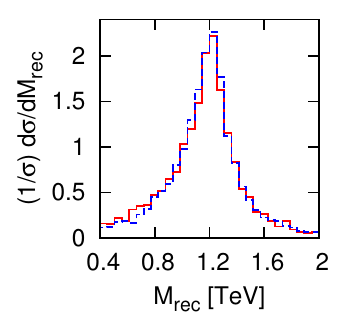}
\includegraphics[width=0.35\linewidth]{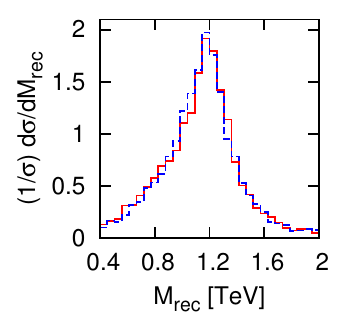}
\caption{
The squark mass ($m_{\tilde{q}}= 1.2~\text{TeV}$) in the process $pp\rightarrow2\tilde{q}\rightarrow 2q+2(\tilde{N}_1)\rightarrow 2q+2(\tilde{G}+\gamma)$ reconstructed with $\slashed{\mathbf{p}}_T$ decomposition and neutralino-jet pairing as described in the text.
The masses of the lightest neutralino and gravitino are $m_{\tilde{N}_1}=100~\text{GeV},\:m_{\tilde{G}}=1~\text{eV}$; the centre of mass energy is $8$~TeV.
Panels on the left (right) show the mass of the squark calculated from the leading (sub-leading) photon in each event.
Upper (lower) panels pair jets with reconstructed neutralinos using criterion $\alpha$ ($\beta$).
The blue dashed line shows events generated by {\tt MadGraph} and {\tt PYTHIA}, with fast detector simulation performed by {\tt PGS}; the red solid line shows events generated by {\tt Herwig++}.
}
\label{fig:1TeV}
\end{figure}
\end{center}

Fig.~\ref{fig:1TeV} shows the results of the analysis.
The initial mass is reconstructed to $10\%$ accuracy for roughly $\tfrac{1}{3}$ of events passing the basic cuts.
Multiplying by the {\tt Prospino~2} production cross-section and the acceptance -- $20\,{\rm fb}\times0.5$ -- one would expect $\mathcal{O}(100)$ events inside this peak in $30\,{\rm fb}^{-1}$ at $8$~TeV.

I obtained an estimate of the expected accuracy of the mass reconstruction in the case of 100 signal events in the following way.
Roughly 5000 signal events were split into samples each containing 100 events.
For each sample the 100 values of the calculated mass were binned, and the mid-point of the modal bin taken to define the position of the peak and thus the reconstructed mass for that sample.
The values of the mass determined from each of the $\sim\!50$ samples then define a probability distribution quantifying how well the mass can be reconstructed from 100 events.
If the events within each sample were binned into {\it too few} bins, this distribution is precise but inaccurate: all samples will agree which is the modal bin, but since it is wide its centre may be far from the true mass.
If the events within each sample were binned into {\it too many} bins, this distribution is accurate but imprecise: the modal bin is narrow, but a low count-per-bin exacerbates statistical fluctuations and may randomly shift the modal bin to one not containing the true mass.
For this example with a true mass of $1.2~\text{TeV}$ I found that, when binning over the range $[0.4,2]~\text{TeV}$, about $50$ bins is appropriate for 100 events\footnote{
Note that 50 bins for 100 events naively suggests 2 events per bin and thus huge relative fluctuations, however the calculated masses are not flatly distributed -- they cluster around a sharp peak by construction.
}. 
With each sample binned thusly, the distribution of the values calculated from all the samples has a root-mean-square deviation from the true mass of roughly $60~\text{GeV}$: determination at the $5\%$ level.
To put this number (very) roughly in context, the ATLAS Collaboration's Technical Design Report~\cite{ATLAS:1999vwa} considered the signal $2\times(\tilde{q}\rightarrow \chi_2^0q\rightarrow \tilde{l}lq \rightarrow \chi_1^0 llq)$ arising as part of a CMSSM (not simplified) model.
Using kinematic edges,the squark mass could be determined to within $3\%$ with $\mathcal{O}(10^6)$ events before cuts; c.f. $5\%$ with only $200$ events before cuts (100 events after cuts) for the signal and method considered here.

As my analysis makes use of hard jets arising from the decay of signal particles, it could in principle be affected by the (higher order) production of additional jets in the hard scattering.
To investigate this I simulated $2\tilde{q}$ and $2\tilde{q}+1{\rm jet}$ production and combined these consistently into a single sample using the MLM matching procedure~\cite{Alwall:2007fs}.
The reconstructed mass distributions are essentially identical to those of simple $2\tilde{q}$ production shown in Fig.~\ref{fig:1TeV}, which follows from the fact that my method is designed to find the two jets that look most like they have been produced by the decay of the squarks, and discard other jets.

\begin{center}
\begin{figure}[!ht]
\centering
\includegraphics[width=0.5\linewidth]{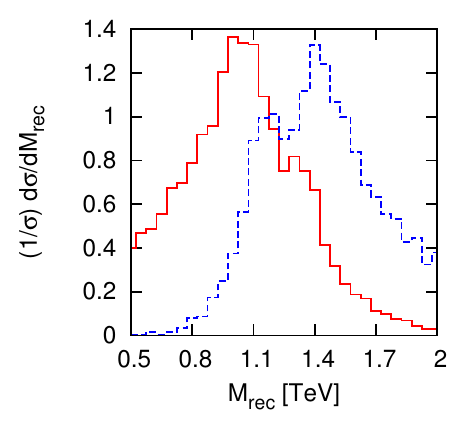}
\caption{
As Fig.~\ref{fig:1TeV} but with one squark from the first two generations having mass $1.1$~TeV and the other seven having mass $1.4$~TeV.
The solid red (dashed blue) line shows the smaller (larger) of the two masses reconstructed in each event.
Only criterion $\alpha$ for jet-neutralino pairing is used.
Events are generated with {\tt Herwig++}.
}
\label{fig:Horns}
\end{figure}
\end{center}

Criterion $\alpha$ can also reconstruct the masses of pair-produced {\it non-degenerate} particles.
In Fig.~\ref{fig:Horns} it is used to analyse the same signal as previously but now with one squark from the first two generations having mass $1.1$~TeV and the other seven having mass $1.4$~TeV.
This unequal splitting is chosen to have large cross-sections for the production of two squarks of {\it different} mass (four lighter squarks and four heavier would merely result in a dominant production of two of the lighter four); nevertheless production of two squarks of the {\it same} mass still has non-zero cross-section.
Thus the distribution of the larger (smaller) of the two masses calculated for each event peaks strongly at $1.4$~TeV ($1.1$~TeV) and weakly at $1.1$~TeV ($1.4$~TeV), with the weak peak resulting from pair-produced degenerate squarks.

\subsection{Discussion}

\begin{center}
\begin{figure}[!ht]
\centering
\includegraphics[width=0.6\linewidth]{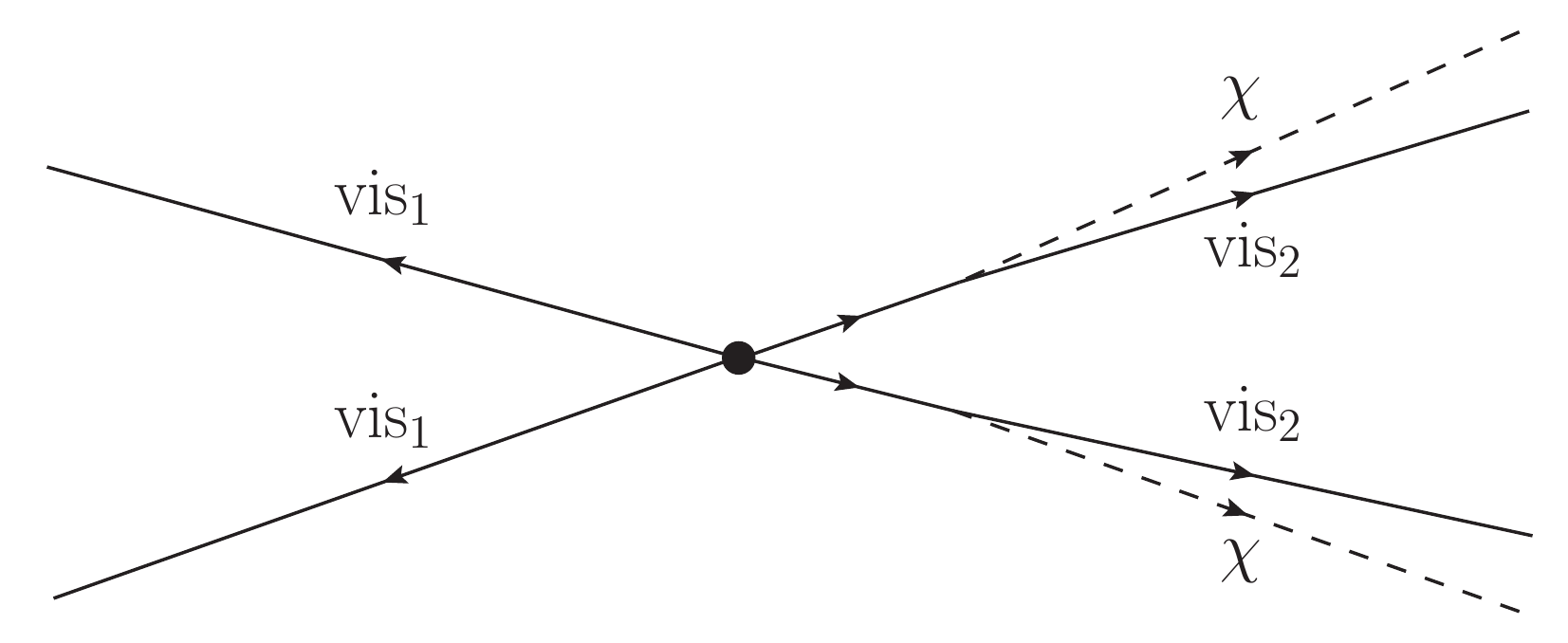}
\caption{
The topology I consider: pair-produced particles each decay into a visible Standard Model particle ${\rm vis}_1$ and a much lighter particle, which is thus boosted; this decays semi-invisibly into ${\rm vis}_2$ and $\chi$.
}
\label{fig:Diagram}
\end{figure}
\end{center}

The final state of the example considered has two jets and two pairs of roughly collinear photons and gravitinos.
The jet could be replaced by any other visible particle -- `vis$_1$' -- the photon too -- `vis$_2$' -- and the gravitino by anything invisible, $\chi$: I show this general topology in Fig.~\ref{fig:Diagram}.  
Provided there are two semi-invisible decays which are boosted (or forced into \mbox{(anti)parallel} behaviour by spin correlations) the same analysis presented here should in theory have some potential for mass reconstruction.
Of course if vis$_1$ and vis$_2$ are objects less clean experimentally than light-flavour jets and photons, such as $b$ quarks or even combinations of particles, the procedure will be more difficult in practice.
Searches for mass peaks in the manner presented, considering various different particle types for vis$_{1,2}$, could discover expected or unexpected resonances.
Below, I outline how the method might be adapted as the topology is distorted and generalised further.

\begin{center}
\begin{figure}[!ht] 
\centering
\includegraphics[width=0.42\linewidth]{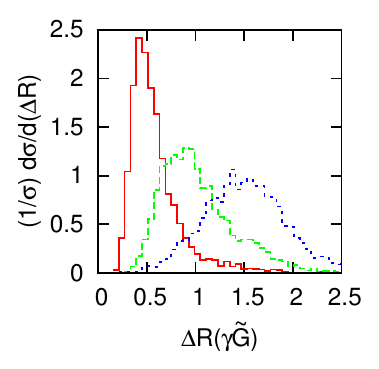}
\includegraphics[width=0.42\linewidth]{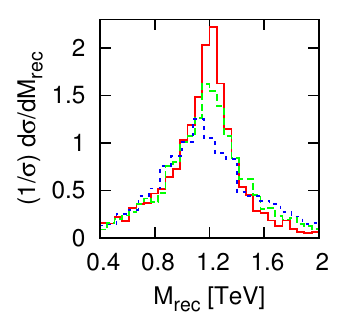}
\caption{
As Fig.~\ref{fig:1TeV}, holding the squark mass at $1.2$~TeV while varying the neutralino mass: $m_{\tilde{N}_1} = 100$,$\,200$,$\,400$~GeV are shown with red solid, green dashed, and blue dotted lines respectively.
The greater $m_{\tilde{N}_1}$, the less collinear its photon and gravitino daughters become, as shown by $\Delta R(\gamma \tilde{G})$ (averaged between the two $\gamma \tilde{G}$ pairs) in the left panel.
This worsens the mass reconstruction: the right panel shows one of the two masses found using one of the two jet-neutralino pairing criteria (all four quantities behave similarly -- see  Fig.~\ref{fig:1TeV}).
Events are generated with {\tt Herwig++}.
}
\label{fig:BoostedOrNot}
\end{figure}
\end{center}

{\it A Less Boosted Intermediate.}
Collinearity of $\chi$ and ${\rm vis}_2$ relies on their common mother particle being boosted; as it becomes less boosted they become less collinear.
I show this effect, and the decreasing sharpness of the mass reconstruction that results, in Fig.~\ref{fig:BoostedOrNot} for my previous gauge-mediation example.
$m_{\tilde{N}_1}$ is increased from $100$ to $400$~GeV for constant $m_{\tilde{q}} = 1.2$~TeV.
If $\tilde{N}_1$ is made heavier still, e.g. $m_{\tilde{N}_1}/m_{\tilde{q}}\rightarrow1$, the increasingly lethargic neutralino gives a less collimated photon-gravitino pair; indeed the two are increasingly back-to-back, and most events fail to meet the requirement that $\slashed{\mathbf{p}}_T$ be in between the two photons.

{\it More Decays Of The Intermediate.}
If ${\rm vis}_2$ is several particles instead of the single photon $\gamma$ I considered, e.g. a lepton pair from a boosted $\tilde{N}_i\rightarrow l^{\pm}l^{\mp}\tilde{N}_1$ decay, by construction they will be collimated and the sum of their four-momenta can be used in place of $p_{\gamma}^{\mu}$ in the analysis.

\begin{center}
\begin{figure}[!ht] 
\centering
\includegraphics[width=0.42\linewidth]{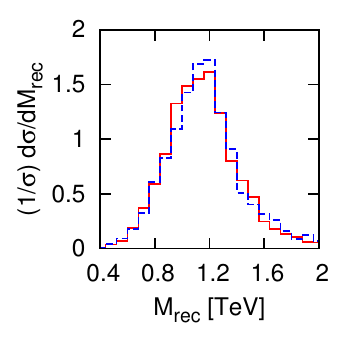}
\includegraphics[width=0.42\linewidth]{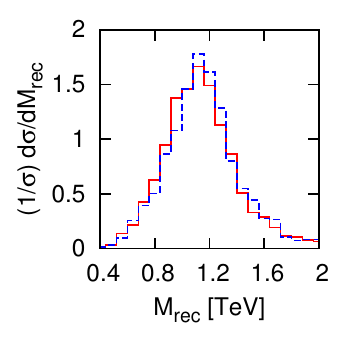}
\caption{
The gluino mass ($m_{\tilde{g}}= 1.2~\text{TeV}$) in the process $pp\rightarrow2\tilde{g}\rightarrow 4q+2(\tilde{N}_1)\rightarrow 4q+2(\tilde{G}+\gamma)$ reconstructed with $\slashed{\mathbf{p}}_T$ decomposition and neutralino-jet pairing criterion $\beta$ as described in the text.
The masses of the lightest neutralino and gravitino are $m_{\tilde{N}_1}=100~\text{GeV},\:m_{\tilde{G}}=1~\text{eV}$ (the masses of other particles are set at $2$~TeV); the centre of mass energy is $8$~TeV.
Panels on the left (right) show the mass of the gluino calculated from the leading (sub-leading) photon in each event.
The blue dashed line shows events generated by {\tt MadGraph} and {\tt PYTHIA}, with fast detector simulation performed by {\tt PGS}; the red solid line shows events generated by {\tt Herwig++}.
}
\label{fig:gluino}
\end{figure}
\end{center}

{\it More Decays Before The Intermediate.}
If the directly pair-produced particles decay to a boosted intermediate and two visible particles rather than one -- via two on-shell steps or a three-body decay -- then each ${\rm vis}_1$ in Fig.~\ref{fig:Diagram} is replaced by two particles which are not collinear.
Criterion $\alpha$ is then not applicable but criterion $\beta$ is, albeit with greater combinatorial ambiguity from the need to pair each reconstructed neutralino with two other visible objects.
In this scenario the boosted intermediate is also less boosted from sharing its energy with more particles, making its semi-invisible decay less collimated.
Despite these difficulties the method is reasonably successful: for Fig.~\ref{fig:gluino} I have generated events for a simplified model with pair-produced gluinos of mass $1.2$~TeV decaying to $q\bar{q}\tilde{N}_1$ ($q$ now denoting a quark of any of the three generations) with the $100$~GeV neutralino decaying to $\gamma \tilde{G}$.
Neutralino-jet pairing is performed with criterion $\beta$ generalised in the obvious way to include four jets rather than two.

{\it Other Combinatoric Complications.}
If ${\rm vis}_1 = {\rm vis}_2$, e.g. if in my former example photons were replaced by jets or jets by photons (but not both of these at once), then there would be a combinatoric ambiguity not just in pairing the reconstructed boosted intermediate with the correct ${\rm vis}_1$ but also in which two particles define the initial $\slashed{\mathbf{p}}_T$ decomposition directions.
The requirement that $\slashed{\mathbf{p}}_T$ be in between the two visible particles onto which it is decomposed eliminates some of the possible decomposition configurations; for the rest, criterion $\beta$ can be generalised to be an optimisation over decomposition configurations as well as pairing possibilities.

{\it More Than Two Invisible Particles.}
With a third $\chi$ in the final state which is expected to be \mbox{(anti)}parallel to one of the first two, our ansatz for the topology still contains only two invisible directions and we can uniquely decompose the observed $\slashed{\mathbf{p}}_T$.
If the two invisible particles that are \mbox{(anti)}parallel have come from the decay of the same particle, we only need to know the sum of their momenta and so we can reconstruct the mass as before.
However if they have come from the decay of two different particles, then we need their individual momenta for mass reconstruction; knowing only their sum, the masses we wish to calculate are under-constrained by one parameter.
Another possibility is $3\chi$ in the signal final state with three different expected directions: there are then three vectors in the transverse plane into which $\slashed{\mathbf{p}}_T$ can be decomposed, with any two of the three giving a unique decomposition.
There are three ways to choose two vectors from the three.
We may have the $\slashed{\mathbf{p}}_T$ in between the two vectors in $0$, $1$ or $2$ of the three ways (neglecting the possibility of exact collinearity between $\slashed{\mathbf{p}}_T$ and one of the vectors).
If $0$, we veto.
If $1$, there is a unique decomposition.
If $2$, $\slashed{\mathbf{p}}_T$ can be expressed as some amount of one of the decompositions plus some amount of the other, with the two coefficients constrained to sum to unity: the masses we wish to calculate are under-constrained by one parameter.
One response, not physically motivated, would be to veto.
Another possibility could be to set the two coefficients based on some other prejudice about the kinematics, such as making an intermediate particle of known mass maximally on-shell.
Which of these three cases ($0$, $1$ or $2$ of the possible decompositions being acceptable) we have will vary on an event by event basis.


\newpage
\begin{appendices}

\section{The Higgs As A Pseudo-Nambu-Goldstone Boson (`Little Higgs')} \label{littleHiggs}

See~\cite{Schmaltz:2005ky} for a review and original references, also~\cite{Contino:2010rs,Cheng:2007bu}.
When a continuous global symmetry $\mathcal{G}$ is spontaneously broken to a subgroup $\mathcal{H}$, Goldstone's Theorem tells us that ${\rm dim}(\mathcal{G}) - {\rm dim}(\mathcal{H})$ Nambu-Goldstone bosons (NGBs) result.
(Mild {\it explicit} breaking, i.e. from small terms in the Lagrangian, results instead in {\it pseudo}-NGBs (pNGBs), which are light but not massless.)
If $\mathcal{G}$ contained a subgroup $\mathcal{F}$ which was gauged, and which gets broken down to $\mathcal{I}\equiv\mathcal{F}\cap \mathcal{H}$, then  ${\rm dim}(\mathcal{F}) - {\rm dim}(\mathcal{I})$ of the NGBs are eaten to give masses to all of the bosons associated with the generators of the (entirely broken) $\mathcal{F}/\mathcal{I}$ coset.
This leaves $({\rm dim}(\mathcal{G}) - {\rm dim}(\mathcal{H})) - ({\rm dim}(\mathcal{F}) - {\rm dim}(\mathcal{I}))$ NGBs.
$\mathcal{I}$ should be (or contain) the Standard Model $SU(2)_L\times U(1)_Y$; and amongst the NGBs we want there to be (a) three degrees of freedom that will ultimately be eaten by the $W$ and $Z$ bosons, as in the Standard Model, but also (b) a degree of freedom corresponding to the Higgs.
In this way we have made the Higgs light; unfortunately we have made it massless, however one thing at a time.

Now let us add a term to the Lagrangian which involves the Higgs, respects $\mathcal{H}$, but breaks $\mathcal{G}$.
The Higgs is now a pNGB rather than a NGB.
However we run straight back into the hierarchy problem: through such a term we have a one-loop quadratic divergence, previously forced to be zero by virtue of the symmetry only being broken at the level of the vacuum and not the Lagrangian.
So instead of explicit breaking with one term, we can use two: we add to the Lagrangian $\lambda_1\mathcal{L}_1 + \lambda_2\mathcal{L}_2$ which only break $\mathcal{G}$ to $\mathcal{H}$ when they are present together, a phenomenon we call {\it collective symmetry breaking}.
With only one of the terms, there is a sufficient amount of symmetry broken only spontaneously and not explicitly that the Higgs is still a NGB.
With both terms it is a pNGB, and we do have quadratic divergences as we must with explicit symmetry breaking, but only at two loops: $\delta m_H^2 \sim (\lambda_1/16\pi^2)(\lambda_2/16\pi^2)\Lambda^2$, since by construction all corrections identically vanish when one coupling vanishes even if the other does not.
The one-loop divergence due to the top loop is cancelled by a similar diagram containing in the loop a fermionic partner of the top -- a partner which is a necessary part of such a setup.

The expected cutoff for naturalness is then postponed to $10$~TeV rather than $1$~TeV.
We must then ask what comes next.
Perhaps nothing more until the Planck scale, if our global symmetry $\mathcal{G}$ was broken spontaneously by a weakly coupled scalar whose vacuum expectation value $v_{\rm \, little\:Higgs}$ is, while larger than $v_{SM} = 174$~GeV, still mysteriously smaller than $M_P$.
There could be a tower of stacked little Higgs models -- each symmetry-breaking scalar being the pNGB of a different symmetry at higher scales -- as discussed in~\cite{Batra:2004ah}.
In this case, while the scalar at the bottom of the stack (which plays the role of the Standard Model Higgs) encounters the quadratically divergent correction at progressively higher orders, increasing numerical factors multiplying the term do not allow escape from the Hierarchy problem.
Alternatively one of the three traditional natural theories may come into play above this cutoff -- extra dimensions, compositeness, or Susy.

\newpage
\section{Optimal Naturalness Beyond Leading Log $\delta m_{H_u}^2$} \label{OptimalNaturalnessBeyondLeadingLogdeltamHu2}
The leading log expression for $\delta m_{H_u}^2$ is obtained by ignoring the scale dependence of $A_t$ and $\MS$; to do better we can integrate $A_t$ and $\MS$ over their varying higher-scale values.
First consider the running of $A_t$ and $\MS$ with arbitrary self and mutual couplings, as well couplings to other particles:
\begin{align} 
\frac{d}{dt}
\begin{pmatrix}
\MS^2(t) 
\\ A^2_t(t)
\end{pmatrix} = &
\begin{pmatrix}
a(t) & b(t) \\
0 & c(t)
\end{pmatrix}
\begin{pmatrix}
\MS^2(t)
\\ A_t^2(t)
\end{pmatrix}
+ \mbox{{\it other} running soft-mass terms} \label{RunningMixingStopsGen_Beta} \\
\therefore \quad 
\begin{pmatrix}
\MS^2(t) 
\\ A^2_t(t)
\end{pmatrix} = &
\begin{pmatrix}
d(t) & e(t) \\
0 & f(t)
\end{pmatrix}
\begin{pmatrix}
\MS^2(0)
\\ A_t^2(0)
\end{pmatrix}
+ \mbox{{\it other} high-scale soft-mass terms,} \label{RunningMixingStopsGen_Integrated}
\end{align}
where $a,b,c$ are running couplings and $d,e,f$ are related to the former by integration, and the lower-left entry of the matrix must vanish since $A_t$ appears in the Lagrangian, not $A_t^2$.
Note that if the {\it other} soft-mass parameters themselves run due to $A_t$ and $\MS$, this feeds back into Eq.~\eqref{RunningMixingStopsGen_Integrated} as corrections to the coefficients $d,e,f$ suppressed by an extra loop factor, which could have an impact but I will neglect this for simplicity.
Integrating the $m_{H_u}^2$ beta function~\eqref{mHuBetaFunction}, keeping just the stop-sector terms as before but now including their scale dependence (and that of the top Yukawa) as in~\eqref{RunningMixingStopsGen_Integrated}, we have
\begin{multline} \label{Integrated_mHu}
\delta m_{H_u}^2(t) = \: \tfrac{3}{8\pi^2}\, \left( 2\MS^2(0) \int_{t'=0}^{t'=t} y_t^2(t') d(t')\:dt' + \right. \\ \left. A_t^2(0) \int_{t'=0}^{t'=t} y_t^2(t') (2e(t')+f(t'))\,dt' \right) \\ + \mbox{{\it other} high-scale soft-mass terms}
\end{multline}
For Lagrange constrained optimisation, Eq.~\eqref{LagrangeMin}, we must differentiate $\delta m_{H_u}^2$ and $\delta m_{h}^2$ with respect to $A_t^2$ and $\MS^2$.
One can either invert Eq.~\eqref{RunningMixingStopsGen_Integrated} and substitute into Eq.~\eqref{Integrated_mHu} to obtain $\delta m_{H_u}^2$ instead as a function of the {\it low}-scale stop parameters, or all derivatives can be taken with respect to the {\it high}-scale stop parameters (using the chain rule for $\delta m_{h}^2$, whose arguments should be evaluated at the low scale).
Both approaches give the same result, as they must:
\begin{multline}
f(t) \left( 2\int_{t'=0}^{t'=t}y_t^2(t') \, d(t') \:dt'\;  + d(t)y_t^2(t)\left( 1+\frac{A_t^2(t)}{2\MS^2(t)} \right) \right) \frac{\partial (\delta m_{h}^2)}{\partial A_t^2} = \\ \left( d(t)\int_{t'=0}^{t'=t}y_t^2(t') (2e(t')+f(t'))\,dt' 
 - 2e(t)\int_{t'=0}^{t'=t}y_t^2(t') \, d(t') \:dt' \right) \frac{\partial (\delta m_{h}^2)}{\partial \MS^2} \label{RGimprovedLagrangeMax_general}
\end{multline}
The leading log relation is recovered for $(d(t),e(t),f(t))=(1,0,1),\: y_t(t)=y_t$.
So what are these functions $d(t),e(t),f(t)$ in the MSSM?
Expressions for the one-loop running parameters can be written down when all Yukawa couplings except that of the top are set to zero~\cite{Ibanez1984511,Essig:2007kh,Carena:1996km}.
$y_t(t)$ and $A_t(t)$ do not require numerical integration if one also sets the $U(1)$ and $SU(2)$ gauge couplings to zero: one finds
\begin{gather}
y_t^2(t) = y_t^2(0) \,\xi^{-16/9}(t)\,G^{-1}(t;\tfrac{-16}{9}) \\
A_t(t)=G^{-1}(t;\tfrac{-16}{9})\left[A_t(0)+\frac{16}{9}M_3(0)\Big(G(t;\tfrac{-16}{9})\xi^{-1}(t)-\,G(t;\tfrac{-25}{9})\Big)\right] \label{runningA}  \\
\mbox{where}\quad\xi(t) = 1+\frac{3}{2\pi}\,\alpha_3(0)t \notag \\
G(t;n) = 1-\frac{3}{4\pi^2}\,y_t^2(0)\,\int_0^t\,dt'\,\xi^{n}(t') \notag 
\end{gather}
From~\eqref{runningA} we can read off that $f(t)=G^{-1}(t;\tfrac{-16}{9})$.
In this same scheme for extracting running parameters, the stop mass necessarily involves numerical integration.
However RG-induced splitting of the stop from the lighter-generation up-type quarks is typically small (and if not the model is necessarily unnatural, as mentioned in Section~\ref{NaturalnessUnderPressure}), so that the running stop is well approximated by its high-scale value plus the gluino-induced term, the latter easily obtained by integrating the one-loop running gluino mass:
\begin{equation}
\MS^2(t) = \MS^2(0)+\frac{8}{9}M_3^2(0)\left( \frac{\alpha^2_3(t)}{\alpha^2_3(0)} -1 \right)
\end{equation}
This gives $d(t)=1,\:e(t)=0$.
Eq.~\eqref{RGimprovedLagrangeMax_general} is then
\begin{multline}
2G^{-1}(t;\tfrac{-16}{9}) \left( \int_{t'=0}^{t'=t}\bigg( \xi^{-16/9}(t')\,G^{-1}(t';\tfrac{-16}{9}) \bigg) \,dt' \right. \\
\left. + \xi^{-16/9}(t)\,G^{-1}(t;\tfrac{-16}{9})\left( 1+\frac{A_t^2(t)}{2\MS^2(t)} \right) \vphantom{\int_{t'=0}^{t'=t}\bigg( \bigg)} \right) \frac{\partial (\delta m_{h}^2)}{\partial A_t^2} = \\
\int_{t'=0}^{t'=t}\bigg( \xi^{-16/9}(t')\,G^{-2}(t';\tfrac{-16}{9})\bigg) \,dt' \; \frac{\partial (\delta m_{h}^2)}{\partial \MS^2}
\label{RGimprovedLagrangeMax_MSSM}
\end{multline}
The integrals can be done analytically and the resulting root, $x_{\text{natural}}$, found; I do not plot it as it is essentially indistinguishable from the one shown in Fig.~\ref{fig:HiggsMassesNLO}, even for very high mediation scales $\Lambda \sim 10^{16}$~GeV.
Thus my attempt at an approximate RG improvement of $\delta m_{H_u}^2$ (resumming all the logs that come with appreciable coupling constant factors) makes no difference to the result obtained from the leading log expression.

An alternative approach to this approximate RG improvement would be to work consistently at Next-to-Leading-Log NLL order for $\delta m_{H_u}^2$.
Barbieri-Giudice fine-tuning measures are given at NLL in~\cite{CahillRowley:2012rv}, from which one can extract the dependence of $\delta m_{H_u}^2$ on any trilinear mixing term or sfermion mass-squared via
\begin{align}
 \delta m_{H_u}^2(A_i) = & \; \int \left( \frac{M_Z^2}{2A_i}  \left.Z_{A_i}\right|_{\tan\beta \rightarrow \infty} \right) \:d A_i \\
\delta m_{H_u}^2(m_{\tilde{f}}^2) = & \; \frac{M_Z^2}{4}  \left.Z_{m_{\tilde{f}}}\right|_{\tan\beta \rightarrow \infty} \label{eq:sfermionMassFT} \\
\mbox{where} \quad Z_{p_i} \equiv & \; \frac{\partial (\log M_Z^2)}{\partial (\log p_i)} \notag
\end{align}
Note that $\delta m_{H_u}^2$ and $m_{\tilde{f}}^2$ both having mass dimension $2$ results in $Z_{m_{\tilde{f}}} \propto m_{\tilde{f}}$, giving the simpler expression~\eqref{eq:sfermionMassFT}. 
$Z_{A_i}$ however can contain further mass-scales beyond $A_i$; indeed for the stop, $Z_{A_t}$ contains terms with $M_1$, $M_3$ and $A_b$. In other words, at NLL $\delta m_{H_u}^2$ depends on $A_t$ not only through an $A_t^2$ term but also through terms $A_t M_1$, $A_t M_3$ and $A_t A_b$. In the spirit of connecting $\delta m_{H_u}^2$ and the Higgs mass to the stop sector in isolation, I will not explore this effect here.

\newpage
\section{Brazil-Band Plots For Dummies} \label{BrazilBandPlots}
Brazil-band plots are used to present exclusion limits for the existence of new particles whose decays are uniquely determined as a function of their mass\footnote{
If $n$ extra parameters were needed to fully specify the decays, $n$ extra dimensions would be needed to plot the exclusion results.
The Higgs boson of the Standard Model and the $W',Z'$ of the Sequential Standard Model, for example, all handily have $n=0$.}.
Before any data is collected, one can calculate the {\it expected} $\sigma_{95\%\,CL} \, / \sigma_{\text{signal}} $ for any given mass $m$ of the new particle.
$\sigma_{\text{signal}}$ is defined as in Section~\ref{ColliderIntroHowToNotSeeSusy} ($\sigma$ unfortunately also denotes uncertainty/error in predictions and measurements).
I will refer interchangeably to numbers of events $N$ and cross-sections $\sigma$ as they differ only by a factor of the integrated luminosity $\mathcal{L}$, through $N=\sigma\mathcal{L}$.
$\sigma_{x\%\,CL}$ is the cross-section excluded at $x\%$ confidence level: roughly, the signal cross-section for which the probability of signal plus background fluctuating as low as the observed value is less than $(100-x)\%$ (a statement to be clarified shortly).
$x=95$ is common in particle physics.
The {\it expected} $\sigma_{95\%\,CL}$ refers to the limit that would be set if the number of events observed coincided with the number of events expected from the background alone.

Let's develop a feel for how the expected $\sigma_{95\%\,CL} \, / \sigma_{\text{signal}} $ will behave.
We predict a certain number of events from the background $N_b$ and from the signal $N_s$, and there is uncertainty in both predictions.
Statistical error, because both follow from probability distributions.
Systematic error, because the experiment may consistently mismeasure something, and the theoretical estimate is made to a finite order in perturbation theory and so is consistently a little off.
Now, if the predicted number of signal events is roughly less than the combined uncertainty in the background and in the signal, then observing only the expected number of background events $N_b$ and no more still wouldn't allow you to sensibly exclude the presence of signal -- it's sufficiently small as to be compatible with $N_b$ events, within the errors.
In this case we have $\sigma_{95\%\,CL} \, / \sigma_{\text{signal}} > 1$.
On the other hand if the predicted number of signal events is roughly greater than the combined uncertainty, then the presence of the signal ought to be visible above the background; in this case observing only $N_b$ events and no more would be compelling evidence for no signal, giving an exclusion: $\sigma_{95\%\,CL} \, / \sigma_{\text{signal}} < 1$.

Also shown in Brazil-band plots are the $\pm\sigma, \pm2\sigma$ uncertainties in the expected exclusion -- the edges of the green and yellow regions respectively.
Where do these come from?
We have a probability distribution for the number of background events (with an expected value $N_b$).
For each possible number of observed events, we would set a different limit, because more (less) events makes the signal look more (less) likely to exist.
Calculating the resulting limit as a function of events then lets us transform the probability distribution for number of background events into a probability distribution for the resulting limit.
I am now in a position to clarify a previous statement -- that the expected limit the limit that would result from observing $N_b$ events -- more specifically it is the median of the distribution I have just described.
The $\pm\sigma$ uncertainties in this expected limit correspond to the $\pm\sigma$ points in this distribution (which doesn't have to be Gaussian -- we just mean the points with the same cumulative probability as the the $\pm\sigma$ points in the standard Gaussian).
In a nut shell: the uncertainty in the expected limit is tied to the uncertainty in the expected number of background events, and observing more (less) events means a weaker (stronger) limit.

Once the experiment has actually been done, one can set {\it observed} $\sigma_{95\%\,CL} \, / \sigma_{\text{signal}} $ limits by seeing how much the signal plus background prediction deviates, relative to the uncertainties, from the observed number of events $N_o$ (i.e. $N_o$ now plays the role that $N_b$ did for expected limits).
Limits that are weaker (stronger) than expected will be set where more (less) events are observed than expected from the background alone.

\begin{figure}[!ht]
\centering
\includegraphics[width=0.9\linewidth]{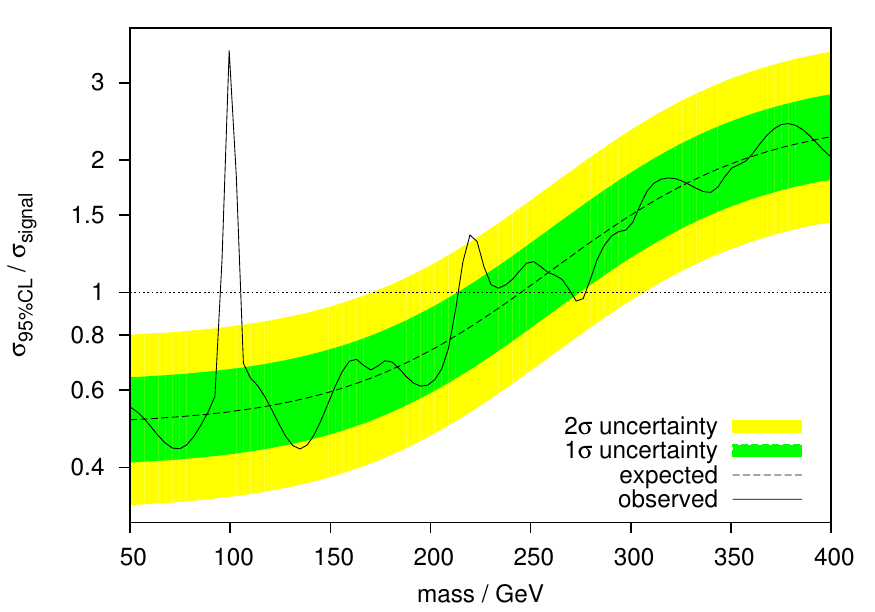}
\caption{A (purely fictional!) Brazil-band plot for a search for a new particle that actually exists, with mass $m=100$~GeV.
A discussion of the relevant features is in the text.}
\label{fig}
\end{figure}

In Fig.~\ref{fig} I show a fictional Brazil-band plot for illustrative purposes.
What information is shown?
We can see that the exclusion expected was the mass range $m<240$~GeV (the `expected' curve is below $1$ here).
The observed exclusion is something like $m<95$~GeV, $105~\rm{GeV}<m<215~\rm{GeV}$, $272~\rm{GeV}<m<278~\rm{GeV}$.
The small region of exclusion around $275$~GeV wasn't expected -- we call this lucky exclusion -- it has arisen due to a negative fluctuation of the background.
The region $215~\rm{GeV}<m<240$ was expected to be excluded, but wasn't: the background has fluctuated upwards, but not dramatically: at most a $2\sigma$ effect.
Close to $100$~GeV however, there is a narrow region where the exclusion is very much worse (i.e. the number of events observed is much higher) than expected.
This is strong evidence for a new particle of roughly that mass!
The width of the excess gets contributions from the mass resolution and bin width of the relevant search channel, and the particle's own fundamental width; being dominated by whichever is largest.
All three of these factors will smear out an otherwise delta-function peak in the invariant mass distribution of the particle's decay products.
Note that all the way along the plot the exclusion (and by implication the background) fluctuates above and below what was expected, but it's only in places where the expected value of $\sigma_{95\%\,CL} \, / \sigma_{\text{signal}} $ is one side of $1$ and the observed value is on the other that the result of exclusion or non-exclusion deviates from expectation.

For completeness I'll mention `blue-band' or signal strength plots, which become useful when we have not just exclusion but also a (tentative) discovery.
The signal strength, often denoted $\mu$, is the factor by which the signal cross-section should be multiplied for best agreement with the data.
There are a number of points worth noting about $\mu$.
\begin{itemize}
\item An exact expression for $\mu$, based on maximising the likelihood function for obtaining all of the observed data, would be complicated.
However for a given mass value $m$ that predicts $N_s$ signal events, with $N_b$ events expected from the background and $N_o$ events observed (with $N_s$, $N_b$ and $N_o$ referring to events in the vicinity of the signal at $m$, not elsewhere), one can think of $\mu$ as being given by $N_o \approx N_b + \mu N_s$.
\item $\mu=0 \Leftrightarrow$ perfect description of the data by the background alone.
$\mu=1 \Leftrightarrow$ perfect description of the data by the background plus signal, with no modification of the signal.
Of course for either of these two statements to be meaningful requires small error bars on $\mu$: $\sigma_\mu < 1$ or preferably $\ll 1$.
\item While physical cross-sections are positive semi-definite, $\mu$ is negative whenever less events are observed than predicted for the background alone.
\item How does $\mu$ relate to the Brazil-band plot?
The approximate relation $N_o \approx N_b + \mu N_s$ shows that $\mu$ is positive (negative) where more (less) events are observed than expected from the background alone.
Therefore
\begin{equation}
 \text{Sign}(\mu) \approx \text{Sign}\left(\frac{\sigma_{95\%\,CL,\text{observed}}}{\sigma_{\text{signal}}} - \frac{\sigma_{95\%\,CL,\text{expected}}}{\sigma_{\text{signal}}}\right),
\end{equation}
or in words, the sign of $\mu$ generally follows the sign of the fluctuation of the observed limit from what was expected.
Knowing that a given mass value {\it is not} excluded, the corresponding $\mu$ can in principle take any value.
All positive values are possible since the lack of exclusion could be due to the presence of a real signal, but with arbitrarily modified strength.
All negative values of possible since we might be dealing with a deficit of observed events, but an insignificant deficit relative to our uncertainties.
Knowing that a given mass value {\it is} excluded, the corresponding $\mu$ must be less than one, since exclusion refers specifically to the $\mu=1$ case.
Therefore we can make one inference going the other way, from blue band to Brazil band: $\mu\geq1$ points are not excluded, $\mu<1$ points may or may not be.
But it's best just to take the relevant information from each respective plot.
\end{itemize}

Earlier I said that $\sigma_{95\%\,CL}$ is the signal cross-section for which the probability of signal plus background fluctuating as low as the observed value, denoted $p_{\rm s+b}$, is less than $5\%$.
However sometimes by chance, $N_o$ will fluctuate so low that even the background-only hypothesis $H_b$ is a poor description of the data.
If this is the case, we should consider the signal plus background hypothesis $H_{s+b}$ excluded only if does a much worse job even than $H_b$ at describing the data.
If it is merely equally poor, the data isn't helping us to distinguish whether or not there is any signal.
Considering the signal to be excluded in this case based on $p_{\rm s+b} < 0.05$ would be a spurious exclusion, and we can protect ourselves against this using the CL$_s$ method~\cite{Junk:1999kv,Read:2002hq}.
$\text{CL}_s \equiv p_{s+b}/(1-p_b)$, with $p_b$ the probability of the background alone contributing at least as many events as observed; and we now require CL$_s < 0.05$ instead of $p_{\rm s+b} < 0.05$.
Then when $H_b$ and $H_{s+b}$ do comparably poor jobs at describing a deficit in events, $p_{s+b} \sim 1-p_b$ and CL$_s$ remains greater than $5\%$.
By construction, the CL$_s$ method prevents cross-section limits becoming too small, and so the $\pm \sigma$ uncertainty in the expected limit will be asymmetric -- smaller in the negative direction than in the positive direction.
This sometimes has the curious effect of counterbalancing the asymmetry induced by plotting $\sigma_{95\%\,CL} \, / \sigma_{\text{signal}} $ on a logarithmic scale!\footnote{
Thanks to the other participants of the Carg\`{e}se International School 2012 for posing this question and to Glen Cowan for answering it!
}

I mentioned that the width of the strong excess in Fig.~\ref{fig} is given by a combination of the experimental mass resolution, bin width, and the hypothesised particle's width.
Indeed the width of any excess or deficit in the plot, significant or not, should be governed by this same combination; call it $\Delta m$.
(Note that $\Delta m$ may change with $m$: the physical width can obviously change, and particularly if we combine different search channels in one plot the mass resolution and bin width can jump whenever we pass between regions where different search channels are most sensitive.)
As an example, consider a deficit of events observed in the bin containing the mass value $M$, so that we have stronger constraints on the particle near $m\sim M$, giving a dip in the Brazil-band plot at that point.
Since the mass of the decay products will be smeared out by $\Delta m$, all points in the mass range $|m-M| \lesssim \Delta m$ will feel this local strong constraint, and the dip in the plot will be $\sim\! \Delta m$ wide.
The same logic applies for a local excess giving a peak in the plot $\sim \!\Delta m$ wide.
(Note that in making Fig.~\ref{fig} I was not careful about giving the random fluctuations a reasonably consistent width.)
Fluctuations a bit broader than $\Delta m$ will occur when two (or more) bins close to each other both have deficits or both have excesses.
Thus the broader the fluctuation relative to $\Delta m$, the less likely it is that the source is a statistical fluctuation: overall we expect correlations over scales $\sim \Delta m$.
For much larger correlations / broader fluctuations, we are likely to conclude that the background was inaccurately modelled in that mass range, so that the (background only) events we observed were consistently above or below the prediction.

\clearpage
\section{Cuts For The ATLAS Collaboration's $1.04~{\rm fb}^{-1}$ Search For Susy With Jets And Missing Energy} \label{cutsAppendix}
The cuts that were chosen for this analysis~\cite{Aad:2011ib}, and which I reproduced in the {\tt RIVET} analysis \mbox{{\it ATLAS\_2011\_S9212183}} are as follows:
\begin{itemize}
 \item Jet candidates are reconstructed using the anti-kt jet clustering algorithm~\cite{Cacciari:2008gp} with a distance parameter of $0.4$.
 Candidates with \linebreak\mbox{$p_T < 20$~GeV} are discarded.
 \item Electron (muon) candidates are required to have \mbox{$p_T > 20$~GeV ($10$~GeV)} and $|\eta| <
2.47 \;(2.4)$.
 \item Jet candidates within a distance $\Delta R = 0.2$ of an electron candidate are discarded, then electron or muon candidates within a distance $\Delta R =0.4$ of surviving jet candidates are discarded.
Next, all jet candidates with $|\eta|>2.8$ are discarded. 
 \item The event is vetoed if there are any electrons or muons with $p_T > 20$~GeV.
 \item Thereafter five separate search channels are defined, each with its own requirements on hadronic activity, $m_{\rm eff}$ and $\slashed{E}_T/m_{\rm eff}$, summarised in \linebreak Table \ref{tab:signal_regions}.
 The {\it effective mass} $m_{\rm eff}$ is calculated as the sum of $\slashed{E}_T$ and the $p_T$ of the two, three or four hardest jets used to define the search channel.
In the high mass channel, all jets with $p_T>40$~GeV are used to compute the $m_{\rm eff}$ value used in the final cut.
The $\Delta\phi$ cut is only applied up to the third leading jet.
 \begin{table}[!hb]
  \begin{center}
  \begin{tabular}{| l l | c|c|c|c|}
 \hline
 & Signal Region &$\geq$ 2-jet & $\geq$ 3-jet & $\geq$ 4-jet & High mass \\
\hline
\hline
& $\slashed{E}_T$ & $>130$ &  $>130$ &  $>130$  & $>130$ \\ 
& Leading jet $p_T$       & $>130$ &  $>130$ &  $>130$  & $>130$ \\ 
\hline
& Second jet      $p_T$   & $>40$ &  $>40$ &  $>40$ & $>80$\\ 
& Third jet      $p_T$       &  --         &  $>40$ &  $>40$ & $>80$\\ 
& Fourth jet      $p_T$       &  --        &     --       &  $>40$ & $>80$\\ 
\hline
&$ \Delta\phi(\textrm{jet},\slashed{\mathbf{p}}_T)_\mathrm{min}$ & $>0.4$ & $>0.4$  & $>0.4$ & $>0.4$   \\ 
& $\slashed{E}_T/m_{\rm eff}$     & $>0.3$     & $>0.25$ & $> 0.25$ & $>0.2$ \\ 
& $m_{\rm eff}$   & $>1000$  &  $>1000$ & $>500/1000$ & $>1100$ \\ 
\hline
\end{tabular}
\caption{Cuts for the five search channels in the ATLAS Collaboration's $1.04~{\rm fb}^{-1}$ jets$+\slashed{E}_T$ search for Susy, lifted straight from the source~\cite{Aad:2011ib} ($m_{\rm eff}$, $\slashed{E}_T$ and $p_T$ in GeV).}
  \label{tab:signal_regions}
  \end{center}
\end{table}
\end{itemize}

\end{appendices}


\bibliographystyle{h-physrev3}
\clearpage
\bibliography{thesis}

\begin{thebibliography}{100}

\bibitem{'tHooft:1979bh}
G.~'t~Hooft,
\newblock NATO Adv.Study Inst.Ser.B Phys. {\bf 59}, 135 (1980).

\bibitem{naturalEWCBnotes}
R.~Kaul,
\newblock Naturalness and electro-weak symmetry breaking,
\newblock
  \url{http://theory.tifr.res.in/stringslhc/talks/naturalness-ewsb.pdf}.

\bibitem{Randall:1999ee}
L.~Randall and R.~Sundrum,
\newblock Phys.Rev.Lett. {\bf 83}, 3370 (1999), hep-ph/9905221.

\bibitem{Contino:2003ve}
R.~Contino, Y.~Nomura, and A.~Pomarol,
\newblock Nucl.Phys. {\bf B671}, 148 (2003), hep-ph/0306259.

\bibitem{Agashe:2004rs}
K.~Agashe, R.~Contino, and A.~Pomarol,
\newblock Nucl.Phys. {\bf B719}, 165 (2005), hep-ph/0412089.

\bibitem{Martin:1997ns}
S.~P. Martin,
\newblock (1997), hep-ph/9709356.

\bibitem{Drees:1996ca}
M.~Drees,
\newblock (1996), hep-ph/9611409.

\bibitem{Signer:2009dx}
A.~Signer,
\newblock J.Phys. {\bf G36}, 073002 (2009), 0905.4630.

\bibitem{PhysRev.159.1251}
S.~Coleman and J.~Mandula,
\newblock Phys. Rev. {\bf 159}, 1251 (1967).

\bibitem{Haag1975257}
R.~Haag, J.~T. \L{}opusza\'{n}ski, and M.~Sohnius,
\newblock Nuclear Physics B {\bf 88}, 257  (1975).

\bibitem{Boulware1979141}
D.~Boulware, S.~Deser, and J.~Kay,
\newblock Physica A: Statistical Mechanics and its Applications {\bf 96}, 141
  (1979).

\bibitem{Hack:2011yv}
T.-P. Hack and M.~Makedonski,
\newblock Phys.Lett. {\bf B718}, 1465 (2013), 1106.6327.

\bibitem{Bertone:2004pz}
G.~Bertone, D.~Hooper, and J.~Silk,
\newblock Phys.Rept. {\bf 405}, 279 (2005), hep-ph/0404175.

\bibitem{Peccei:1977hh}
R.~Peccei and H.~R. Quinn,
\newblock Phys.Rev.Lett. {\bf 38}, 1440 (1977).

\bibitem{Peccei:1977ur}
R.~Peccei and H.~R. Quinn,
\newblock Phys.Rev. {\bf D16}, 1791 (1977).

\bibitem{:2012gk}
ATLAS Collaboration, G.~Aad {\em et~al.},
\newblock Phys.Lett. {\bf B716}, 1 (2012), 1207.7214.

\bibitem{:2012gu}
CMS Collaboration, S.~Chatrchyan {\em et~al.},
\newblock Phys.Lett. {\bf B716}, 30 (2012), 1207.7235.

\bibitem{Giudice:2011cg}
G.~F. Giudice and A.~Strumia,
\newblock Nucl.Phys. {\bf B858}, 63 (2012), 1108.6077.

\bibitem{Giudice:1998xp}
G.~F. Giudice, M.~A. Luty, H.~Murayama, and R.~Rattazzi,
\newblock JHEP {\bf 9812}, 027 (1998), hep-ph/9810442.

\bibitem{Wells:2003tf}
J.~D. Wells,
\newblock (2003), hep-ph/0306127.

\bibitem{ArkaniHamed:2004fb}
N.~Arkani-Hamed and S.~Dimopoulos,
\newblock JHEP {\bf 0506}, 073 (2005), hep-th/0405159.

\bibitem{Giudice:2004tc}
G.~Giudice and A.~Romanino,
\newblock Nucl.Phys. {\bf B699}, 65 (2004), hep-ph/0406088.

\bibitem{Fox:2005yp}
P.~J. Fox {\em et~al.},
\newblock (2005), hep-th/0503249.

\bibitem{Fayet:1974jb}
P.~Fayet and J.~Iliopoulos,
\newblock Phys.Lett. {\bf B51}, 461 (1974).

\bibitem{Dimopoulos1981150}
S.~Dimopoulos and H.~Georgi,
\newblock Nuclear Physics B {\bf 193}, 150  (1981).

\bibitem{SakaiMSSM}
N.~Sakai,
\newblock Zeitschrift für Physik C Particles and Fields {\bf 11}, 153 (1981).

\bibitem{Bechtle:2012jw}
P.~Bechtle {\em et~al.},
\newblock (2012), 1211.1955.

\bibitem{Arbey:2012bp}
A.~Arbey, M.~Battaglia, A.~Djouadi, and F.~Mahmoudi,
\newblock (2012), 1211.4004.

\bibitem{Bechtle:2013gu}
P.~Bechtle {\em et~al.},
\newblock (2013), 1301.2345.

\bibitem{PhysRevD.20.403}
S.~Ferrara, L.~Girardello, and F.~Palumbo,
\newblock Phys. Rev. D {\bf 20}, 403 (1979).

\bibitem{Barbieri198863}
R.~Barbieri and G.~Giudice,
\newblock Nuclear Physics B {\bf 306}, 63  (1988).

\bibitem{Allanach:2012qd}
B.~Allanach and M.~Parker,
\newblock (2012), 1211.3231.

\bibitem{Kitano:2005wc}
R.~Kitano and Y.~Nomura,
\newblock Phys.Lett. {\bf B631}, 58 (2005), hep-ph/0509039.

\bibitem{Papucci:2011wy}
M.~Papucci, J.~T. Ruderman, and A.~Weiler,
\newblock JHEP {\bf 1209}, 035 (2012), 1110.6926.

\bibitem{ATLAS-CONF-2012-109}
CERN Report No. ATLAS-CONF-2012-109, 2012 (unpublished).

\bibitem{Dreiner:2012gx}
H.~K. Dreiner, M.~Kramer, and J.~Tattersall,
\newblock Europhys.Lett. {\bf 99}, 61001 (2012), 1207.1613.

\bibitem{Belanger:2012mk}
G.~Belanger, M.~Heikinheimo, and V.~Sanz,
\newblock JHEP {\bf 1208}, 151 (2012), 1205.1463.

\bibitem{Allanach:2012vj}
B.~Allanach and B.~Gripaios,
\newblock JHEP {\bf 1205}, 062 (2012), 1202.6616.

\bibitem{Hedri:2013pvl}
S.~E. Hedri, A.~Hook, M.~Jankowiak, and J.~G. Wacker,
\newblock (2013), 1302.1870.

\bibitem{Curtin:2012rm}
D.~Curtin, R.~Essig, and B.~Shuve,
\newblock (2012), 1210.5523.

\bibitem{Thaler:2010tr}
J.~Thaler and K.~Van~Tilburg,
\newblock JHEP {\bf 1103}, 015 (2011), 1011.2268.

\bibitem{Gallicchio:2010sw}
J.~Gallicchio and M.~D. Schwartz,
\newblock Phys.Rev.Lett. {\bf 105}, 022001 (2010), 1001.5027.

\bibitem{Kribs:2012gx}
G.~D. Kribs and A.~Martin,
\newblock Phys.Rev. {\bf D85}, 115014 (2012), 1203.4821.

\bibitem{Chatrchyan:2013lya}
CMS Collaboration, S.~Chatrchyan {\em et~al.},
\newblock (2013), 1303.2985.

\bibitem{CMS-PAS-SUS-12-023}
CERN Report No. CMS-PAS-SUS-12-023, 2012 (unpublished).

\bibitem{ATLAS-CONF-2013-037}
CERN Report No. ATLAS-CONF-2013-037, 2013 (unpublished).

\bibitem{ATLAS-CONF-2012-166}
CERN Report No. ATLAS-CONF-2012-166, 2012 (unpublished).

\bibitem{Dimopoulos:1995mi}
S.~Dimopoulos and G.~Giudice,
\newblock Phys.Lett. {\bf B357}, 573 (1995), hep-ph/9507282.

\bibitem{Cohen:1996vb}
A.~G. Cohen, D.~Kaplan, and A.~Nelson,
\newblock Phys.Lett. {\bf B388}, 588 (1996), hep-ph/9607394.

\bibitem{Brust:2011tb}
C.~Brust, A.~Katz, S.~Lawrence, and R.~Sundrum,
\newblock JHEP {\bf 1203}, 103 (2012).

\bibitem{Craig:2012di}
N.~Craig, M.~McCullough, and J.~Thaler,
\newblock JHEP {\bf 1206}, 046 (2012), 1203.1622.

\bibitem{Craig:2012hc}
N.~Craig, S.~Dimopoulos, and T.~Gherghetta,
\newblock JHEP {\bf 1204}, 116 (2012), 1203.0572.

\bibitem{Wymant:2012zp}
C.~Wymant,
\newblock Phys.Rev. {\bf D86}, 115023 (2012), 1208.1737.

\bibitem{Hall:2011aa}
L.~J. Hall, D.~Pinner, and J.~T. Ruderman,
\newblock JHEP {\bf 1204}, 131 (2012), 1112.2703.

\bibitem{Baer:2011ab}
H.~Baer, V.~Barger, and A.~Mustafayev,
\newblock (2011), 1112.3017.

\bibitem{Heinemeyer:2011aa}
S.~Heinemeyer, O.~Stal, and G.~Weiglein,
\newblock Phys.Lett. {\bf B710}, 201 (2012), 1112.3026.

\bibitem{Arbey:2011ab}
A.~Arbey, M.~Battaglia, A.~Djouadi, F.~Mahmoudi, and J.~Quevillon,
\newblock Phys.Lett. {\bf B708}, 162 (2012), 1112.3028.

\bibitem{Draper:2011aa}
P.~Draper, P.~Meade, M.~Reece, and D.~Shih,
\newblock Phys.Rev. {\bf D85}, 095007 (2012), 1112.3068.

\bibitem{Carena:2011aa}
M.~Carena, S.~Gori, N.~R. Shah, and C.~E. Wagner,
\newblock JHEP {\bf 1203}, 014 (2012), 1112.3336.

\bibitem{Cao:2011sn}
J.~Cao, Z.~Heng, D.~Li, and J.~M. Yang,
\newblock Phys.Lett. {\bf B710}, 665 (2012), 1112.4391.

\bibitem{Kang:2012sy}
Z.~Kang, J.~Li, and T.~Li,
\newblock JHEP {\bf 1211}, 024 (2012), 1201.5305.

\bibitem{Desai:2012qy}
N.~Desai, B.~Mukhopadhyaya, and S.~Niyogi,
\newblock (2012), 1202.5190.

\bibitem{Cao:2012fz}
J.-J. Cao, Z.-X. Heng, J.~M. Yang, Y.-M. Zhang, and J.-Y. Zhu,
\newblock JHEP {\bf 1203}, 086 (2012), 1202.5821.

\bibitem{Lee:2012sy}
H.~M. Lee, V.~Sanz, and M.~Trott,
\newblock JHEP {\bf 1205}, 139 (2012), 1204.0802.

\bibitem{Christensen:2012ei}
N.~D. Christensen, T.~Han, and S.~Su,
\newblock Phys.Rev. {\bf D85}, 115018 (2012), 1203.3207.

\bibitem{Brummer:2012ns}
F.~Brummer, S.~Kraml, and S.~Kulkarni,
\newblock JHEP {\bf 1208}, 089 (2012), 1204.5977.

\bibitem{Badziak:2012rf}
M.~Badziak, E.~Dudas, M.~Olechowski, and S.~Pokorski,
\newblock (2012), 1205.1675.

\bibitem{CahillRowley:2012rv}
M.~W. Cahill-Rowley, J.~L. Hewett, A.~Ismail, and T.~G. Rizzo,
\newblock (2012), 1206.5800.

\bibitem{Arbey:2012dq}
A.~Arbey, M.~Battaglia, A.~Djouadi, and F.~Mahmoudi,
\newblock (2012), 1207.1348.

\bibitem{Baer:2012up}
H.~Baer, V.~Barger, P.~Huang, A.~Mustafayev, and X.~Tata,
\newblock (2012), 1207.3343.

\bibitem{Antusch:2012gv}
S.~Antusch, L.~Calibbi, V.~Maurer, M.~Monaco, and M.~Spinrath,
\newblock (2012), 1207.7236.

\bibitem{Martin:1993zk}
S.~P. Martin and M.~T. Vaughn,
\newblock Phys.Rev. {\bf D50}, 2282 (1994), hep-ph/9311340.

\bibitem{Meade:2008wd}
P.~Meade, N.~Seiberg, and D.~Shih,
\newblock Prog.Theor.Phys.Suppl. {\bf 177}, 143 (2009), 0801.3278.

\bibitem{Gamberini1990331}
G.~Gamberini, G.~Ridolfi, and F.~Zwirner,
\newblock Nuclear Physics B {\bf 331}, 331  (1990).

\bibitem{PhysRevD.46.3981}
R.~Arnowitt and P.~Nath,
\newblock Phys. Rev. D {\bf 46}, 3981 (1992).

\bibitem{deCarlos:1993yy}
B.~de~Carlos and J.~Casas,
\newblock Phys.Lett. {\bf B309}, 320 (1993), hep-ph/9303291.

\bibitem{Carena:1995bx}
M.~S. Carena, J.~Espinosa, M.~Quiros, and C.~Wagner,
\newblock Phys.Lett. {\bf B355}, 209 (1995), hep-ph/9504316.

\bibitem{Brignole:1992uf}
A.~Brignole,
\newblock Phys.Lett. {\bf B281}, 284 (1992).

\bibitem{Frank:2006yh}
M.~Frank {\em et~al.},
\newblock Journal of High Energy Physics {\bf 2007}, 047 (2007).

\bibitem{Degrassi:2002fi}
G.~Degrassi, S.~Heinemeyer, W.~Hollik, P.~Slavich, and G.~Weiglein,
\newblock Eur.Phys.J. {\bf C28}, 133 (2003), hep-ph/0212020.

\bibitem{Heinemeyer:1998np}
S.~Heinemeyer, W.~Hollik, and G.~Weiglein,
\newblock Eur.Phys.J. {\bf C9}, 343 (1999), hep-ph/9812472.

\bibitem{Heinemeyer:1998yj}
S.~Heinemeyer, W.~Hollik, and G.~Weiglein,
\newblock Comput.Phys.Commun. {\bf 124}, 76 (2000), hep-ph/9812320.

\bibitem{ATLAS-CONF-2013-014}
CERN Report No. ATLAS-CONF-2013-014, 2013 (unpublished).

\bibitem{CMS-PAS-HIG-12-045}
CERN Report No. CMS-PAS-HIG-12-045, 2012 (unpublished).

\bibitem{Degrassi:2012ry}
G.~Degrassi {\em et~al.},
\newblock (2012), 1205.6497.

\bibitem{Alekhin:2012py}
S.~Alekhin, A.~Djouadi, and S.~Moch,
\newblock (2012), 1207.0980.

\bibitem{Bezrukov:2012sa}
F.~Bezrukov, M.~Y. Kalmykov, B.~A. Kniehl, and M.~Shaposhnikov,
\newblock (2012), 1205.2893.

\bibitem{Allanach:2004rh}
B.~Allanach, A.~Djouadi, J.~Kneur, W.~Porod, and P.~Slavich,
\newblock JHEP {\bf 0409}, 044 (2004), hep-ph/0406166.

\bibitem{Haisch:2012re}
U.~Haisch and F.~Mahmoudi,
\newblock JHEP {\bf 1301}, 061 (2013), 1210.7806.

\bibitem{Kang:2012ra}
Z.~Kang, T.~Li, T.~Liu, C.~Tong, and J.~M. Yang,
\newblock (2012), 1203.2336.

\bibitem{Craig:2012xp}
N.~Craig, S.~Knapen, D.~Shih, and Y.~Zhao,
\newblock (2012), 1206.4086.

\bibitem{Agashe:2012zq}
K.~Agashe, Y.~Cui, and R.~Franceschini,
\newblock (2012), 1209.2115.

\bibitem{PhysRevD.86.035023}
D.~Albornoz~V\'asquez {\em et~al.},
\newblock Phys. Rev. D {\bf 86}, 035023 (2012).

\bibitem{Ellwanger:2009dp}
U.~Ellwanger, C.~Hugonie, and A.~M. Teixeira,
\newblock Phys.Rept. {\bf 496}, 1 (2010), 0910.1785.

\bibitem{Vilenkin:1984ib}
A.~Vilenkin,
\newblock Phys.Rept. {\bf 121}, 263 (1985).

\bibitem{Barbieri:2006bg}
R.~Barbieri, L.~J. Hall, Y.~Nomura, and V.~S. Rychkov,
\newblock Phys.Rev. {\bf D75}, 035007 (2007), hep-ph/0607332.

\bibitem{Gherghetta:2012gb}
T.~Gherghetta, B.~von Harling, A.~D. Medina, and M.~A. Schmidt,
\newblock JHEP {\bf 1302}, 032 (2013), 1212.5243.

\bibitem{Vasquez:2010ru}
D.~A. Vasquez, G.~Belanger, C.~Boehm, A.~Pukhov, and J.~Silk,
\newblock Phys.Rev. {\bf D82}, 115027 (2010), 1009.4380.

\bibitem{AlbornozVasquez:2011js}
D.~Albornoz~Vasquez, G.~Belanger, and C.~Boehm,
\newblock Phys.Rev. {\bf D84}, 095008 (2011), 1107.1614.

\bibitem{AlbornozVasquez:2012px}
D.~Albornoz~Vasquez, G.~Belanger, J.~Billard, and F.~Mayet,
\newblock Phys.Rev. {\bf D85}, 055023 (2012), 1201.6150.

\bibitem{Aalseth:2010vx}
CoGeNT collaboration, C.~Aalseth {\em et~al.},
\newblock Phys.Rev.Lett. {\bf 106}, 131301 (2011), 1002.4703.

\bibitem{Bernabei:2010mq}
DAMA Collaboration, LIBRA Collaboration, R.~Bernabei {\em et~al.},
\newblock Eur.Phys.J. {\bf C67}, 39 (2010), 1002.1028.

\bibitem{Belanger:2010gh}
G.~Belanger {\em et~al.},
\newblock Comput.Phys.Commun. {\bf 182}, 842 (2011), 1004.1092.

\bibitem{Ellwanger:2006rn}
U.~Ellwanger and C.~Hugonie,
\newblock Comput.Phys.Commun. {\bf 177}, 399 (2007), hep-ph/0612134.

\bibitem{Ellwanger:2004xm}
U.~Ellwanger, J.~F. Gunion, and C.~Hugonie,
\newblock JHEP {\bf 0502}, 066 (2005), hep-ph/0406215.

\bibitem{Ellwanger:2005dv}
U.~Ellwanger and C.~Hugonie,
\newblock Comput.Phys.Commun. {\bf 175}, 290 (2006), hep-ph/0508022.

\bibitem{Komatsu:2008hk}
WMAP Collaboration, E.~Komatsu {\em et~al.},
\newblock Astrophys.J.Suppl. {\bf 180}, 330 (2009), 0803.0547.

\bibitem{Skordis:2005xk}
C.~Skordis, D.~Mota, P.~Ferreira, and C.~Boehm,
\newblock Phys.Rev.Lett. {\bf 96}, 011301 (2006), astro-ph/0505519.

\bibitem{Aprile:2011hi}
XENON100 Collaboration, E.~Aprile {\em et~al.},
\newblock Phys.Rev.Lett. {\bf 107}, 131302 (2011), 1104.2549.

\bibitem{Abdo:2010ex}
Fermi-LAT Collaboration, A.~Abdo {\em et~al.},
\newblock Astrophys.J. {\bf 712}, 147 (2010), 1001.4531.

\bibitem{Strigari:2006rd}
L.~E. Strigari, S.~M. Koushiappas, J.~S. Bullock, and M.~Kaplinghat,
\newblock Phys.Rev. {\bf D75}, 083526 (2007), astro-ph/0611925.

\bibitem{Boehm:2002yz}
C.~Boehm, T.~Ensslin, and J.~Silk,
\newblock J.Phys. {\bf G30}, 279 (2004), astro-ph/0208458.

\bibitem{Boehm:2010kg}
C.~Boehm, J.~Silk, and T.~Ensslin,
\newblock (2010), 1008.5175.

\bibitem{Grellscheid:2011ij}
D.~Grellscheid, J.~Jaeckel, V.~V. Khoze, P.~Richardson, and C.~Wymant,
\newblock JHEP {\bf 1203}, 078 (2012), 1111.3365.

\bibitem{Beenakker:1996ed}
W.~Beenakker, R.~Hopker, and M.~Spira,
\newblock (1996), hep-ph/9611232.

\bibitem{Beenakker:1996ch}
W.~Beenakker, R.~Hopker, M.~Spira, and P.~Zerwas,
\newblock Nucl.Phys. {\bf B492}, 51 (1997), hep-ph/9610490.

\bibitem{Beenakker:1999xh}
W.~Beenakker {\em et~al.},
\newblock Phys.Rev.Lett. {\bf 83}, 3780 (1999), hep-ph/9906298.

\bibitem{Spira:2002rd}
M.~Spira,
\newblock p. 217 (2002), hep-ph/0211145.

\bibitem{Plehn:2004rp}
T.~Plehn,
\newblock Czech.J.Phys. {\bf 55}, B213 (2005), hep-ph/0410063.

\bibitem{Bahr:2008pv}
M.~Bahr {\em et~al.},
\newblock Eur.Phys.J. {\bf C58}, 639 (2008), 0803.0883.

\bibitem{Gieseke:2011na}
S.~Gieseke {\em et~al.},
\newblock (2011), 1102.1672.

\bibitem{Buckley:2010ar}
A.~Buckley {\em et~al.},
\newblock (2010), 1003.0694.

\bibitem{Das:2012rr}
D.~Das, U.~Ellwanger, and A.~M. Teixeira,
\newblock JHEP {\bf 1204}, 067 (2012), 1202.5244.

\bibitem{PhysRevD.87.074004}
M.~Spannowsky and C.~Wymant,
\newblock Phys. Rev. D {\bf 87}, 074004 (2013), 1301.0345.

\bibitem{Ellwanger:2011aa}
U.~Ellwanger,
\newblock JHEP {\bf 1203}, 044 (2012), 1112.3548.

\bibitem{King:2012is}
S.~King, M.~Muhlleitner, and R.~Nevzorov,
\newblock Nucl.Phys. {\bf B860}, 207 (2012), 1201.2671.

\bibitem{Gunion:2012zd}
J.~F. Gunion, Y.~Jiang, and S.~Kraml,
\newblock Phys.Lett. {\bf B710}, 454 (2012), 1201.0982.

\bibitem{Brooijmans:2012yi}
G.~Brooijmans {\em et~al.},
\newblock (2012), 1203.1488.

\bibitem{Bechtle:2008jh}
P.~Bechtle, O.~Brein, S.~Heinemeyer, G.~Weiglein, and K.~E. Williams,
\newblock Comput.Phys.Commun. {\bf 181}, 138 (2010), 0811.4169.

\bibitem{Bechtle:2011sb}
P.~Bechtle, O.~Brein, S.~Heinemeyer, G.~Weiglein, and K.~E. Williams,
\newblock Comput.Phys.Commun. {\bf 182}, 2605 (2011), 1102.1898.

\bibitem{Me}
\url{http://www.ippp.dur.ac.uk/~hndv85/}.

\bibitem{Lisanti:2009uy}
M.~Lisanti and J.~G. Wacker,
\newblock Phys.Rev. {\bf D79}, 115006 (2009), 0903.1377.

\bibitem{Bellazzini:2009xt}
B.~Bellazzini, C.~Csaki, A.~Falkowski, and A.~Weiler,
\newblock Phys.Rev. {\bf D80}, 075008 (2009), 0906.3026.

\bibitem{Falkowski:2010hi}
A.~Falkowski, D.~Krohn, L.-T. Wang, J.~Shelton, and A.~Thalapillil,
\newblock Phys.Rev. {\bf D84}, 074022 (2011), 1006.1650.

\bibitem{Chen:2010wk}
C.-R. Chen, M.~M. Nojiri, and W.~Sreethawong,
\newblock JHEP {\bf 1011}, 012 (2010), 1006.1151.

\bibitem{Forshaw:2007ra}
J.~Forshaw, J.~Gunion, L.~Hodgkinson, A.~Papaefstathiou, and A.~Pilkington,
\newblock JHEP {\bf 0804}, 090 (2008), 0712.3510.

\bibitem{Englert:2011iz}
C.~Englert, T.~S. Roy, and M.~Spannowsky,
\newblock Phys.Rev. {\bf D84}, 075026 (2011), 1106.4545.

\bibitem{Lewis:2012pf}
I.~Lewis and J.~Schmitthenner,
\newblock JHEP {\bf 1206}, 072 (2012), 1203.5174.

\bibitem{Englert:2012wf}
C.~Englert, M.~Spannowsky, and C.~Wymant,
\newblock Phys.Lett. {\bf B718}, 538 (2012), 1209.0494.

\bibitem{Hall:1981bc}
L.~J. Hall and M.~B. Wise,
\newblock Nucl.Phys. {\bf B187}, 397 (1981).

\bibitem{Ellwanger:2010nf}
U.~Ellwanger,
\newblock Phys.Lett. {\bf B698}, 293 (2011), 1012.1201.

\bibitem{ATLAS-CONF-2013-012}
CERN Report No. ATLAS-CONF-2013-012, 2013 (unpublished).

\bibitem{Marumi}
M.~Kado,
\newblock Status of atlas higgs results,
\newblock
  \url{https://lpsc.in2p3.fr/Indico/getFile.py/access?contribId=0&sessionId=0&%
resId=0&materialId=slides&confId=861}.

\bibitem{CMS-PAS-HIG-13-001}
CERN Report No. CMS-PAS-HIG-13-001, 2013 (unpublished).

\bibitem{Belanger:2008nt}
G.~Belanger, C.~Hugonie, and A.~Pukhov,
\newblock JCAP {\bf 0901}, 023 (2009), 0811.3224.

\bibitem{Nelson:1993nf}
A.~E. Nelson and N.~Seiberg,
\newblock Nucl.Phys. {\bf B416}, 46 (1994), hep-ph/9309299.

\bibitem{Shih:2007av}
D.~Shih,
\newblock JHEP {\bf 0802}, 091 (2008), hep-th/0703196.

\bibitem{O'Raifeartaigh:1975pr}
L.~O'Raifeartaigh,
\newblock Nucl.Phys. {\bf B96}, 331 (1975).

\bibitem{Kang:2012fn}
Z.~Kang, T.~Li, and Z.~Sun,
\newblock (2012), 1209.1059.

\bibitem{Jaeckel200983c}
J.~Jaeckel,
\newblock Nuclear Physics A {\bf 820}, 83c  (2009).

\bibitem{Jaeckel:2010ni}
J.~Jaeckel and A.~Ringwald,
\newblock Ann.Rev.Nucl.Part.Sci. {\bf 60}, 405 (2010), 1002.0329.

\bibitem{Beringer:1900zz}
Particle Data Group, J.~Beringer {\em et~al.},
\newblock Phys.Rev. {\bf D86}, 010001 (2012).

\bibitem{Intriligator:2006dd}
K.~A. Intriligator, N.~Seiberg, and D.~Shih,
\newblock JHEP {\bf 0604}, 021 (2006), hep-th/0602239.

\bibitem{Buican:2008ws}
M.~Buican, P.~Meade, N.~Seiberg, and D.~Shih,
\newblock JHEP {\bf 03}, 016 (2009), 0812.3668.

\bibitem{Cheung:2007es}
C.~Cheung, A.~L. Fitzpatrick, and D.~Shih,
\newblock JHEP {\bf 07}, 054 (2008), 0710.3585.

\bibitem{Dine:1993yw}
M.~Dine and A.~E. Nelson,
\newblock Phys.Rev. {\bf D48}, 1277 (1993), hep-ph/9303230.

\bibitem{Dine:1994vc}
M.~Dine, A.~E. Nelson, and Y.~Shirman,
\newblock Phys.Rev. {\bf D51}, 1362 (1995), hep-ph/9408384.

\bibitem{Dine:1995ag}
M.~Dine, A.~E. Nelson, Y.~Nir, and Y.~Shirman,
\newblock Phys.Rev. {\bf D53}, 2658 (1996), hep-ph/9507378.

\bibitem{Giudice:1997ni}
G.~Giudice and R.~Rattazzi,
\newblock Nucl.Phys. {\bf B511}, 25 (1998), hep-ph/9706540.

\bibitem{Dimopoulos:1996ig}
S.~Dimopoulos and G.~Giudice,
\newblock Phys.Lett. {\bf B393}, 72 (1997), hep-ph/9609344.

\bibitem{Giudice:1998bp}
G.~Giudice and R.~Rattazzi,
\newblock Phys.Rept. {\bf 322}, 419 (1999), hep-ph/9801271.

\bibitem{PhysRevD.79.035002}
L.~M. Carpenter, M.~Dine, G.~Festuccia, and J.~D. Mason,
\newblock Phys. Rev. D {\bf 79}, 035002 (2009).

\bibitem{Carpenter:2008he}
L.~M. Carpenter,
\newblock (2008), 0812.2051.

\bibitem{Carone:1995kp}
C.~D. Carone and H.~Murayama,
\newblock Phys.Rev. {\bf D53}, 1658 (1996), hep-ph/9510219.

\bibitem{Jaeckel:2011ma}
J.~Jaeckel, V.~V. Khoze, and C.~Wymant,
\newblock JHEP {\bf 1104}, 126 (2011), 1102.1589.

\bibitem{Jaeckel:2011qj}
J.~Jaeckel, V.~V. Khoze, and C.~Wymant,
\newblock JHEP {\bf 1105}, 132 (2011), 1103.1843.

\bibitem{Allanach:2001kg}
B.~Allanach,
\newblock Comput.Phys.Commun. {\bf 143}, 305 (2002), hep-ph/0104145.

\bibitem{Carena:2010gr}
M.~Carena, P.~Draper, N.~R. Shah, and C.~E. Wagner,
\newblock Phys.Rev. {\bf D82}, 075005 (2010), 1006.4363.

\bibitem{ATLAS:1999vwa}
{ATLAS Collaboration},
\newblock {ATLAS: Detector and physics performance technical design report.
  Volume 2}, 1999.

\bibitem{FridmanRojas:2012yh}
I.~Fridman-Rojas and P.~Richardson,
\newblock (2012), 1208.0279.

\bibitem{Englert:1964et}
F.~Englert and R.~Brout,
\newblock Phys.Rev.Lett. {\bf 13}, 321 (1964).

\bibitem{Higgs:1964ia}
P.~W. Higgs,
\newblock Phys.Lett. {\bf 12}, 132 (1964).

\bibitem{Higgs:1964pj}
P.~W. Higgs,
\newblock Phys.Rev.Lett. {\bf 13}, 508 (1964).

\bibitem{Deser:1977uq}
S.~Deser and B.~Zumino,
\newblock Phys.Rev.Lett. {\bf 38}, 1433 (1977).

\bibitem{Cremmer:1978iv}
E.~Cremmer {\em et~al.},
\newblock Phys.Lett. {\bf B79}, 231 (1978).

\bibitem{Abel:2010vba}
S.~Abel, M.~J. Dolan, J.~Jaeckel, and V.~V. Khoze,
\newblock JHEP {\bf 1012}, 049 (2010), 1009.1164.

\bibitem{Aad:2011ib}
ATLAS Collaboration, G.~Aad {\em et~al.},
\newblock Phys.Lett. {\bf B710}, 67 (2012), 1109.6572.

\bibitem{Barate:2003sz}
LEP Working Group for Higgs boson searches, ALEPH Collaboration, DELPHI
  Collaboration, L3 Collaboration, OPAL Collaboration, R.~Barate {\em et~al.},
\newblock Phys.Lett. {\bf B565}, 61 (2003), hep-ex/0306033.

\bibitem{Akula:2011zq}
S.~Akula {\em et~al.},
\newblock Phys.Lett. {\bf B699}, 377 (2011), 1103.1197.

\bibitem{Jaeckel:2011wp}
J.~Jaeckel, V.~V. Khoze, T.~Plehn, and P.~Richardson,
\newblock Phys.Rev. {\bf D85}, 015015 (2012), 1109.2072.

\bibitem{Dolan:2011ie}
M.~J. Dolan, D.~Grellscheid, J.~Jaeckel, V.~V. Khoze, and P.~Richardson,
\newblock JHEP {\bf 1106}, 095 (2011), 1104.0585.

\bibitem{TheGrid}
\url{http://www.phenogrid.dur.ac.uk/}.

\bibitem{Cacciari:2011ma}
M.~Cacciari, G.~P. Salam, and G.~Soyez,
\newblock Eur.Phys.J. {\bf C72}, 1896 (2012), 1111.6097.

\bibitem{Grajek:2013ola}
P.~Grajek, A.~Mariotti, and D.~Redigolo,
\newblock (2013), 1303.0870.

\bibitem{Dermisek:2006ey}
R.~Dermisek and H.~D. Kim,
\newblock Phys.Rev.Lett. {\bf 96}, 211803 (2006), hep-ph/0601036.

\bibitem{Ellis:2008mc}
J.~R. Ellis, J.~Giedt, O.~Lebedev, K.~Olive, and M.~Srednicki,
\newblock Phys.Rev. {\bf D78}, 075006 (2008), 0806.3648.

\bibitem{Barger:1987re}
V.~D. Barger, T.~Han, and J.~Ohnemus,
\newblock Phys.Rev. {\bf D37}, 1174 (1988).

\bibitem{Lester:1999tx}
C.~Lester and D.~Summers,
\newblock Phys.Lett. {\bf B463}, 99 (1999), hep-ph/9906349.

\bibitem{Rogan:2010kb}
C.~Rogan,
\newblock (2010), 1006.2727.

\bibitem{Allanach:2000kt}
B.~Allanach, C.~Lester, M.~A. Parker, and B.~Webber,
\newblock JHEP {\bf 0009}, 004 (2000), hep-ph/0007009.

\bibitem{Han:2005mu}
T.~Han,
\newblock p. 407 (2005), hep-ph/0508097.

\bibitem{Kats:2011qh}
Y.~Kats, P.~Meade, M.~Reece, and D.~Shih,
\newblock JHEP {\bf 1202}, 115 (2012), 1110.6444.

\bibitem{Cheung:2010mc}
C.~Cheung, Y.~Nomura, and J.~Thaler,
\newblock JHEP {\bf 1003}, 073 (2010), 1002.1967.

\bibitem{Argurio:2011hs}
R.~Argurio, Z.~Komargodski, and A.~Mariotti,
\newblock Phys.Rev.Lett. {\bf 107}, 061601 (2011), 1102.2386.

\bibitem{Argurio:2011gu}
R.~Argurio {\em et~al.},
\newblock JHEP {\bf 1206}, 096 (2012), 1112.5058.

\bibitem{Baryakhtar:2012rz}
M.~Baryakhtar, N.~Craig, and K.~Van~Tilburg,
\newblock JHEP {\bf 1207}, 164 (2012), 1206.0751.

\bibitem{Macesanu:2002db}
C.~Macesanu, C.~McMullen, and S.~Nandi,
\newblock Phys.Rev. {\bf D66}, 015009 (2002), hep-ph/0201300.

\bibitem{Cohen:2011aa}
T.~Cohen, A.~Hook, and B.~Wecht,
\newblock Phys.Rev. {\bf D85}, 115004 (2012), 1112.1699.

\bibitem{Plehn:1999xi}
T.~Plehn, D.~L. Rainwater, and D.~Zeppenfeld,
\newblock Phys.Rev. {\bf D61}, 093005 (2000), hep-ph/9911385.

\bibitem{Goncalves-Netto:2013nla}
D.~Goncalves-Netto, D.~Lopez-Val, K.~Mawatari, I.~Wigmore, and T.~Plehn,
\newblock (2013), 1303.0845.

\bibitem{CMSsearch}
{CMS Collaboration},
\newblock Physics Analysis Summary CMS-PAS-SUS-12-018.

\bibitem{Alwall:2011uj}
J.~Alwall, M.~Herquet, F.~Maltoni, O.~Mattelaer, and T.~Stelzer,
\newblock JHEP {\bf 1106}, 128 (2011), 1106.0522.

\bibitem{Sjostrand:2006za}
T.~Sjostrand, S.~Mrenna, and P.~Z. Skands,
\newblock JHEP {\bf 0605}, 026 (2006), hep-ph/0603175.

\bibitem{PGS}
{J. Conway {\it et al}},
\newblock
  \url{http://www.physics.ucdavis.edu/~conway/research/software/pgs/pgs4-gener%
al.htm}.

\bibitem{Cacciari:2008gp}
M.~Cacciari, G.~P. Salam, and G.~Soyez,
\newblock JHEP {\bf 0804}, 063 (2008), 0802.1189.

\bibitem{Aad:2012zza}
ATLAS Collaboration, G.~Aad {\em et~al.},
\newblock Phys.Lett. {\bf B718}, 411 (2012), 1209.0753.

\bibitem{Lester:2011nj}
C.~G. Lester,
\newblock JHEP {\bf 1105}, 076 (2011), 1103.5682.

\bibitem{Kribs:2009yh}
G.~D. Kribs, A.~Martin, T.~S. Roy, and M.~Spannowsky,
\newblock Phys.Rev. {\bf D81}, 111501 (2010), 0912.4731.

\bibitem{Alwall:2007fs}
J.~Alwall {\em et~al.},
\newblock Eur.Phys.J. {\bf C53}, 473 (2008), 0706.2569.

\bibitem{Schmaltz:2005ky}
M.~Schmaltz and D.~Tucker-Smith,
\newblock Ann.Rev.Nucl.Part.Sci. {\bf 55}, 229 (2005), hep-ph/0502182.

\bibitem{Contino:2010rs}
R.~Contino,
\newblock (2010), 1005.4269.

\bibitem{Cheng:2007bu}
H.-C. Cheng,
\newblock (2007), 0710.3407.

\bibitem{Batra:2004ah}
P.~Batra and D.~E. Kaplan,
\newblock JHEP {\bf 0503}, 028 (2005), hep-ph/0412267.

\bibitem{Ibanez1984511}
L.~Ibanez and C.~L\'{o}pez,
\newblock Nuclear Physics B {\bf 233}, 511  (1984).

\bibitem{Essig:2007kh}
R.~Essig and J.-F. Fortin,
\newblock JHEP {\bf 0804}, 073 (2008), 0709.0980.

\bibitem{Carena:1996km}
M.~S. Carena, P.~H. Chankowski, M.~Olechowski, S.~Pokorski, and C.~Wagner,
\newblock Nucl.Phys. {\bf B491}, 103 (1997), hep-ph/9612261.

\bibitem{Junk:1999kv}
T.~Junk,
\newblock Nucl.Instrum.Meth. {\bf A434}, 435 (1999), hep-ex/9902006.

\bibitem{Read:2002hq}
A.~L. Read,
\newblock J.Phys. {\bf G28}, 2693 (2002).

\end{thebibliography}

\end{document}